\documentclass[10pt]{iopart}
\pdfoutput=1
\usepackage{fullpage}
\usepackage{iopams}
\usepackage{setstack}
\usepackage{braket}
\usepackage[pdftex,breaklinks=true,bookmarks=false,colorlinks=true,citecolor=violet,urlcolor=blue,linkcolor=black]{hyperref}

\usepackage{leftidx}
\usepackage{tikz}
\usetikzlibrary{arrows}
\usetikzlibrary[shapes.geometric]
\usetikzlibrary{decorations}

\newcommand{\hilb}{\mathcal{H}}
\newcommand{\lbf}{-\log \left|\langle A\cup B|A\otimes B\rangle\right|^2}

\begin{document}
 
 \title[Logarithmic corrections to the free energy from sharp corners with angle $2\pi$]{Logarithmic corrections to the free energy from sharp corners with angle $2\pi$}
 %Semi-universal logarithmic corrections to the corner free energy, and quantum fidelity in critical one-dimensional systems
 \author{Jean-Marie St\'ephan$^{1}$ and J\'er\^ome Dubail$^{2}$}
 
 \address{$^1$ Physics Department, University of Virginia, Charlottesville, VA 22904-4714}
 \address{$^2$ Department of Physics, Yale University, P.O. Box 208120, New Haven, CT 06520-8120, USA}
 
 \eads{\mailto{jean-marie.stephan@virginia.edu}, \mailto{jerome.dubail@yale.edu}}

 %%%%%%%%%%%%%%%%%%%%%%%%%%%%%%%%%%%%%%%%%%%%%%%%%
\begin{abstract}
%%%%%%%%%%%%%%%%%%%%%%%%%%%%%%%%%%%%%%%%%%%%%%%%%
We study subleading corrections to the corner free energy in classical two-dimensional critical systems, focusing on a generic boundary perturbation by the stress-tensor of the underlying conformal field theory (CFT). In the particular case of an angle $2\pi$, we find that there is an unusual correction of the form $L^{-1}\log L$, where $L$ is a typical length scale in the system. This correction also affects the one-point function of an operator near the corner. The prefactor can be seen as semi-universal, in the sense that it depends on a \emph{single} non-universal quantity, the extrapolation length. Once this ultraviolet cutoff is known, the term is entirely fixed by the geometry of the system, and the central charge of the CFT. Such a corner appears for example in the bipartite fidelity of a one-dimensional quantum system at criticality, which allows for several numerical checks in free fermions systems. We also present an exact result in the XX and Ising chains that confirms this analysis. Finally, we 
consider applications to the time evolution of the (logarithmic) Loschmidt echo and the entanglement entropy following a local quantum quench. Due to subtle issues in analytic continuation, we find that the logarithmic term in imaginary time transforms into a time-dependent $L^{-2}$ correction for the entanglement entropy, and a $L^{-2}\log L$ term for the Loschmidt echo. 
 
%%%%%%%%%%%%%%%%%%%%%%%%%%%%%%%%%%%%%%%%%%%%%%%%%
\end{abstract}
%%%%%%%%%%%%%%%%%%%%%%%%%%%%%%%%%%%%%%%%%%%%%%%%%

\date{\today}
\maketitle

{\sffamily\tableofcontents}
\vfill\eject

\hypersetup{linkcolor=red}

\section{Introduction}
\label{sec:intro}

\subsection{Universality and free energies}

Phase transitions in statistical models at equilibrium are among the most studied phenomena in physics, with many important results that have accumulated over the XX$^{th}$ century. Our theoretical understanding of these phenomena largely relies on lattice models, some of which may be solved analytically \cite{Baxter}---or at least studied numerically---as well as on field theoretic results \cite{zinnjustin}. Critical points fall into different universality classes, which are caracterized by the large scale properties of the correlation functions of local observables. The free energy of these systems can also reveal interesting information about their underlying universality class \cite{Affleck_c,Cardy_c}, or about the renormalization group flows between different critical points  \cite{Zamo_c,CardyA,KomargodskiSchwimmer,Klebanov}. Quantitatively, our understanding of critical phenomena is most advanced in two dimensions, both from the lattice and the field theoretic points of view, thanks to lattice integrability \cite{Baxter} and conformal invariance in the continuum \cite{BPZ}.

An interesting development over the past twenty years has been the application of the techniques initially developed in the field of classical statistical models to {\it quantum} many-body systems. There, some quantities that are typically of interest are the ones that characterize the {\it entanglement} in the ground-state. Prominent examples of such quantities are the von Neumann entanglement entropy, the R\'enyi entropies \cite{Holzhey_ee,Vidal_ee,CalabreseCardy_ee}, as well as other types of fidelity measures \cite{ZanardiPaunkovic,venuti2009universal,Bipartite_fidelity,zanardi2012entanglement}. In many situations, thanks to the classical-to-quantum correspondence, such quantities, defined for the ground state of a quantum system in $d$ spatial dimensions, can be reformulated as free energies of a {\it classical} system in $d+1$ spatial dimensions \cite{Holzhey_ee,CalabreseCardy_ee,Bipartite_fidelity}. It turns out that these classical statistical systems are usually defined on domains that include conical singularities, and/or corners if the systems have boundaries. For the convenience of the reader, some of these aspects are reviewed in the \ref{sec:allthat}.

\begin{figure}[htbp]
	\begin{center}
		\includegraphics[width=0.7\textwidth]{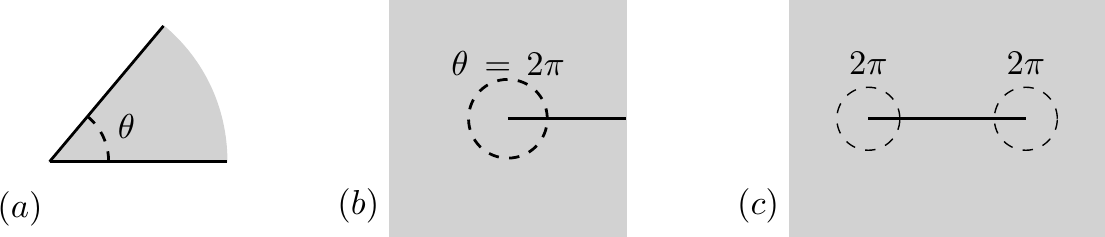}
	\end{center}
	\caption{Fluctuating degrees of freedom in the shaded area are described by a field theory. The boundary has sharp corners. (a) a corner with internal angle $\theta$. (b) the case $\theta = 2\pi$, which will be of special interest in that paper. (c) an example of a configuration with two sharp corners with angles $2\pi$: the two ends of a slit.}
	\label{fig:some_corners}
\end{figure}

Motivated by these recent developments, the purpose of this paper is to study classical statistical models with a boundary, in the presence of one or several corners (see e.g. Fig.~\ref{fig:some_corners}). For critical two-dimensional systems described by a conformal field theory, it has long been known \cite{CardyPeschel} that the free energy contains a piece of order $O(\log L)$, where $L$ is the typical size of the system, with a {\it universal} prefactor proportional to the central charge. This phenomenon has quantum-mechanical counterparts: in \cite{Bipartite_fidelity} we showed that the same universal term appears in the (logarithmic) {\it bipartite fidelity} (LBF in short) of the ground state of a quantum critical system with dynamic exponent $z=1$, or in non-equilibrium situations, after a local quantum quench (see also the \ref{sec:allthat}). We will come back to quantum systems and bipartite entanglement later; for now, let us formulate our arguments in the language of classical statistical systems at equilibrium.

The study of statistical systems in the presence of corner singularities is a topic that is interesting in its own right \cite{CardyPeschel,Rectangle,Rectangle2,Jesper,Jesper2,WuIzmailianGuo,KlebanVassileva,CardyIgloi}. In this paper, we focus on the subleading corrections to the corner free energies in two-dimensional critical systems, beyond the leading order of Cardy and Peschel \cite{CardyPeschel}. In general these subleading corrections are generated by irrelevant bulk or boundary perturbations to the action of the conformal field theory (CFT), and they take the form of $O(L^{-\alpha})$ terms, with some exponent $\alpha$ (not necessarily an integer). Here we focus on the arguably most generic boundary perturbation, which corresponds to the stress-tensor of the CFT itself. When the  corner has an internal angle $2\pi$, and \emph{only} in that case, we show that the stress-tensor generates an interesting subleading contribution to the corner free energy of order $O(L^{-1}
 \log L)$. A similar correction also appears in the correlation functions near the corner. Such corrections may be regarded as {\it semi-universal}: they are of the form
\begin{equation}
	\delta F_{\rm corner} \, = \, c ~ f_{\rm geom.} ~ \frac{\xi}{L} ~\log (L/a), 
\end{equation}
where $c$ is the central charge, $f_{\rm geom.}$ is a dimensionless geometric factor that is easily calculable (we will give a closed formula for this factor below). The only length scale that appears here, apart from the typical size of the system $L$ and the UV cutoff $a$ (lattice spacing), is the {\it extrapolation length} $\xi$. This extrapolation length is a concept that is ubiquitous in the study of surface critical phenomena \cite{Diehl}. The importance of this length has been emphasized lately in the context of global quantum quenches \cite{CC_globalquench}, entanglement spectra in fractional quantum hall systems \cite{DRR}, and also, although in a slightly different form, in quantum impurity problems such as the Kondo screening cloud \footnote{we argue below that the extrapolation length is nothing but a coupling constant associated with the stress-tensor, viewed as a boundary perturbation. As such, it is formally analogous to the ``size of the screening cloud'' in \cite{Kondoscreening,Eriksson_Johannesson_Kondo}, where these boundary perturbations were considered in the calculation of the entanglement entropy. Note also that, in a similar context, subleading corrections to the entanglement entropy of order $O(\log L/L)$ have been obtained in \cite{Corrections_ee2,Eriksson_Johannesson_Kondo}. Let us emphasize here that those corrections do not come from the stress-tensor, and are not corner singularities; instead, they are generated by an operator of scaling dimension $3/2$ \cite{Corrections_ee2}, when such an operator exists in the spectrum of the model. These corrections \cite{Corrections_ee2,Eriksson_Johannesson_Kondo} are {\it different} from the ones we study below, they will not appear in our analysis.}. Its appearance in logarithmic corrections to the corner free energy provides a practically reliable way of measuring this length in numerical simulations; it also affects various computable quantities in one-dimensional quantum systems, such as the bipartite fidelity \cite{Bipartite_fidelity} or the 
emptiness formation probability \cite{EFP}. Below, we will also explore the consequences of the appearance of this correction in time-dependent problems following a local quench, and we will provide extensive numerical and exact computations in support of our results in free fermion systems.

\subsection{Why is $O(L^{-1}\log L)$ interesting?}
\label{sec:why}
A simple way of seeing why such corrections are robust is to write down an expansion of the free energy, as a function of the dominant length scale $L$. In the following we will encounter several times the following form
\begin{equation}\label{eq:somefreenrj}
 F(L)=f_2 L^2+f_1 L +b_0 \log L+f_0+b_{-1} \frac{\log L}{L}+\frac{f_{-1}}{L}+o(1/L).
\end{equation}
$f_2$ and $f_1$ are the bulk and surface free energies in two dimensions. They are sensitive to the microscopic details of the lattice model. Universal properties, associated to the correlations at large scales, appear in the subleading corrections. Typically, in applications to entanglement measures in quantum systems, the contribution of the extensive part of the free energy, $f_2 L^2 + f_1 L$, disappears because of the normalization of the reduced density matrix \cite{CalabreseCardy_ee,Bipartite_fidelity}. This is the case both for the entanglement entropy and for the LBF. Then, in these applications, the first term of interest is the one proportional to $\log L$. This logarithmic term has a dimensionless prefactor $b_0$ which does not depend on any cutoff: it is universal. Of course, non-universal short length scales do appear when one computes the free energies of lattice models; their possible effect can be guessed by formally replacing the dominant length $L$ by $L+a$ 
 in all the terms\footnote{There may be more than one UV cutoff $a$ in the different 
 terms in the expansion, for simplicity here we look at the effect of a simple shift $L\rightarrow L+a$. It is also possible to 
perform the same analysis with more general power-law terms added to the free energy.} in (\ref{eq:somefreenrj}), and expanding again in $L$. We get
\begin{equation}
 F^\prime(L)=f_2 L^2+(f_1+2af_2)L+b_0\log L+f_0+a f_1+a^2 f_2+b_{-1}\frac{\log L}{L}+\frac{f_{-1}+a b_0}{L}+o(1/L).
\end{equation}
We see that the coefficient $b_0$ remains unaffected, as expected. The same thing happens for the $O(L^{-1}\log L)$ term. For this reason, even though $b_{-1}$ has the dimension of a length, it is a quantity that is more robust than $f_1,f_0$ or $f_{-1}$. More precisely, $b_{-1}$ must be of the form $b_{-1}=\xi\times \left({\rm universal \;factor}\right)$, for a {\it single} length scale $\xi$. Once $\xi$ is fixed, then the coefficient $b_{-1}$ is uniquely determined by the global geometry of the domain. As mentioned earlier, we will see that $\xi$ is nothing but the extrapolation length \cite{Diehl}, and the universal factor $f_{\rm geom.} = b_{-1}/(\xi c)$ is calculable with CFT methods (here $c$ is the central charge). Of course the argument presented here is qualitative; it needs to be supported by field-theoretic as well as lattice calculations. This is the goal of this paper.

\subsection{Organization of the manuscript}
The paper is organized as follows. In Sec.~\ref{sec:loglsl_theory} we explain the generic appearance of a $L^{-1}\log L$ term when there is a corner with angle $2\pi$. We provide explicit and general formulae for the free energy as well as the one-point function of a primary operator. We then apply these result to the ``pants'' geometry that corresponds to the logarithmic bipartite fidelity (LBF) in Sec.~\ref{sec:lbf_application}. Interestingly the correction takes a simple form, that we test numerically with high precision in free fermionic systems. We also consider the effect of boundary changing operators, and provide an exact calculation of this term for an XX chain divided into two equal halves. Sec.~\ref{sec:time_evolution} deals with the study of the entanglement entropy and the Loschmidt echo following a local quench. The corresponding imaginary time domain has the shape of ``double pants''. There is still a $L^{-1}\log L$ correction in this case, but intriguingly, it
  disappears after analytic continuation to real time. However, we show that a slight modification of the Loschmidt echo---dubbed ``detector'' in \cite{SD_localquench}---preserves the correction. We finish with some concluding remarks in Sec.~\ref{sec:conclusion}.

Some additional information is provided in three long appendices. \ref{sec:allthat} summarizes the connections between the bipartite fidelity, corner free energies, X-ray edge singularities, as well as some closely related recent attempts at measuring entanglement. The other two are more technical. In \ref{sec:app_bcc} we generalize the original calculation of the universal LBF finite-size function presented in \cite{Bipartite_fidelity}, to account for possible changes in boundary conditions. Finally, we explain in \ref{sec:exact} the exact calculation of the LBF for an XX chain cut into two halves, and how the asymptotic expansion allows to recover the CFT results.

\section{Criticality, corner free energy, and the extrapolation length in 2d statistical models}
\label{sec:loglsl_theory}
In this section we focus on a generic statistical system defined on a lattice. We imagine that this system is at a critical point, such that the long-distance physics is captured by a Euclidean field theory that is conformally invariant in the bulk. The lattice spacing $a$ will play the role of a UV cutoff. We are interested in a system with a boundary.

\subsection{The extrapolation length as a boundary perturbation by the stress-tensor}
In the study of critical phenomena, the concept of extrapolation length
is used very often in the presence of a boundary in the system \cite{Diehl}. The
idea is the following. Take a $d$-dimensional system on a lattice
$(a\mathbb{N})\times (a\mathbb{Z})^{d-1}$, with the plane $(0,x_2,x_3,\dots,x_d)$ playing the role of the boundary. If the system is critical, it is described at large scale by some scale-invariant field theory in the bulk, with a scale-invariant boundary condition at the surface. In many cases, in order to relate lattice observables to their field theory counterpart, one needs to make the assumption that, in the continuum, the boundary of the system is at position $x_1 = -\xi$ rather than at $x_1=0$. The length $\xi >0$ is dubbed ``extrapolation length''; it is usually of the order of the lattice spacing $a$, and it plays the role of a UV cutoff when one deals with quantities that are sensitive to the presence of the surface.
$$
\begin{tikzpicture}
	\filldraw[gray!40] (-2,0) rectangle (2,3);
	\draw[thick] (-2,0) -- (2,0);
	\draw[thick,dashed] (-2,0.5) -- (2,0.5);
	\draw[<->] (0,0.05) -- (0,0.45);
	\draw (0.25,0.25) node{$\xi$};
\end{tikzpicture}
$$
For instance, consider the expectation value of some local observable on the lattice at position $(x_1,0,\dots,0)$. When $x_1 \gg a$, its behavior is captured by a one-point function $\left< \mathcal{\phi}(x_1,0,\dots,0) \right>$ of the corresponding local operator $\mathcal{\phi}$ in the field theory. This one-point function is fixed by scale-invariance, and by the value of the extrapolation length $\xi$: $\left< \mathcal{\phi}(x_1,0,\dots,0) \right> \, \propto \, 1/|x_1 + \xi|^{h}$, where $h$ is the scaling dimension of the operator $\mathcal{\phi}$. Since this holds in the regime $x_1 \gg a \sim \xi$, $\left<\mathcal{\phi}(x_1,0\dots,0) \right> \, \simeq \, 1/|x_1|^h (1 -h \xi/x_1 + \dots)$, so the presence of the extrapolation length $\xi>0$ determines the form of the leading correction to scaling for this one-point function.

In general, lattice effects may be tackled by adding perturbations by local operators to the field theory action $S$. The perturbations may live in the bulk or at the surface of the system. One possible perturbing operator at the surface is the stress-tensor $T_{\mu\nu}$. Of course, other perturbations can appear at the surface, and should be taken into account, but let us start with the stress-tensor only. The perturbation by the component $T_{\perp \perp} = T_{11}$ along the $d-1$-dimensional boundary can be written as
\begin{equation}
	\label{eq:variationS1}
S \, \longrightarrow \, S + \frac{\xi}{2\pi} \int d^{d-1}{\bf x}_{\parallel} \, T_{\perp \perp} \, .
\end{equation}
The coupling $\xi$ has the dimension of a length; it is nothing but the extrapolation length itself. To see this, recall that the definition of the stress-tensor is that, under an infinitesimal transformation $x_\mu \mapsto x_\mu + \varepsilon_\mu (x)$, the variation of the action is
\begin{equation}
	\label{eq:variationS2}
	\delta S \, = \, -\frac{1}{2\pi} \int d^d {\bf x} \, T_{\mu \nu} \partial_\mu \varepsilon_\nu ,
\end{equation}
where the factor $2\pi$ is a normalization convention (it is the standard convention in CFT). Now consider a transformation that moves the boundary of the system from $x_1=0$ down to $x_1 = -\xi$, for instance $\varepsilon_1 (x) = \xi$ for $x_1 \leq 0$ and $\varepsilon_1 = 0$ for $x_1>0$ (and $\varepsilon_2(x)=0$ everywhere). Then the only non-vanishing component of the tensor $\partial_\mu \varepsilon_\nu$ is $\partial_1 \varepsilon_1(x) = -\xi \delta(x_1)$, and it gives back the expression (\ref{eq:variationS1}) when it is inserted in (\ref{eq:variationS2}). Thus, the appearance of the extrapolation length $\xi$ is due to a perturbation by the stress-tensor along the boundary\footnote{We are grateful to Nick Read for making this simple but crucial observation. This equivalence between the idea of an extrapolation length and the perturbation by the stress-tensor along the boundary plays an important role in \cite{DRR}.}.

It is natural to ask what other boundary perturbations may occur. In general, this depends on the system, and on the local operators that are available in the field theory. First, one has to identify which conformal boundary condition describes the system in the scaling limit. Second, only operators that are irrelevant along the surface (in some cases marginal ones could also be present) can appear as perturbations, since {\it relevant} operators would drive the system towards a {\it different} conformal boundary condition under the Renormalization Group (RG) flow. Then the determination of the possible surface perturbations boils down to a standard RG analysis based on symmetry criteria and comparison of scaling dimensions of the operators present in the field theory. In this paper we will be mostly interested in the subleading corrections due to the stress-tensor; to study these corrections, we may assume that we are in a situation such that the stress-tensor is the least  
 irrelevant boundary perturbation.

\subsection{Leading order of the corner free energy at criticality: the Cardy-Peschel formula}
\label{sec:cardypeschel}
From now on, we restrict our discussion to the two-dimensional case. In two dimensions, as pointed out by Cardy and Peschel \cite{CardyPeschel}, there is a universal contribution to the free energy of a critical system in a domain with a corner with internal angle $\theta$. For completeness, in this section we quickly recall their result and its derivation; this is independent from the extrapolation length and boundary perturbations described in the previous section. The expert reader may skip this discussion and go directly to the next section, which contains new results about the extrapolation length and subleading corrections to the corner free energy.
\begin{figure}[htbp]
\centering
\begin{tikzpicture}
	\filldraw[gray!40] (3,0) -- (2.2,2) -- (0,0) -- cycle;
	\draw[thick] (3,0) -- (0,0) -- (2.2,2);
	\draw[thick] (0.8,0) arc (0:180-138:0.8);
	\draw (0.9,0.4) node{$\theta$};
	\draw[thick,dashed] (2,0) arc (0:180-138:2);
	\draw[<->] (2,-0.2) -- (0,-0.2);
	\draw (1,-0.4) node{$L$};
\end{tikzpicture}
\caption{Sharp corner with internal angle $\theta$. $L$ is a typical length in the system.}
\label{fig:sharpcorner}
\end{figure}
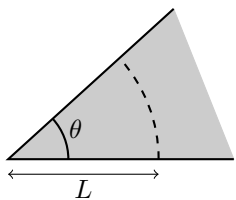\\
The Cardy-Peschel term can be computed as follows. The mapping $z \mapsto w(z) = z^{\theta/\pi}$ maps the upper half-plane onto the corner shown in Fig~\ref{fig:sharpcorner}. The holomorphic component of the stress-tensor in the corner is
\begin{eqnarray}\nonumber
T(w) & =  &  \left(\frac{dw}{dz} \right)^{-2} \left[T(z) -\frac{c}{12} \{w,z \}\right] \\
&=& \frac{z^{2(1-\theta/\pi)}}{(\theta/\pi)^2} \left[ T(z) - \frac{c}{12} \frac{1-(\theta/\pi)^2}{2 z^2} \right].
\end{eqnarray}
Here we have used the transformation law of the stress tensor \cite{BigYellowBook},
\begin{equation}\label{eq:transfo_T}
T(w) = \left(\frac{dw}{dz} \right)^{-2} \left[ T(z) -\frac{c}{12} \{ w,z\} \right], 
\end{equation}
 and the fact that $\braket{T(z)}=0$ in the upper-half plane. $\{w,z\}$ denotes the Schwarzian derivative:
 \begin{equation}
  \{w,z\}=\frac{d^3w/dz^3}{dw/dz}-\frac{3}{2}\left(\frac{d^2w/dz^2}{dw/dz}\right)^2.
 \end{equation}
The generator of dilatations in the corner is then given by the following integral over the dashed line in Fig.~\ref{fig:sharpcorner}
\begin{eqnarray}\nonumber
	\hat{D} & = & \frac{1}{2\pi i} \int w \,T(w) dw \, +\, {\rm c.c} \\\nonumber
	& = & \frac{1}{2 \pi i} \int_C z^{\theta/\pi} \frac{z^{1-\theta/\pi}}{\theta/\pi} \left[T(z)-\frac{c}{24}\frac{1-(\theta/\pi)^2}{ z^2} \right] dz \\
	& = & \frac{\pi}{\theta} \hat{L}_0+ \frac{c}{24} \left(\frac{\theta}{\pi}-\frac{\pi}{\theta}\right),
\end{eqnarray}
where $C$ is a contour encircling the origin in the complex plane (we have used $T(z) = \overline{T}(\bar{z})$ on the real axis, and the standard boundary CFT trick that the non-chiral theory on the half-plane is like the chiral theory on the plane \cite{Cardy1984}). Now, the free energy $F_{\rm corner}$ scales as
\begin{equation}
e^{-F_{\rm corner}} \,=\, Z_{\rm corner} \, = \, \left<L^{-\hat{D}}\right> .
\end{equation}
When there is no operator at the corner, $\left< \hat{L}_0 \right>=0$, and one finds the Cardy-Peschel term
\begin{equation}
F_{\rm corner} \, = \, \frac{c}{24} \left(\frac{\theta}{\pi} - \frac{\pi}{\theta} \right) \log L .
\end{equation}
When the boundary condition is different on the two sides of the corner, one must insert a {\it boundary condition changing operator} \cite{Cardy1989} at the corner, which is a primary operator with a conformal dimension $h_{bcc}$. In that case, $\left\langle \hat{L}_0 \right\rangle=h_{bcc}$, and the Cardy-Peschel formula becomes  
\begin{equation}\label{eq:cardypeschel_bcc}
F_{\rm corner} \, = \, \left[\frac{\pi}{\theta}h_{bcc} + \frac{c}{24} \left(\frac{\theta}{\pi} - \frac{\pi}{\theta} \right)\right] \log L  .
\end{equation}

\subsection{Extrapolation length in the corner}

\begin{figure}[htbp]
\begin{center}
\begin{tikzpicture}
	\filldraw[gray!40] (3,0) -- (2.2,2) -- (0,0) -- cycle;
	\filldraw[gray!40] (3,0) rectangle (0,-0.3);
	\filldraw[gray!40] (3,0) rectangle (0,-0.4);
	\draw[thick] (3,-0.4) -- (0,-0.4);
	\draw[<->] (3.2,0) -- (3.2,-0.4);
	\draw (3.4,-0.2) node{$\xi$};
	\filldraw[gray!40] (2.2,2) -- (0,0) -- (-0.3,0.3) -- (1.9,2.3) -- cycle;
	\draw[thick] (-0.3,0.3) -- (1.9,2.3);
	\draw[<->] (2.3,2.1) -- (2,2.4);
	\draw (2.3,2.4) node{$\xi$};
	\draw[thick,dashed] (3,0) -- (0,0) -- (2.2,2);
	\draw[thick] (0.8,0) arc (0:180-138:0.8);
	\draw (0.9,0.4) node{$\theta$};
	\filldraw[blue!50] (0,0) circle (0.5) node[black]{?};
\end{tikzpicture}
\end{center}
	\caption{How can one adapt the idea of the extrapolation length to a domain with a corner of angle $\theta$?}
	\label{fig:corner_extrapolation}
\end{figure}
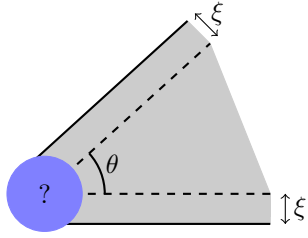

The question is now: what kind of subleading corrections to the corner free energy should we expect? Can we adapt the idea of an extrapolation length to a domain with a corner? In the picture shown in Fig.~\ref{fig:corner_extrapolation}, this does not look completely obvious: the effect of the extrapolation length should be a shift of the sides of the domain, but it is not clear what one should do close to the corner.
The question makes more sense when it is formulated in terms of the perturbation by the stress-tensor along the surface,
\begin{equation}
S \longrightarrow S + \frac{\xi}{2\pi} \int T_{\perp \perp} d l ,
\end{equation}
which gives a correction to the corner free energy $F_{\rm corner} \rightarrow F_{\rm corner }+ \delta F_{\rm corner}$. Here $d l$ is a line element along each of the sides of the domain. We want to evaluate the correction at the first order in $\xi$ for the corner shown in Fig.~\ref{fig:corner_extrapolation},
\begin{equation}
\delta F_{\rm corner} \, =\, - \frac{i\xi}{2\pi i} \int_{C_1} dw\, \left<T(w)\right> - e^{i \theta} \frac{i \xi}{2\pi i} \int_{C_2} dw \, \left<T(w)\right> \,+ \, {\rm c.c.}
\end{equation}
Here $C_1$ is the horizontal side of the corner in Fig. \ref{fig:corner_extrapolation} and $C_2$ is the image of $C_1$ under the rotation of angle $\theta$ (the other side on the figure). The factor $e^{i \theta}$ comes from the Jacobian of that rotation. As we shall see shortly, one needs to be careful with UV and IR cutoffs in the integrals. But first, let us evaluate $\left< T(w)\right>$. Again, it can be computed using a conformal mapping from the upper half-plane to the domain with the corner, and using the transformation law of the stress-tensor. However, unlike the Cardy-Peschel term, one expects the subleading corrections to be sensitive to the full geometry of the domain, not only to the sum of the different angles. Therefore, we need to work with a slightly more precise form of the conformal mapping from the upper half-plane to the domain with the corner shown in Fig. \ref{fig:corner_extrapolation}. More precisely, we are interested in a generic simply connected dom
 ain with a corner of internal angle $\theta$. We take a conformal mapping $z \mapsto w(z)$ from the upper half-plane to this generic domain with a corner located at $w(0)$. Such a mapping always exists, thanks to the Riemann mapping theorem \cite{ahlfors}. Close to the corner, the mapping takes the form:
\begin{equation}\label{eq:right_tranfs1}
w(z)-w(0) \, = \,z^{\theta/\pi} ~f(z),
\end{equation}
where $f(z)$ is a function that is analytic around $z=0$. Using the $SL(2,\mathbb{R})$ reparametrizations of the upper half-plane\footnote{These are the special conformal transformations $\zeta(z)=\frac{az+b}{cz+d}$, with $a,b,c,d \,\in \mathbb{R}$ and $ad-bc=1$. They leave the upper half-plane invariant, so if $z \mapsto w(z)$ maps the upper-half plane to the desired geometry, $z\mapsto w(\zeta(z))$ also does.}, $f(z)$ can be chosen such that it has
the following expansion around $z=0$: 
\begin{equation}\label{eq:right_tranfs2}
f(z) \, = \, 1 + f_2~ z^2 +  f_3 ~z^3  + \dots
\end{equation}
Such a function $f(z) \neq cst$ appears as soon as the domain is not exactly a single infinite corner $\left\{ w \in \mathbb{C}; \; {\rm arg\,} z \in [0,\theta] \right\} $, but differs from it at some scale $L$. For instance, one can think of some polygon with sides that have a length of order $L$. The typical scale $L$ appears in the mapping $z \mapsto z^{\theta/\pi} f(z)$  as the coefficient $f_2$: typically, we have $L \sim |f_2|^{-2\pi/\theta}$. 

Now, with the mapping (\ref{eq:right_tranfs1},\ref{eq:right_tranfs2}), we can compute the expectation value of the component of the stress-tensor $\left< T(w)\right>$. Again, we use the transformation law (\ref{eq:transfo_T}) to relate $\left< T(w)\right>$ to the Schwarzian derivative of the conformal mapping $z \mapsto w(z)$, and to the expectation value of the stress-tensor in the upper half-plane, $\left< T(z)\right>$, which is zero (here we assume that there is no boundary condition changing operator on the boundary). After some simple algebra, we find that the correction to the free energy takes the form
\begin{eqnarray}\fl
	\nonumber
\delta F_{\rm corner} &=& \frac{\xi ~c}{12 \pi} \int_{-z_{IR}}^{-z_{UV}} dz \left( \frac{dw}{dz} \right)^{-1} \{w,z\} \, +\,  \frac{\xi ~c}{12 \pi} \int_{ z_{UV}}^{z_{IR}} dz \left( \frac{dw}{dz} \right)^{-1} \{w,z\} \\\fl
 &=& \xi \int_{ - z_{IR}}^{-z_{UV}} \left[ \frac{B}{z^{1+\theta/\pi}} + f_2 \frac{C}{z^{\theta/\pi-1}} + \dots \right]dz \, + \,\xi \int_{z_{UV}}^{z_{IR}} \left[ \frac{B}{z^{1+\theta/\pi}} + f_2 \frac{C}{z^{\theta/\pi-1}} + \dots \right]dz.
\label{eq:intermediate1}
\end{eqnarray}
Here, each of the two integrals corresponds to one side of the corner. On each side, the variable $w$ is integrated between an UV cutoff of order $\sim a$ (lattice spacing) and an IR cutoff which is the system size, of order $\sim L$. The values of $z_{UV}$ and $z_{IR}$ are the inverse images of these two cutoffs under the mapping $z \mapsto w(z)$; we are thus integrating over the variable $z$ between $z_{UV}>0$ and $z_{IR} > z_{UV}$ for one side of the corner, and between $-z_{IR}$ and $-z_{UV}$ for the other side. Note that, since $w \sim z^{\theta/\pi}$, the UV and IR bounds for the variable $z$ are respectively of order $z_{UV} \sim a^{\frac{\pi}{\theta}}$ and $z_{IR }\sim L^{\frac{\pi}{\theta}}$. The precise values of $z_{IR}, z_{UV}$ depend on the details of the system. Here, what is important is that the first term in (\ref{eq:intermediate1}) gives a contribution of the form $A + B' \xi/L$ where $A, B'$ (and $B$ and $C$) are numerical constants of order $O(1)$. The second term also gives such a contribution, {\it unless} $\theta = 2\pi$. This is the reason why $\theta = 2\pi$ is so special: when $\theta < 2\pi$, the corrections to the free energy are algebraic, with coefficients that are some mixture of the different cutoffs, and that are not particularly meaningful. As emphasized in the introduction, the case when there is a {\it logarithmic} correction is much more interesting; this happens when $\theta = 2\pi$.

\subsection{Logarithmic correction in the case $\theta = 2\pi$}
When $\theta =2\pi$, we see that the second term in (\ref{eq:intermediate1}) gives a logarithm piece when one integrates from an UV cutoff of order $\sim a^{\frac{\pi}{\theta}} = \sqrt{a}$ to an IR cutoff of order $\sim L^{\frac{\pi}{\theta}} = \sqrt{L}$. A proper evaluation of the Schwarzian derivative $\{w,z \}$ of the mapping (\ref{eq:right_tranfs2}) leads to the following term:
\begin{eqnarray}
	\nonumber
\delta F_{\rm corner} &=& A + \frac{\xi~c}{12 \pi} \int_{\sim\sqrt{a}}^{\sim\sqrt{L}} \frac{3f_2}{2}\frac{dz}{z}  + \dots\\
 & = & A \,  +\,  \frac{\xi~c~f_2}{16 \pi} \log L/a  \, + \, O(1/L). \label{eq:logF}
\end{eqnarray}
The first term $A$ is a non-universal constant of order $O(1)$ that depends on $a$ and $\xi$. The second term is of order $O(L^{-1}\log L)$; it is the one we are interested in. It depends on a single length scale, the extrapolation length $\xi$. In that sense, it is ``semi-universal'': once $\xi$ is known, then this term is fixed entirely by the geometry of the sample, which is encoded in the single parameter $f_2$ (more precisely, $L \times f_2$, which is a dimensionless parameter). Note that such a statement does {\rm not} hold, for instance, for the correction of order $O(1)$, that is typically a ratio of {\rm several} short length scales like $\xi$ and $a$. Eq.~(\ref{eq:logF}) is one of the central analytical results of the paper; we will show later that this logarithmic term is calculable in lattice systems, and we will find that it agrees remarkably well with numerics.

To conclude this section, let us recast our result (\ref{eq:logF}) in a more general form, valid for any simply connected domain with several corner of angle $2\pi$ (we will later consider specific examples of such domains). This more general form can be obtained by relaxing the condition (\ref{eq:right_tranfs2}), and by repeating the same arguments as above. The expressions involved in the calculation are slightly more complicated, but the arguments remain identical. For a given conformal mapping $z \mapsto w(z)$ from the upper half-plane to the domain with the corners, we find the following correction to the free energy:
\begin{equation}
	\delta F_{\rm corner} \,  = \,A' \,+\, c~f_{\rm geom.}~ \frac{\xi}{L} \log (L/a) \, + \, O(1/L).
\end{equation}
The constant $A'$ is of order $O(1)$; the interesting term is the second one. The geometric factor is a dimensionless parameter defined by
\begin{equation}\label{eq:loglsl_full}
 f_{\rm geom.} \,  = \, \frac{L}{24\pi}\, \sum_{2\pi \, {\rm corners}} {\rm Res}\, \left[ e^{i \alpha_c} \left(\frac{dw}{dz}\right)^{-1}\!\!\{w,z\}\,,\,z=z_c\right].
\end{equation}
The sum runs over the different corners with internal angles $2\pi$, which are located at the positions $w(z_c)$, with the $z_c$'s along the boundary of the upper half-plane. The phase factor $e^{i \alpha_c}$ appears after some algebra when one relaxes the condition (\ref{eq:right_tranfs2}); $\alpha_c$ is defined as 
\begin{equation}
\alpha_c \, =\,\arg \left(\left.\frac{d^2w}{dz^2}\right|_{z=z_c}\right).
\end{equation}
We will discuss several examples of such geometric factors for specific domains below.

\subsection{Correction to the one-point function}
\label{sec:corr_onepoint}
The correction $S \rightarrow S+ \delta S$ due to the extrapolation length (the stress-tensor) must also affect the one-point function of a primary operator $\phi(w,\overline{w})$ close to the corner,
\begin{equation}
	\left< \phi(w,\overline{w}) \right>_{\delta S} \, = \,  \left< \phi(w) \right>_0 \, + \, \delta \left< \phi(w,\overline{w}) \right>.
\end{equation}
\begin{figure}[htbp]
\begin{center}
\begin{tikzpicture}
	\filldraw[gray!40] (3,0) -- (2.2,2) -- (0,0) -- cycle;
	\filldraw[gray!40] (3,0) rectangle (0,-0.3);
	\filldraw[gray!40] (3,0) rectangle (0,-0.4);
	\draw[thick] (3,-0.4) -- (0,-0.4);
	\draw[<->] (3.2,0) -- (3.2,-0.4);
	\draw (3.4,-0.2) node{$\xi$};
	\filldraw[gray!40] (2.2,2) -- (0,0) -- (-0.3,0.3) -- (1.9,2.3) -- cycle;
	\draw[thick] (-0.3,0.3) -- (1.9,2.3);
	\draw[<->] (2.3,2.1) -- (2,2.4);
	\draw (2.3,2.4) node{$\xi$};
	\draw[thick,dashed] (3,0) -- (0,0) -- (2.2,2);
	\filldraw[blue!50] (0,0) circle (0.5) node[black]{?};
	\filldraw (1.8,1) circle (2pt) node[below]{$\phi(w,\overline{w})$};
\end{tikzpicture}
\end{center}
	\caption{How does the extrapolation length in the corner modify the behaviour of the one-point function?}
	\label{fig:corner_one_point}
\end{figure}
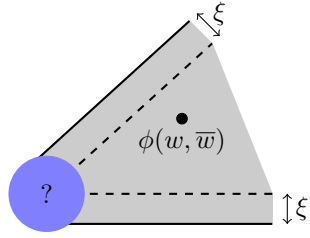
The one-point function without perturbation $\left< \phi(w,\overline{w}) \right>_0$ can be computed using the inverse of the conformal mapping $z \mapsto w(z)$ \cite{BigYellowBook}. In this section we compute the correction $\delta \left< \phi(w,\overline{w}) \right>$ at the first order in the extrapolation length $\xi$. The variation of the correlation function is, at the first order in $\delta S$,
\begin{eqnarray}
	\delta \left< \phi(w, \overline{w}) \right> & =& \frac{\left< \phi(w,\overline{w}) e^{-\delta S} \right>_0}{\left< e^{-\delta S}\right>_0} \,= \, - \left< \delta S \, \phi(w,\overline{w}) \right>_0 \, + \, \left< \delta S \right>_0 \left< \phi(w,\overline{w}) \right>_0 .
\end{eqnarray}
From now on, all the expectation values that will appear are taken with the {\it unperturbed } action $S$, so we drop the subscript '$0$' in $\left< \,.\, \right>_0$. The terms on the right-hand-side involve $\delta S$, which is a sum of two integrals along the two sides of the corner, $C_1$ and $C_2$. Let us focus on the first side $C_1$. Again, we have to integrate over the variable $w$ from the UV cutoff ($\sim a$) to the IR cutoff ($\sim L$), and we use a conformal mapping from the upper half-plane to our domain, $z \mapsto w(z)$, with $w(z) \simeq z^{\theta/\pi}$ for $z\rightarrow 0$:
\begin{eqnarray}\nonumber
\delta \braket{\phi(w,\overline{w})}_{C_1}&=& \frac{i \xi}{2\pi i} \int_{\sim a}^{\sim L} dw' \left[ \left<T(w') \phi(w,\overline{w}) \right> - \left<T(w') \right> \left< \phi(w,\overline{w}) \right> \right]  \, + \, c.c. \\
	&=&  \frac{\xi}{2\pi} \left| \frac{dw}{dz}\right|^{-2h} \int_{\sim a^{\frac{\pi}{\theta}}}^{\sim L^{\frac{\pi}{\theta}}} dz' \left( \frac{dw'}{dz'} \right)^{-1}  \left<T(z') \phi(z,\overline{z}) \right>  \, + \, c.c.
\end{eqnarray}
In the second line, we have used the transformation law of the stress-tensor (\ref{eq:transfo_T}), the transformation law of the primary operator $\phi(w,\overline{w})$ \cite{BigYellowBook},
\begin{equation}
\phi(w,\overline{w}) \, = \, \left| \frac{dw}{dz} \right|^{-2h}  \phi(z,\overline{z}),
\end{equation}
and the fact that the terms with the Schwarzian derivative cancel in the difference $\left<T \phi \right> - \left< T\right>\left< \phi\right>$. The contribution of the other side of the corner $C_2$ is similar. Summing these two contributions, we see that the correction to the one-point function must be of the following form:
\begin{equation}\label{eq:onepoint_corr_full}\fl
		\frac{\delta \left< \phi(w,\overline{w}) \right>}{\left<\phi(w,\overline{w}) \right>} \,=\, \frac{\xi}{2\pi} \! \int_{\sim a^{\frac{\pi}{\theta}}}^{\sim L^{\frac{\pi}{\theta}}}\! dz'  {z'}^{1- \theta/\pi}  \frac{\left<T(z') \phi(z,\overline{z}) \right>}{\left< \phi(z,\overline{z}) \right>}
  +  \frac{\xi}{2\pi} \!\int^{\sim -a^{\frac{\pi}{\theta}}}_{\sim -L^{\frac{\pi}{\theta}}}\! dz'  {z'}^{1- \theta/\pi} \frac{ \left<T(z') \phi(z,\overline{z}) \right> }{\left< \phi(z,\overline{z})\right>} \, + \, c.c.  
\end{equation}
Like in the case of the free energy, if $\theta < 2\pi$, this will ultimately lead to algebraic corrections of order $O(1)$, $O(1/L)$, and so on. When $\theta =2\pi$, however, things are again more interesting, in the sense that we find a logarithmic divergence. So, from now on, we focus again on the case $\theta = 2\pi$. We notice that $\left< T(z') \phi(z,\overline{z})\right>$ is analytic at $z'=0$ (since $z$ is not at the origin), and we expand
\begin{equation}
\left(\frac{dw'}{dz'}\right)^{-1} =\frac{1}{2z'} +O(z'),
\end{equation}
so we find a logarithmic divergence coming from the pole at $z'=0$. We have
\begin{eqnarray}\nonumber
		\frac{\delta \left< \phi(w,\overline{w}) \right>}{\left<\phi(w,\overline{w}) \right>} & = & \frac{\xi}{2\pi } \int^{\sim\sqrt{L}}_{\sim \sqrt{a}} dz' \, \frac{1}{2z'} \frac{\left<T(0) \phi(z,\overline{z}) \right>}{\left<\phi(z,\overline{z})\right>}   + \,  \frac{\xi}{2\pi} \int^{-\sqrt{a}}_{\sim -\sqrt{L}} dz' \, \frac{1}{2z'} \frac{\left<T(0) \phi(z,\overline{z}) \right>}{\left< \phi(z,\overline{z})\right>} \, +\,  c.c. \\\fl
 & = &  \frac{\xi}{4\pi} \log (L/a)  \times \frac{\left<T(0) \phi(z,\overline{z}) \right>}{\left<\phi(z,\overline{z})\right>}  +c.c+O(1/L).
\end{eqnarray}
To evaluate $\left<T(0) \phi (z, \overline{z}) \right>/\left< \phi(z,\overline{z}) \right>$ we use the conformal Ward identity \cite{BigYellowBook},
\begin{eqnarray}
	\left<T(0) \phi(z,\overline{z}) \right> & = &  \left[ \frac{h}{z^2} - \frac{1}{z}\partial_z \right]\left< \phi(z,\overline{z}) \right>,
\end{eqnarray}
together with the known form of the one-point function in the upper half-plane: $\left< \phi(z,\overline{z}) \right> \, \propto \, 1/(z - \overline{z})^{2h}$. This leads to
\begin{eqnarray}\nonumber
	\frac{\left<T(0) \phi(z,\overline{z}) \right>}{\left< \phi(z,\overline{z})\right>} & = &  \frac{h}{z^2} \left(\frac{2z}{z-\overline{z}} +1\right) .
\end{eqnarray}
Adding this expression and its complex conjugate, we find that the correction $\delta \left<\phi(w,\overline{w}) \right>$ due to the stress-tensor at the boundary is of the form:
\begin{equation}
	\label{eq:one_point}
	\frac{\delta \left< \phi(w, \overline{w}) \right>}{\left< \phi(w,\overline{w}) \right>} \, = \, -\frac{\xi ~h}{\pi}  \frac{({\rm Im\,}z)^2}{|z|^4} ~\log (L/a)    +  O(1/L).
\end{equation}
Note that, since $|z|^2 \sim L$, the logarithmic term that we find here is of order $\log(L/a)/L$. Formula (\ref{eq:one_point}) is our second main analytic result. It shows that the extrapolation length generates a logarithmic correction to the one-point function in a domain with a corner of angle $2\pi$. If the domain has several such angles, then the total correction is simply the sum of the ones associated to each corner. We will provide checks of the formula (\ref{eq:one_point}) in Sec.~\ref{sec:onepoint_examples}. Let us conclude here with the following remark: our assumption in this section was that the conformal mapping $z \mapsto w(z)$ form the upper half-plane has the expansion close to the corner with angle $2\pi$: $w(z) \, \simeq \, z^2 + \dots$. This form is invariant under a composition with a special conformal mapping of the form $z \mapsto z/(b z +1)$, with $b \in \mathbb{R}$. This means that our result (\ref{eq:one_point}) must also be invariant under such a reparametrization. This does not 
seem obvious in (\ref{eq:one_point}); however one can check that ${\rm Im \,} z /|z|^2$ is indeed invariant under the reparametrization $z \mapsto z/(b z+1)$, $b \in \mathbb{R}$.

\section{Application to the bipartite fidelity and related quantities}
\label{sec:lbf_application}

In this section we discuss the consequences of the general results of Sec.~\ref{sec:loglsl_theory} on the bipartite fidelity of $1d$ spin chains \cite{Bipartite_fidelity}. This quantity is an overlap defined with respect to a bipartition of the system. For a given bipartition of the sites of a spin chain $A \cup B$, the bipartite fidelity is the overlap between the ground state $\left| A \cup B \right>$ of the Hamiltonian of the full spin chain, and the ground state $\left| A \otimes B\right>$ of the Hamiltonian where all the interactions between the subsystems $A$ and $B$ have been switched off. We will focus on the logarithm of this overlap, dubbed ``bipartite logarithmic fidelity'', or LBF:
\begin{equation}
 \mathcal{F}_{A,B}=-\log \left|\braket{A\cup B|A\otimes B}\right|^2.
\end{equation}
Below, the bipartition of the sites of the chain is chosen as follows: $A$ corresponds to the first $\ell$ sites, and $B$ to the $L-\ell$ remaining sites (see Fig. \ref{fig:lbf_cut}). $\ket{A\cup B}$ is then the ground state of the total chain of length $L$, and $\ket{A\otimes B}=\ket{A}\otimes \ket{B}$, where $\ket{A}$ (resp. $\ket{B}$) is the ground state of the chain $A$ (resp. $B$) of length $\ell$ (resp. $L-\ell$). Some of our motivations for the study of the LBF are explained in \ref{sec:allthat}; it is a quantity which can be studied analytically by various methods, including CFT methods \cite{Bipartite_fidelity} and lattice integrability techniques \cite{Weston1,Weston2}.

We start with the most generic case (Sec.~\ref{sec:lbf_generic}), before generalizing to possible changes in boundary conditions (Sec.~\ref{sec:lbf_bcc}). The system considered here will be an open chain cut into two parts, as shown in Fig.~\ref{fig:lbf_cut}. 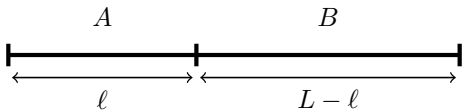
\begin{figure}[htbp]
\centering
 \begin{tikzpicture}
  \draw[ultra thick] (0,0) -- (6,0);
  \draw[ultra thick] (0,-0.15) -- (0,0.15);
  \draw[ultra thick] (2.5,-0.15) -- (2.5,0.15);
  \draw[ultra thick] (6,-0.15) -- (6,0.15);
  \draw (1.25,0.5) node {$A$};
  \draw (4.25,0.5) node {$B$};
  \draw[<->,semithick] (0.05,-0.3)-- (2.45,-0.3);
  \draw[<->,semithick] (2.55,-0.3)-- (5.95,-0.3);
  \draw (1.25,-0.6) node {$\ell$};
  \draw (4.25,-0.6) node {$L-\ell$};
 \end{tikzpicture}
 \caption{Open chain of length $L$ cut into two subsystems $A$ and $B$ of respective length $\ell$ and $L-\ell$.}
 \label{fig:lbf_cut}
\end{figure}

\subsection{Shape dependence in the generic case}
\label{sec:lbf_generic}
Let us first recall some basic facts about the LBF, following our previous study \cite{Bipartite_fidelity}. In imaginary time, the fidelity we are interested in can be expressed as a ratio of partition functions
\begin{equation}
 \left|\langle A\cup B|A\otimes B\rangle\right|^2=\frac{\left({\cal Z}_{A,B}\right)^2}{{\cal Z}_{A\otimes B} {\cal Z}_{A\cup B}}.
\end{equation}
The corresponding geometries are shown in Fig.~\ref{fig:freenrj_lbf}. In terms of free energies $f=-\log {\cal Z}$, the result reads
\begin{equation}\label{eq:lbf_freenrj}
 \mathcal{F}_{A,B}=2f_{A,B}-f_{A\otimes B}-f_{A\cup B}.
\end{equation}
Each of these free energies contribute to a bulk free energy ($\propto L^2$) and a line free energy ($\propto L$), but these two contributions are canceled by the linear contribution (\ref{eq:lbf_freenrj}). In the end the leading (logarithmic) contribution comes from the Cardy-Peschel $2\pi$ corner in $f_{A,B}$ (see Fig.~\ref{fig:freenrj_lbf}).
\begin{figure}[ht]
 \centering
 \begin{tikzpicture}
 \draw [color=gray!40,fill=gray!40] (0,0) -- (4,0) -- (4,2) -- (0,2) -- (0,0);
\draw[color=red] (2,0.75) node {$2\pi$}; 
 \draw[semithick,color=red] (1.75,1.2) arc[radius=0.25, start angle=180, end angle=0];
  \draw[semithick,color=red,<-] (1.75,1.17) arc[radius=0.25, start angle=-173, end angle=10];
  \draw [ultra thick] (0,0) -- (4,0);
  \draw [ultra thick] (0,1.2) -- (2,1.2);
  \draw [ultra thick] (0,2) -- (4,2);
  \draw [<->] (2.6,0.05) -- (2.6,1.15);
  \draw (3.15,0.6) node {$L-\ell$};
  \draw [<->] (2.6,1.25) -- (2.6,1.95);
  \draw (3.15,1.6) node {$\ell$};
  \draw (2,-0.4) node {$f_{A,B}=-\log {\cal Z}_{A,B}$};
  \begin{scope}[xshift=5.5cm]
  \draw [thick,dotted] (-0.75,-0.6) -- (-0.75,2.2);
  \draw [color=gray!40,fill=gray!40] (0,0) -- (4,0) -- (4,1.15) -- (0,1.15) -- (0,0);
  \draw [color=gray!40,fill=gray!40] (0,1.25) -- (4,1.25) -- (4,2) -- (0,2) -- (0,1.25);
  \draw [ultra thick] (0,0) -- (4,0);
  \draw [ultra thick] (0,1.15) -- (4,1.15);
  \draw [ultra thick] (0,1.25) -- (4,1.25);
  \draw [ultra thick] (0,2) -- (4,2);
  \draw [<->] (2.6,0.05) -- (2.6,1.1);
  \draw (3.15,0.6) node {$L-\ell$};
  \draw [<->] (2.6,1.3) -- (2.6,1.95);
  \draw (3.15,1.6) node {$\ell$};
    \draw (2,-0.4) node {$f_{A\otimes B}=-\log {\cal Z}_{A\otimes B}$};
  \end{scope}
  \begin{scope}[xshift=11cm]
  \draw [thick,dotted] (-0.75,-0.6) -- (-0.75,2.2);
  \draw [color=gray!40,fill=gray!40] (0,0) -- (4,0) -- (4,2) -- (0,2) -- (0,0);
  \draw [ultra thick] (0,0) -- (4,0);
  \draw [ultra thick] (0,2) -- (4,2);
  \draw [<->] (2.6,0.05) -- (2.6,1.95);
  \draw (3.1,1) node {$L$};
    \draw (2,-0.4) node {$f_{A \cup B}=-\log {\cal Z}_{A\cup B}$};
  \end{scope}
 \end{tikzpicture}
\caption{Free energies $f_{A,B}$, $f_{A\otimes B}$ and $f_{A\cup B}$ involved in the calculation of the LBF. The leading logarithmic contribution comes from the corner with angle $\theta=2\pi$ in $f_{A,B}$.}
\label{fig:freenrj_lbf}
\end{figure}
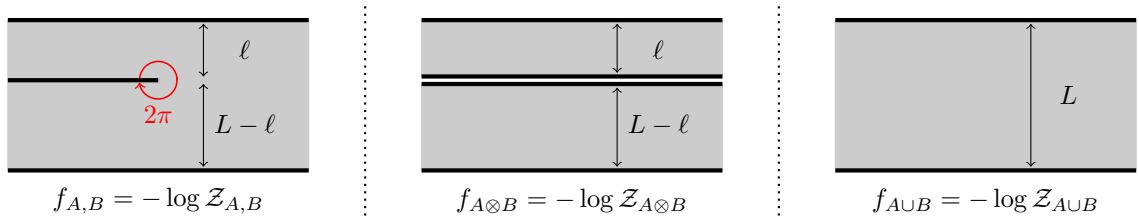\\
Let us introduce the aspect ratio 
\begin{equation}
x=\frac{\ell}{L},
\end{equation}
which we keep fixed in the following. We expect the LBF to scale as
\begin{equation}\label{eq:scaling}
 \mathcal{F}(x,L)=\frac{c}{8}\log L+f(x)+g(x) \frac{\log L}{L}+O(1/L).
\end{equation}
The leading logarithmic can be obtained as an application of the Cardy-Peschel formula (\ref{eq:cardypeschel_bcc}): it is therefore a signature of the underlying critical behaviour. Note also that when approaching criticality, this term  is replaced by $(c/8)\log \lambda$ \cite{Weston2}, where $\lambda$ is the correlation length, in the regime $L\gg\lambda \gg a$. $f(x)$ is universal finite-size function, and has been computed in \cite{Bipartite_fidelity}. It is given by
\begin{equation}\label{eq:f}
 f(x)=\frac{c}{24}\left[\left(2x-1+\frac{2}{x}\right)\log (1-x)+\left(1-2x+\frac{2}{1-x}\right)\log x\right]+{\rm cst.}
\end{equation}
We are, however, more interested in the subleading term proportional to $L^{-1}\log L$. The conformal transformation from the upper half-plane is  \cite{Bipartite_fidelity}
\begin{equation}
 w(z)=\frac{L}{\pi}\left[x\log (z+1)+(1-x)\log \left(\frac{zx}{1-x}-1\right)\right],
\end{equation}
and has expansion
\begin{equation}\label{eq:conf_expansion}
w(z)=iL(1-x)-\frac{Lx}{2\pi(1-x)}z^2+O(z^3)
\end{equation}
around $z=0$. It is then straightforward to apply the Eq.~(\ref{eq:loglsl_full}) derived in Sec.~\ref{sec:loglsl_theory}. Since the coefficient of $z^2$ in (\ref{eq:conf_expansion}) is negative, the phase factor is $e^{i\alpha}=-1$. We get the simple result
\begin{equation}\label{eq:loglsl_xx}
 g(x)=\frac{\xi\, c}{24}\left(\frac{1}{x(1-x)}-1\right).
\end{equation}
$\xi$ is the extrapolation length. Notice that the precise forms of Eqs.~(\ref{eq:scaling},\ref{eq:f},\ref{eq:loglsl_xx}) rely on the technical assumption that the (conformal) boundary condition be the same everywhere. This assumption will be relaxed in the following subsection (\ref{sec:lbf_bcc}). 

\begin{figure}[ht]
\begin{center}
 \includegraphics[width=10cm]{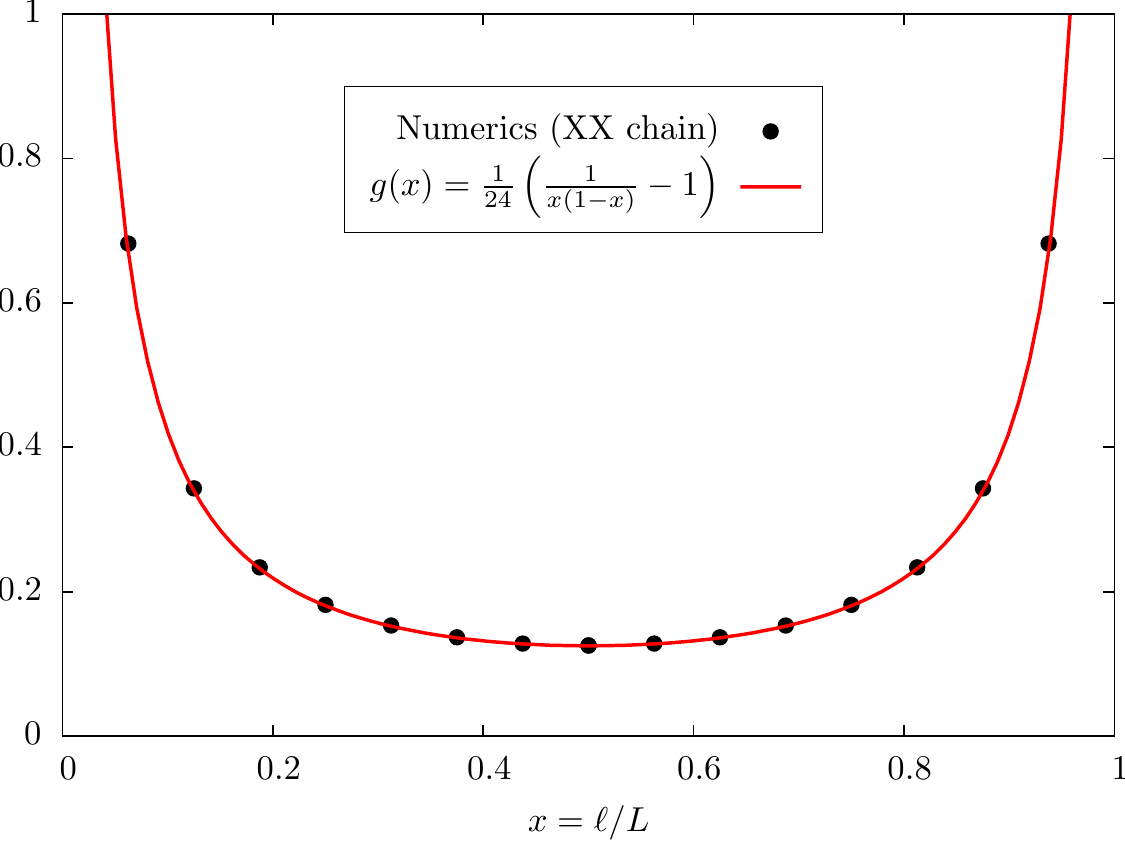}
\end{center}
\caption{Numerical extraction of the $\log(L)/L$ term $g(x)$ in the XX chain at half-filling. The result is compared with the CFT prediction of Eq.~(\ref{eq:loglsl_xx}) with $c=1$ and $\xi=1$.}
\label{fig:loglsl}
 \end{figure}

Let us now check this formula in the XX chain at half-filling, a simple critical model with central charge $c=1$. In general $\xi$ is a non-universal quantity, it depends on the specifics of the lattice model. Here it can be obtained from the standard bosonization argument that for a chain of $L$ sites between $x=1$ and $x=L$, the boundary conditions on the free bosonic field $\varphi$ have to be (see e.g. \cite{EggertAffleck}) $\varphi(0)=\varphi(L+1)=0$. We therefore have $\xi=1$. 
As is well known, this model can be solved exactly using a mapping onto free fermions \cite{lieb1961two}, and the overlap we are interested in can be expressed as a determinant:
\begin{equation}
 \mathcal{F}(x,L)=-2\log\left(\det_{1\leq i,j\leq L/2} M_{ij}\right).
\end{equation}
The matrix elements are given in Sec.~(\ref{sec:exact}), we refer to it for the derivation. Using standard numerical routines, it is then straightforward to evaluate $\mathcal{F}$ numerically for large system sizes. The numerical results are shown in Fig.~\ref{fig:loglsl}.
\\
 In practice we extract the $L^{-1}\log L$ term by fitting the linear combination
\begin{equation}\label{eq:det_xx}
 \mathcal{F}(x,L)-\frac{1}{8}\log L-f(x)
\end{equation}
to the form $a_0+b_1L^{-1}\log L+a_1 L^{-1}$ for system sizes $L=512,1024,2048,4096,8192$ and fixed $x$.  
 The data agrees perfectly with the CFT prediction (\ref{eq:loglsl_xx}), confirming our analysis. Interestingly, the shape is convex-up, as opposed to the concave-down form of the constant term $f(x)$. Let us finish with two remarks regarding the extrapolation length $\xi$
\begin{itemize}
 \item There exist a known exact mapping from an XX chain of length $L$ at half-filling to two critical Ising chains of length $L/2$ \cite{PeschelSchotte,Turban}. As a consequence \cite{Bipartite_fidelity}, $\mathcal{F}_{\rm XX}(x,L)=2\mathcal{F}_{\rm Ising}(x,L/2)$. This result agrees with the well known value of the central charge $c=1/2$ and an extrapolation length $\xi=1/2$.
 \item It is possible to change the extrapolation length, while keeping the central charge constant. To do so, one can change the hoping amplitude at the two boundaries of the XX chain: $H=-\sum_j  t_{j}\left (c_{j+1}^\dag c_j +h.c\right)$, with e.g. $t_{j}=1$ for $j=2,3,\ldots,L-2$ and $t_1=t_{L-1}=t\neq 1$. We also checked numerically that our results are compatible with this.
\end{itemize}
\subsection{The effect of boundary changing operators}
\label{sec:lbf_bcc}
It is also interesting to consider a slightly more general LBF, defined as 
\begin{equation}
 \mathcal{F}_{\alpha\beta\gamma\delta}=-\log \big|\leftidx{_{\alpha\delta\!}}{\left\langle A\cup B||A\right\rangle}{_{\alpha\beta}}\left.\otimes| B\right\rangle_{\gamma\delta} \big|^2,
\end{equation}
where we have imposed different conformal boundary conditions $\alpha\delta$, $\alpha \beta$, $\gamma\delta$ on the states $|A\cup B\rangle$, $|A\rangle$, $|B\rangle$, as is shown in Fig.~\ref{fig:chain_bcc}.
\begin{figure}[ht]
\centering
 \begin{tikzpicture}
  \draw [ultra thick] (0,0) -- (6,0);
  \draw [ultra thick] (0,-0.15) -- (0,0.15);
  \draw [ultra thick] (2.4,-0.15) -- (2.4,0.15);
  \draw [ultra thick] (6,-0.15) -- (6,0.15);
  \draw [semithick,<->] (0.05, -0.4) -- (2.35,-0.4);
    \draw [semithick,<->] (2.45, -0.4) -- (5.95,-0.4);
    \draw (1.2,-0.7) node {$\ell$};
    \draw (4.2,-0.7) node {$L-\ell$};
    \draw[color=red] (-0.2,0.3) node {${\bf \alpha}$};
    \draw[color=red] (2.2,0.3) node {${\bf \beta}$};
    \draw[color=red] (2.6,0.265) node {${\bf \gamma}$};
    \draw[color=red] (6.2,0.3) node {${\bf \delta}$};
 \end{tikzpicture}
\caption{Bipartition of the chain with different boundary conditions imposed everywhere. Notice that the change between $\beta$ and $\gamma$ directly affects the leading Cardy-Peschel term.}
\label{fig:chain_bcc}
\end{figure}
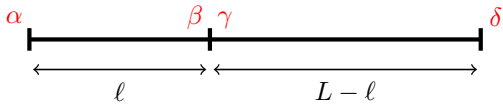\\
The scaling form of the LBF is changed to 
\begin{equation}
 \mathcal{F}=\left(\frac{c}{8}+h_{\beta\gamma}\right)\log L+f_{\alpha\beta\gamma\delta}(x)+g_{\alpha\beta\gamma\delta}(x)\frac{\log L}{L}+O(1/L),
\end{equation}
where the leading term is only affected by the change in boundary condition from $\beta$ to $\gamma$, and is obtained from the Cardy-Peschel formula, Eq.~(\ref{eq:cardypeschel_bcc}). It is possible to generalize the result established in Sec.~\ref{sec:loglsl_theory} to take into account such changes in boundary condition. The $L^{-1}\log L$ contribution we are interested in is modified to
\begin{equation}\label{eq:loglsl_bcc}
 g_{\alpha\beta\gamma\delta}(x)=\xi\times \left[h_{\delta\alpha}-\frac{c}{24}+\left(\frac{c}{24}-h_{\alpha\beta}\right)\frac{1}{x}+\left(\frac{c}{24}-h_{ \gamma\delta}\right)\frac{1}{1-x}
 \right],
\end{equation}
where $h_{ij}$ is the dimension of the operator changing the boundary condition from $i$ to $j$. 
Notice that also the result of \cite{Bipartite_fidelity} for $f_{\alpha\beta\gamma\delta}(x)$ is modified in a non-trivial way. The exact expression for $f_{\alpha\beta\gamma\delta}$, as well as the derivation of both results, is given in \ref{sec:app_bcc}.

Such changes in conformal boundary condition can be achieved by explicitly changing the boundary condition on the lattice (e.g. reversing a spin), but are sometimes only apparent  in the effective field theory description. We will consider the two cases in Sec.~\ref{sec:lbf_isingbcc} and Sec.~\ref{sec:lbf_filling}, to check the main result of this section, Eq.~(\ref{eq:loglsl_bcc}).
\subsubsection{Reversed spins in the Ising chain}
\label{sec:lbf_isingbcc}
\begin{figure}[ht]
 \begin{center}
  \includegraphics[width=10cm]{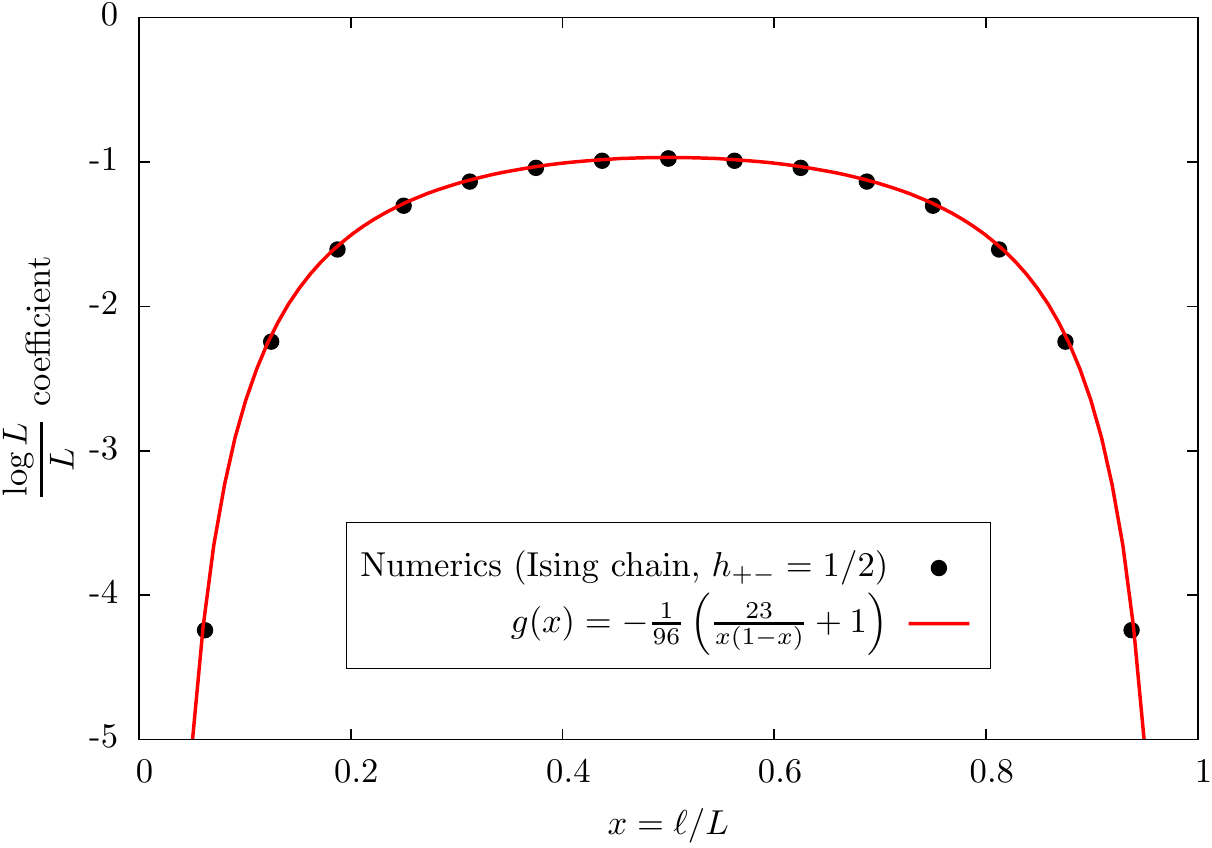}
 \end{center}
 \caption{Numerical extraction of the $\log L/L$ term in an Ising chain in transverse field with boundary conditions $+--+$, and comparison with the CFT prediction of Eq.~(\ref{eq:loglsl_ictf})}
 \label{fig:ising}
\end{figure}

Let us consider the first excited state of an open Ising chain in transverse field \cite{lieb1961two}. We use the convention where the spins are measured in the basis of the eigenstates of the transverse field $\sigma_j^z$. As is well known, it is twofold degenerate with the spins at both ends pointing in opposite directions, $\sigma_1^x=\pm 1=-\sigma_L^x$. Excited states therefore naturally change the boundary conditions. For example, one can study the following modified LBF, with $\alpha=\delta=\uparrow$ and $\beta=\gamma=\downarrow$,
 \begin{equation}
 \mathcal{F}_{\uparrow\downarrow\downarrow\uparrow}=-\log \big|\leftidx{_{\uparrow\uparrow\!}}{\left\langle A\cup B||A\right\rangle}{_{\uparrow\downarrow}}\left.\otimes| B\right\rangle_{\downarrow\uparrow} \big|^2.
\end{equation}
In this case we have $h_{\alpha\beta}=h_{\gamma\delta}=h_{\uparrow\downarrow}$ and $h_{\delta\alpha}=0$, so that we get
\begin{equation}\label{eq:loglsl_ictf}
g(x)=\left[\left(\frac{c}{24}-h_{\uparrow\downarrow}\right)\frac{1}{x(1-x)} -\frac{c}{24}\right]\frac{\xi \log L}{L}.
\end{equation}
For the Ising universality class, we have $c=1/2$, $h_{\uparrow\downarrow}=1/2$ \cite{Cardy_bccoperatorcontent}, and we have previously found $\xi=1/2$. We present numerical checks of this formula in Fig.~\ref{fig:ising}, using the same method as in the previous section. 
\\
The results are again in excellent agreement with our prediction. Notice that the boundary condition changing operators have a huge impact on the result, since they change in this case the sign of $g(x)$.
\subsubsection{Changing the filling fraction in a Luttinger liquid}
\label{sec:lbf_filling}
Let us now turn our attention to a slightly more subtle case, where the change in boundary condition is not obvious in the underlying lattice model, but rather emergent in the continuum limit. The prototype Hamiltonian we have in mind is that of an XXZ chain in magnetic field 
\begin{equation}
 H=\sum_{j=1}^{L-1}\left(\sigma_j^x\sigma_{j+1}^x+\sigma_j^y\sigma_{j+1}^y+\Delta \sigma_j^z\sigma_{j+1}^z\right)-h\sum_{j=1}^{L}\sigma_j^z.
\end{equation}
The magnetization sector $M=\frac{1}{L}\sum_{j} \sigma_j^z$ in which the ground state lies is a non-trivial function of the magnetic field $h$. We are interested here in situations where $M\neq 0$ and therefore $h\neq 0$. As has been noted in Refs~\cite{Fermiedge2,Fermiedge3,XXZ_bethenum2}, non trivial magnetization sectors may induce changes in boundary condition. To see this, it is most convenient to consider the special case $\Delta=0$, that is equivalent to free-fermions hoping on the line. The magnetization is related to the fermion density by $M=(\rho-1/2)$, and the ground-state energy is given by
\begin{equation}
 E_0(L)=-\sum_{m=1}^{\rho L} \cos \left(\frac{m\pi}{L+1}\right)
\end{equation}
up to an unimportant constant. The Fermi momentum is $k_F=\rho\pi$ and the Fermi speed $v_F=\sin k_F$. The low-energy physics can be accessed \cite{Affleck_c,Cardy_c} using the Euler-Maclaurin formula
\begin{equation}
 E_0(L)=\alpha L+\beta -\frac{\pi v_F}{24 L}\left[c-24 h_\rho\right]+O(1/L^2).
\end{equation}
$c=1$ is the central charge. $h_\rho$ is the dimension of a boundary changing operator, given here by
\begin{equation}
 h_\rho=\frac{1}{2}\left(\rho-\frac{1}{2}\right)^2.
\end{equation}
It is proportional to the so-called ``phase shift'' emphasized in the bosonization literature. Its presence means that although the boson height field obeys a Dirichlet boundary condition at both ends of the chain, there is still a height difference between the two ends, that needs to be taken into account. For general $\Delta$ the phase shift is to our knowledge  not known analytically (see however Ref.~\cite{XXZ_bethe1} for closely related calculations). We refer to \cite{XXZ_bethenum2} for a more thorough description.

\begin{figure}[ht]
 \begin{center}
  \includegraphics[width=8.2cm]{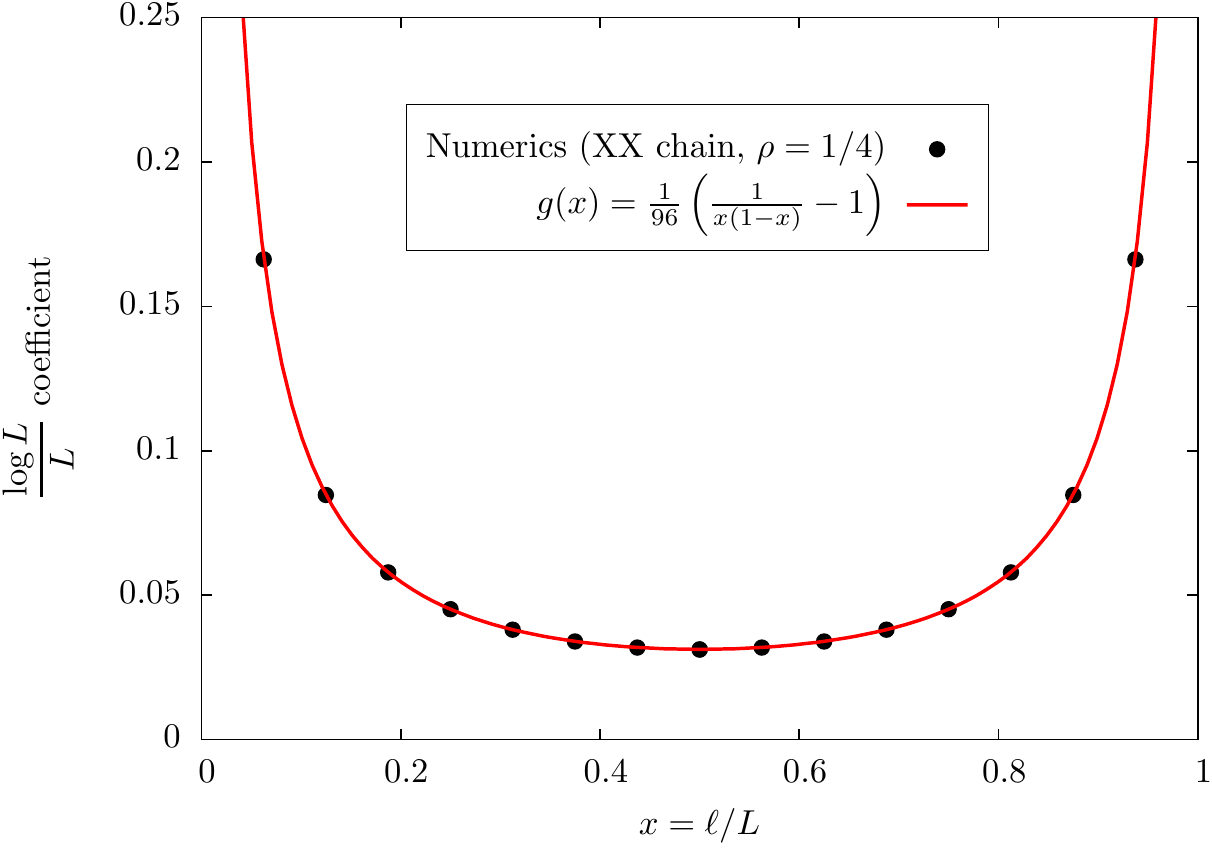}\hfill
  \includegraphics[width=8.2cm]{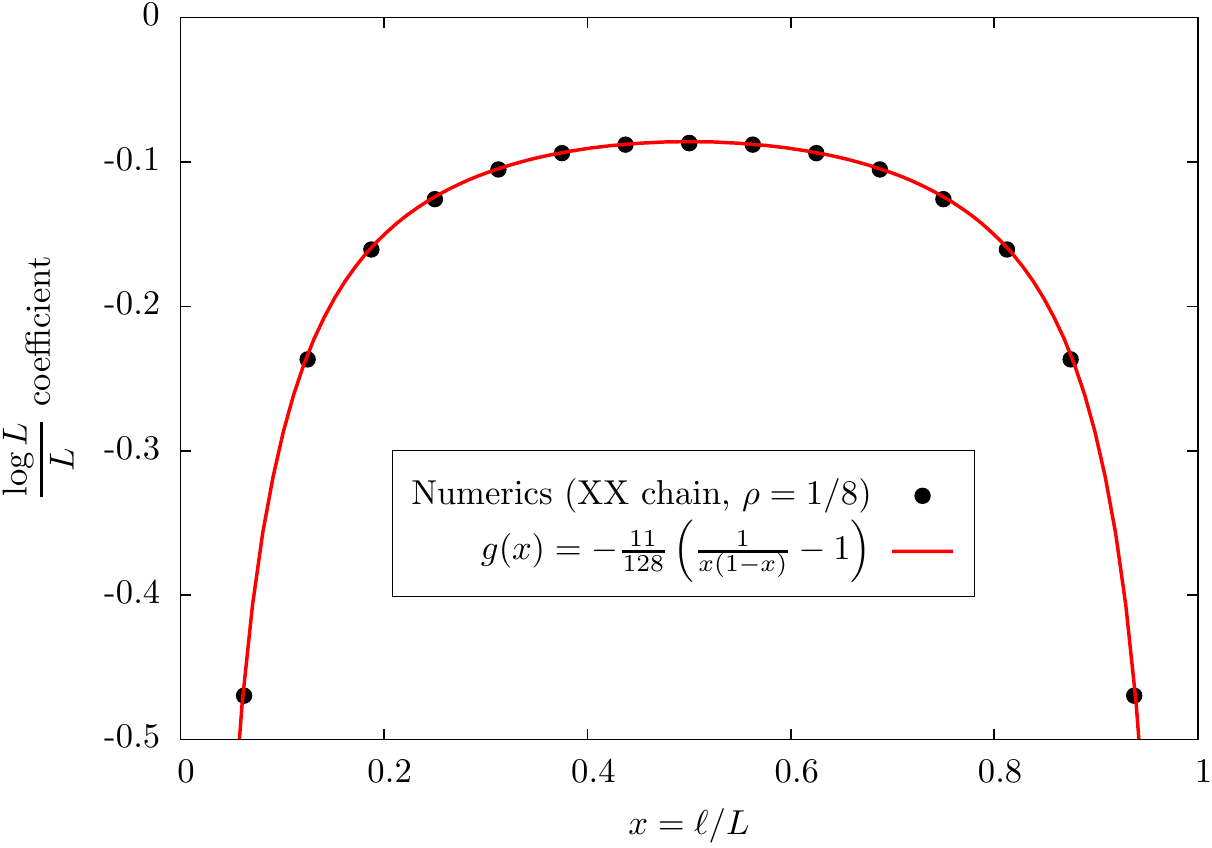}
 \end{center}
 \caption{Numerical extraction of the $\log L/L$ term in the XX chain away from half filling, and comparison with the CFT prediction of Eq.~(\ref{eq:loglsl_xxrho}). \emph{Left:} Filling fraction $\rho=1/4$.\emph{Right:} Filling fraction $\rho=1/8$. In this case the dimension of the boundary operator is large enough to inverse the shape.}
 \label{fig:filling}
\end{figure}

For the ground-state $\ket{A\cup B}$, this means one can set $\varphi(0)=0$, and therefore $\varphi(L)=\Delta \varphi$. Similarly for $\ket{A}$ (resp. $\ket{B}$) we have $\varphi(0)=0$ and $\varphi(\ell^{-})=\Delta \varphi$ (resp. $\varphi(\ell^+)=0$ and $\varphi(L)=\Delta \varphi$). We have then to take into account four boundary changing operators with dimension $h_{\alpha\beta}=h_{\beta\gamma}=h_{\gamma\delta}=h_{\delta\alpha}=h_\rho$ in the calculation of the LBF. Using Eq.~(\ref{eq:loglsl_bcc}), we find that the $L^{-1}\log L$ term is modified to
\begin{equation}\label{eq:loglsl_xxrho}
 g_\rho(x)=\xi \times \left(\frac{c}{24}-h_\rho\right)\left(\frac{1}{x(1-x)}-1\right).
\end{equation}
We performed checks of this formula in the $\Delta=0$, case. The results are presented in Fig.~\ref{fig:filling}, for two filling fractions $\rho=1/4$ and $\rho=1/8$.
\\
The numerics show once again a perfect agreement with our formulas. We did not perform any numerical checks for general $\Delta$, but we expect our formulas to be valid, with the more general
\begin{equation}
 h_\rho=\frac{R^2}{2}\left(\frac{\delta}{\pi}\right)^2.
\end{equation}
$R$ is the compactification radius\footnote{At half-filling it is known exactly \cite{XXZ_bethe1}, $R=\sqrt{2-(2/\pi) \arccos \Delta}$}, obtained by solving numerically the integral equation derived from the Bethe-ansatz \cite{XXZ_bethe1,XXZ_bethenum1,XXZ_bethenum2,XXZ_bethenum3}. $\delta/\pi$ is the phase shift \cite{XXZ_bethenum2}. 

Let us now come back to the orthogonality exponent. As we have seen the leading behavior is given by $\mathcal{F}\sim (c/8+h_\rho)\log L$, so that the LBF can be used to measure $h_\rho$, without having to know a priori the Fermi speed $v_F$. However the (single interval) entanglement entropy is not sensitive to this, and always scale as $S\sim (c/6)\log L$. One might wonder why this is case. Such a behavior can be traced back to the fact that the replicated geometry used to calculate the EE doesn't really have any boundary, rather twists that allow to jump from one cylinder to the next. The LBF involves at least an explicit boundary (the cut), and the possible appearance of $h_\rho$ is intimately connected to the fact that for $c=1$ compact theories, Dirichlet stands for an (infinite) parameter family of conformal boundary conditions \cite{OshikawaAffleck1,OshikawaAffleck2}: in bosonization language the forward scattering is exactly marginal. For $c<1$ rational theories the number of conformal boundary conditions is finite, and a natural ``free '' boundary condition should not generate any change in boundary 
condition at the cut\footnote{Except of course if one explicitly forces the matter, as we did in \ref{sec:lbf_isingbcc}.}.  

\subsection{An exact result for the symmetric cut}
\label{sec:lbf_exact}
For a symmetric cut $x=\ell/L=1/2$, it was shown \cite{PhDStephan} that the determinant (\ref{eq:matelements}) can be computed explicitly. Let us note $N=\rho L$ the number of fermions in the chain, $\theta_k=k\pi/(L/2+1)$ and $\phi_l=l\pi/(L+1)$. Then the fidelity is given by 
\begin{equation}\label{eq:lbfxx_1}
 \mathcal{F}=-2\log \left|D_N\right|,
\end{equation}
with
\begin{equation}\label{eq:lbfxx_2}\fl
 D_N=
 \frac{
 \prod_{k=1}^{N/2}\sin^2 \theta_k\prod_{l=1}^{N}\sin\left(\frac{l\pi+\phi_l}{2}\right)
 \prod_{1\leq k<l\leq N/2}\left(\cos \theta_k-\cos \theta_l\right)^2 \prod_{1\leq k<l\leq N}^{(k-l){\rm \,even}}\left(\cos \phi_k-\cos \phi_l\right)
 }{\left[(L/2+1/2)(L/2+1)\right]^{N/2}
  \prod_{k=1}^{N/2}\prod_{l=1}^{N} \left(\cos\theta_k-\cos \phi_l\right)
 }.
\end{equation}
From this formula it is possible to perform a systematic asymptotic expansion, and prove the CFT result at $x=1/2$. The details of the expansion, as well as the derivation of Eqs.~(\ref{eq:lbfxx_1},\ref{eq:lbfxx_2}) are given in \ref{sec:exact}. The final result reads
\begin{equation}
 \mathcal{F}(L)=\sum_{k=0}^{\infty}\left(\gamma_k\log L+\mu_k\right)L^{-k}.
\end{equation}
The equality is meant in the sense of asymptotic series. The $\gamma_k$ are given by
\begin{eqnarray}
 \gamma_0&=&\frac{1}{8}+h_\rho,\\
 \gamma_k&=&(-1)^{k+1}\left(\frac{1}{24}-h_\rho\right)\left(2^{k+1}-1\right)\quad,\quad k\geq 1.
\end{eqnarray}
$\gamma_0$ is nothing but the Cardy-Peschel term, and $\gamma_1$ is the $L^{-1}\log L$ term. Both match perfectly the CFT prediction. The non universal $\mu_k$ terms are much harder to obtain from the exact solution, and less interesting to us. We only computed $\mu_0$ in the appendix. It is quite striking that all the expressions for the $\gamma_k$ are very simple. Such an observation is consistent with the general idea that terms protected by a logarithm are less dependent on the numerous UV cutoff present in the systems. We did not attempt to calculate them using CFT for $k\geq 2$. The main reason for that is that they are slightly less robust than the $L^{-1}\log L$; in particular they are not semi-universal, in the sense of Sec.~\ref{sec:why}. Note finally that this exact result also applies to the Ising chain, because $\mathcal{F}_{XX}(x,L)=\mathcal{F}_{Ising}(x,L/2)$, due to the exact mapping of Refs.~\cite{PeschelSchotte,Turban}.

\subsection{Correction to the one-point function: two examples}
\label{sec:onepoint_examples}
\begin{figure}[htbp]
	\centering
	(a)
	\begin{tikzpicture}[scale=0.7]
		\filldraw[gray!50] (-3.5,-3) rectangle (3.5,3);
		\draw[very thick] (-2,0) -- (-0.2,0);
		\draw[->] (-4.2,-3.8) -- (3,-3.8) node[right]{$x$};
		\draw[->] (-4.2,-3.8) -- (-4.2,3) node[left]{$y$};
		\draw (-4.1,0) -- (-4.3,0) node[left]{$0$};
		\draw (-2,-3.7) -- (-2,-3.9) node[below]{$0$};
		\filldraw (2,0) circle (3pt) node[below]{$\phi(w,\overline{w})$};
		\draw (2.2,-2.4) node{$w=x+iy$};
		\draw[<->] (-2,-0.2) -- (-0.2,-0.2);
		\draw (-1.1,-0.5) node{$L$};
		\draw[<->] (0,0) -- (1.6,0);
		\draw (0.9,0.5) node{$\ell$};
	\end{tikzpicture}
	\qquad (b) 
	\begin{tikzpicture}[scale=0.7]
		\filldraw[gray!50] (-4,-2) rectangle (4,2);
		\draw[thick] (-4,2) -- (4,2) (-4,-2) -- (4,-2);
		\draw[thick] (0,0) -- (4,0);
		\draw[->] (-4.5,-2.5) -- (3,-2.5) node[right]{$x$};
		\draw[->] (-4.5,-2.5) -- (-4.5,2) node[left]{$y$};
		\draw (-4.4,0) -- (-4.6,0) node[left]{$0$};
		\draw (0,-2.4) -- (0,-2.6) node[below]{$0$};
		\filldraw (0,1.3) circle (3pt) node[right]{$\phi(w,\overline{w})$};
		\draw (2.7,-1.5) node{$w=x+iy$};
		\draw[<->] (-3,-1.9) -- (-3,1.9);
		\draw[<->] (0.5,-1.9) -- (0.5,-0.1);
		\draw (-2.6,0) node{$L$};
		\draw (1,-0.7) node{$L/2$};
	\end{tikzpicture}
	\caption{There is a correction of order $O(L^{-1}\log L)$ in the one-point function. Here are the two geometries we use as examples.}
	\label{fig:slit_one_point}
\end{figure}
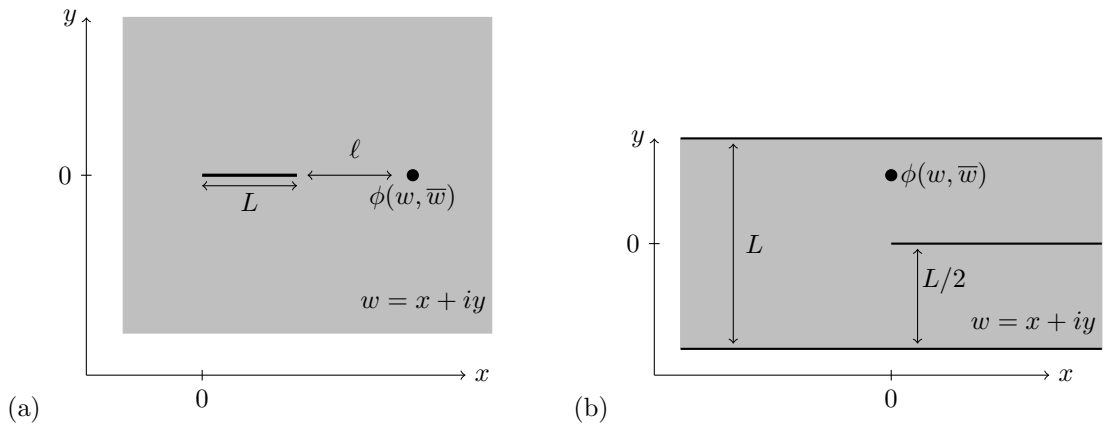

In this section we provide numerical checks of our formula (\ref{eq:one_point}) for the $O(L^{-1}\log L)$ correction to the one-point function close to a corner. The two geometries we consider are pictured in Fig.~\ref{fig:slit_one_point}.
\subsubsection{One-point function in the plane with a slit}
Our first example is about the one-point function $\left< \phi(w,\overline{w}) \right>$ in the geometry shown in Fig. \ref{fig:slit_one_point}(a). The domain is the complex plane minus a segment of length $L$: $\mathbb{C}\setminus [0,L]$. 
To compute the one-point function $\left< \phi(w,\overline{w})\right>$ in conformal field theory, we first need a mapping $z \mapsto w(z)$ from the upper half-plane to $\mathbb{C}\setminus [0,L]$. We chose the following one:
\begin{equation}
	\label{eq:mapping_slit}
	w(z) \, = \, \frac{z^2}{1+\frac{z^2}{L}}. 
\end{equation}
Using the inverse of this mapping, and the fact that the one-point function in the upper half-plane is proportional to $1/|{\rm Im\,} z(w) |^{2h}$, we find that
the one-point function in $\mathbb{C} \setminus [0,L]$ is, at the leading order:
\begin{eqnarray}
	\left< \phi(w,\overline{w}) \right> & =& \left| \frac{dz}{dw} \right|^{2h} \left< \phi(z,\overline{z}) \right> \, \propto \,   \left(\frac{1}{ \left| \frac{w}{L}(1-\frac{w}{L}) \right|^{\frac{1}{2}} ~ {\rm Im \,} \sqrt{w(L-\overline{w})} } \right)^{2h}.
\end{eqnarray}
Now we turn to the subleading correction of order $O(\log L/L)$. The mapping (\ref{eq:mapping_slit}) has the behaviour we want at small $z$,
\begin{equation}
	w(z) \, \simeq \, z^2(1 - \frac{1}{L}z^2+ \dots),
\end{equation}
so we can apply our formula (\ref{eq:one_point}) directly. We find that there is a subleading contribution $\delta \left<\phi(w,\overline{w}) \right>$ coming from the corner at position $w =0$:
\begin{eqnarray}
	\frac{\delta \left< \phi(w,\overline{w})\right>}{\left< \phi(w,\overline{w}) \right>} & =& - \frac{\xi ~h}{\pi} \frac{{\rm Im\,} \sqrt{\frac{w}{L} (1- \frac{\overline{w}}{L})}}{ \left| w/L \right|^2} \times  \frac{\log(L/a)}{L} \, + \, O(1/L).
\end{eqnarray}
There is another contribution coming from the corner at position $w=L$. This can be obtained simply by symmetry, by substituting $w \rightarrow L-\overline{w}$. In total, the subleading correction to the one-point function coming from the two corners is:
\begin{equation}
	\frac{\delta \left< \phi(w,\overline{w})\right>}{\left< \phi(w,\overline{w}) \right>} =  - \frac{\xi ~h}{\pi} \left[ \left({\rm Im\,} \sqrt{\frac{w}{L} (1- \frac{\overline{w}}{L})}\right)^2  \left( \frac{1}{ \left| w/L \right|^2}   +  \frac{1}{ \left| 1-w/L \right|^2} \right) \right] \times  \frac{\log(L/a)}{L} \, + \, \dots
\end{equation}
When $w$ lies on the real axis, at a distance $\ell>0$ from the right corner, the results simplify into
\begin{eqnarray}\label{eq:efp_cft1}
 \left\langle \phi\right\rangle&\propto&\left(\frac{1}{\pi \frac{\ell}{L}\left[1+\frac{\ell}{L}\right]}\right)^{2h}L^{-2h},\\\nonumber\\\label{eq:efp_cft2}
  \frac{\delta \left< \phi\right>}{ \left< \phi\right>}&=&- \frac{\xi ~h}{\pi} ~ \left(2 + \frac{1}{\frac{\ell}{L} (1+ \frac{\ell}{L}) } \right)  \frac{\log (L/a)}{L} \,+\, O(1/L).
\end{eqnarray}
It is possible to check this formula in the Ising chain in transverse field at the critical point, by studying the quantity
\begin{equation}
 \braket{\sigma_{L+\ell}}_{\rm slit}=2\braket{P_{L+\ell}}_{\rm slit}-1,
\end{equation}
where $P_i=(1+\sigma_i)/2$ is the projector onto the state with up spin $i$. We measure the spins in the basis of the eigenstates of the transverse field Pauli matrices. Upon performing a Jordan-Wigner transformation, $P_i$ projects onto to a state where the site $i$ is occupied by a fermion. A lattice realization of the slit can be obtained by setting
\begin{eqnarray}\label{eq:efp_onepoint_ictf1}
 \langle P_{L+\ell}\rangle_{\rm slit}&=&\frac{\bra{\Omega} P_{L+\ell}\prod_{j=1}^{L} P_j\ket{\Omega}}{\bra{\Omega} \prod_{j=1}^{L} P_j \ket{\Omega}},
\end{eqnarray}
where $|\Omega\rangle$ is the ground-state of the chain. The sequence of $L$ up spins is expected \cite{EFP} to renormalize to a free conformal invariant boundary condition. Note that the denominator in (\ref{eq:efp_onepoint_ictf1}) is nothing but the emptiness formation probability \cite{Franchini}; we refer to Ref.~\cite{EFP} for a detailed CFT treatment. Both terms in the ratio (\ref{eq:efp_onepoint_ictf1}) can be obtained as simple determinants, using free fermions techniques \cite{Franchini}. In general we have 
\begin{equation}
 \bra{\Omega}  \prod_{i=1}^{n} P_{x_i}\ket{\Omega}=\det_{1\leq i,j\leq n}\left(\frac{\delta_{ij}}{2}+\frac{1}{2\pi (x_i-x_j+1/2)}\right),
\end{equation}
which allows for fast numerical computations on very large system sizes.  For a fixed aspect ratio $\ell/L$, we observe that the one point function scales as
\begin{equation}
 \braket{\sigma_{L+\ell}}_{\rm slit}-\frac{2}{\pi}=\frac{f(\ell/L)}{L}\left(1+g(\ell/L)\frac{\log L}{L}\right)+O(1/L^2),
\end{equation}
so that the leading term is proportional to $L^{-1}$. By direct comparison with (\ref{eq:efp_cft1}) we recover the dimension $h=1/2$ of the fermionic field.  
\begin{figure}[htbp]
\includegraphics[width=7.8cm]{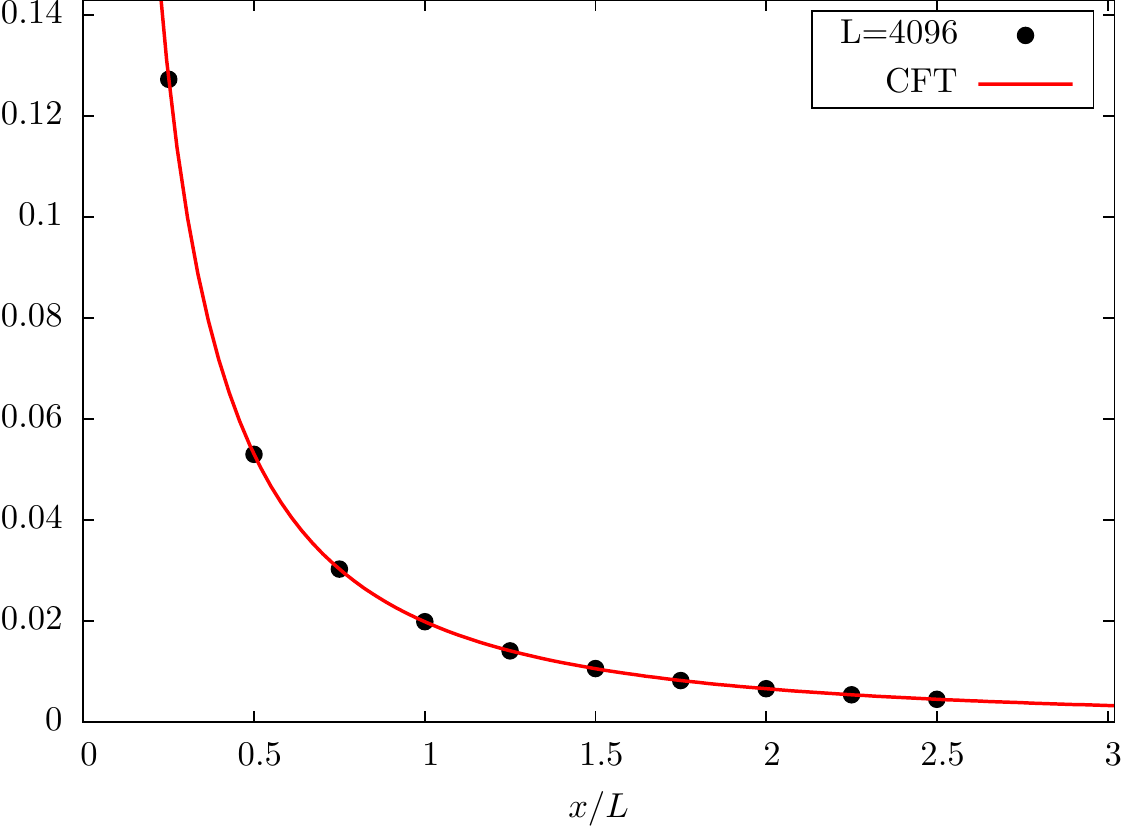}
\hfill
\includegraphics[width=7.8cm]{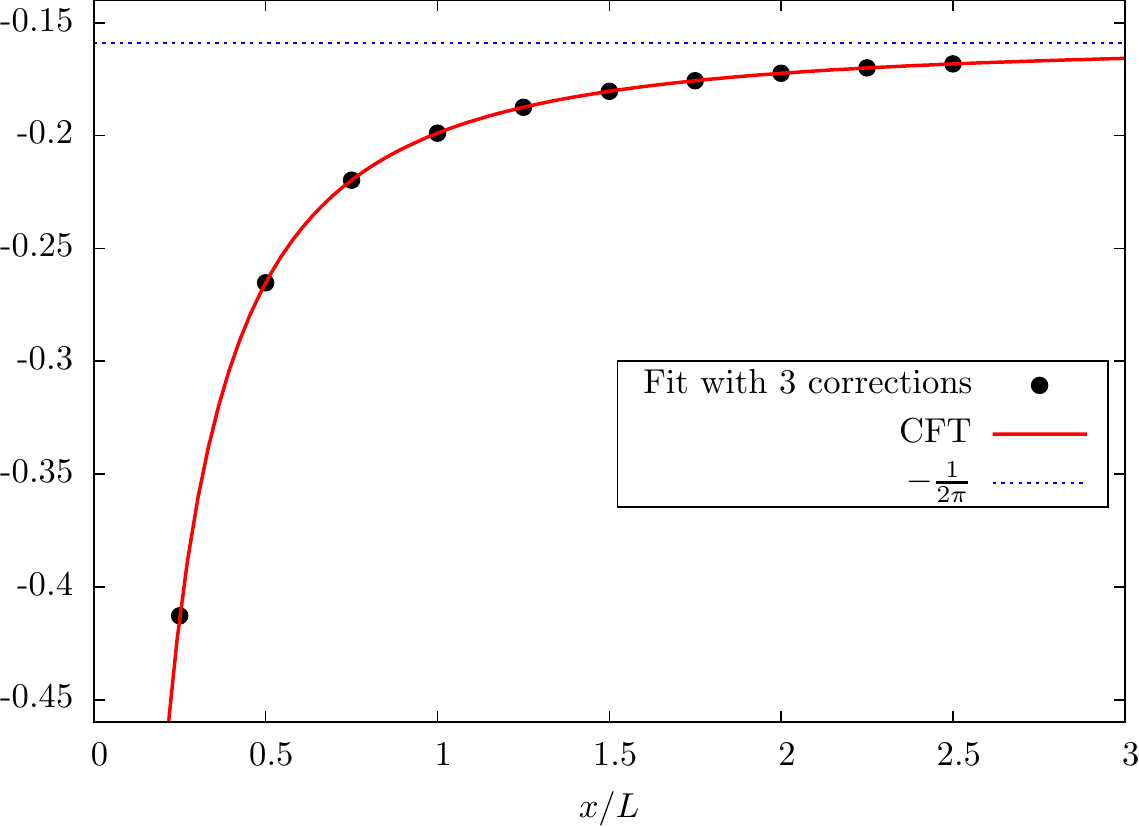}
 \caption{\emph{Left:} Leading term in the one-point function. \emph{Right:} Subleading $L^{-1}\log L$ term in the one-point function. In both cases the numerics match perfectly the CFT calculation.}
 \label{fig:onepoints_efp}
\end{figure}
The numerical results for $f(\ell/L)$ are shown in Fig.~\ref{fig:onepoints_efp} on the left, for $L=4096$. We observe that the conformal scaling of (\ref{eq:efp_cft1}) is perfectly obeyed. We also notice that the proportionality coefficient is very close to $1/4$, a result that is probably exact. The identification of the subleading term requires some additional care. In practice, we fit $\braket{\sigma_{L+\ell}}_{\rm slit}-\frac{2}{\pi}$ to the form $a_1L^{-1}+b_2L^{-2}\log L+a_2L^{-2}$ for very large system sizes, and plot the ratio $b_2/a_1$. The data, shown in Fig.~\ref{fig:onepoints_efp} on the right matches perfectly the CFT calculation (\ref{eq:efp_cft2}) with the expected values $h=1/2$ and $\xi=1/2$. 

\subsubsection{One-point function in the 'pants' geometry}

Our second example is the one corresponding to the domain shown in Fig. \ref{fig:slit_one_point}(b). A conformal mapping from the half-plane to this ``pants geometry'' is
\begin{equation}
	w(z) \, = \, - \frac{L}{2\pi} \log \left(1 - \frac{2\pi z^2}{L} \right).
\end{equation}
With the inverse of this mapping, we get the leading order of the one-point function, exactly as we did previously for the slit:
\begin{eqnarray}
	\left< \phi(w,\overline{w}) \right> & \propto &   \left[\frac{1}{L ~\left| \sinh \frac{\pi w}{L} \right|^{\frac{1}{2}} ~ \left( {\rm Im\,} \sqrt{1-e^{-\frac{2\pi w}{L}}}  \right)} \right]^{2h} .
\end{eqnarray}
Again, we have chosen the mapping such that it has the following expansion for small $z$:
\begin{equation}
	w(z) \, \simeq \, z^2 \left(1 + \frac{\pi}{L} z^4 + \dots\right) ,
\end{equation}
so we can apply directly our formula (\ref{eq:one_point}) for the subleading correction due to the stress-tensor. We find
\begin{eqnarray}
	\frac{\delta \left< \phi(w,\overline{w})\right>}{ \left< \phi(w,\overline{w})\right>} &=& -\frac{\xi~h}{2}   \left( e^{\frac{\pi}{L} {\rm Re \, }w} ~  \frac{{\rm Im \,} \sqrt{1-e^{-\frac{2\pi w}{L}}}}{ \left| \sinh\frac{\pi w}{L} \right| }  \right)^2~ \frac{\log (L/a)}{L} \, + \, O(1/L)
\end{eqnarray}
We now proceed to check these results numerically in the Ising chain for $w$ on the imaginary axis, at a distance $y>0$ from the the corner (see Fig.~\ref{fig:slit_one_point}). With $h=1/2$ and $\xi=1/2$, the above formulas simplify into
\begin{eqnarray}
\left\langle \phi(w,\overline{w})\right\rangle&\propto&\frac{1}{L \sin \frac{\pi y}{L}\cos\left(\frac{\pi}{4}+\frac{\pi y}{2L}\right)},\\\nonumber\\
 \frac{\delta \left< \phi(w,\overline{w})\right>}{ \left< \phi(w,\overline{w})\right>} &=& -\frac{1}{8}  \left( \frac{1}{\sin \frac{\pi y}{L} }-1 \right)  ~ \frac{\log (L/a)}{L} \, + \, O(1/L).
\end{eqnarray}
\begin{figure}[htbp]
 \includegraphics[width=8cm]{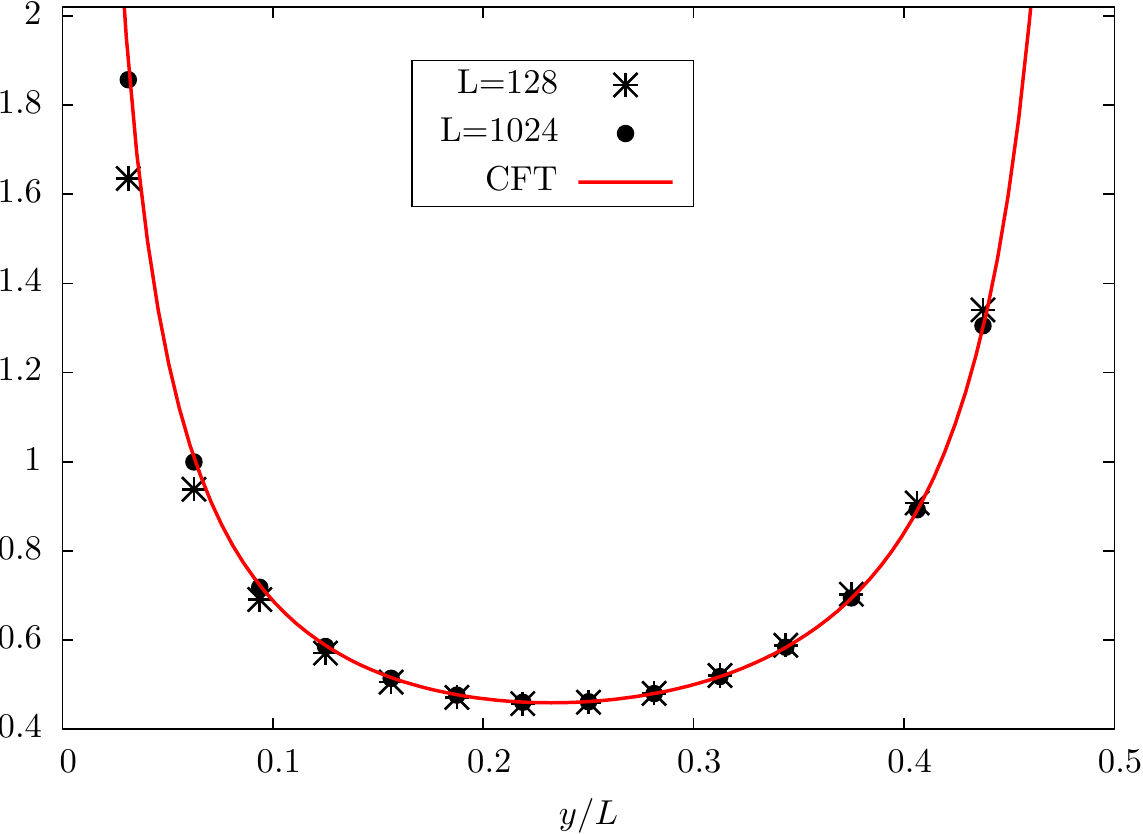}
 \hfill
 \includegraphics[width=8cm]{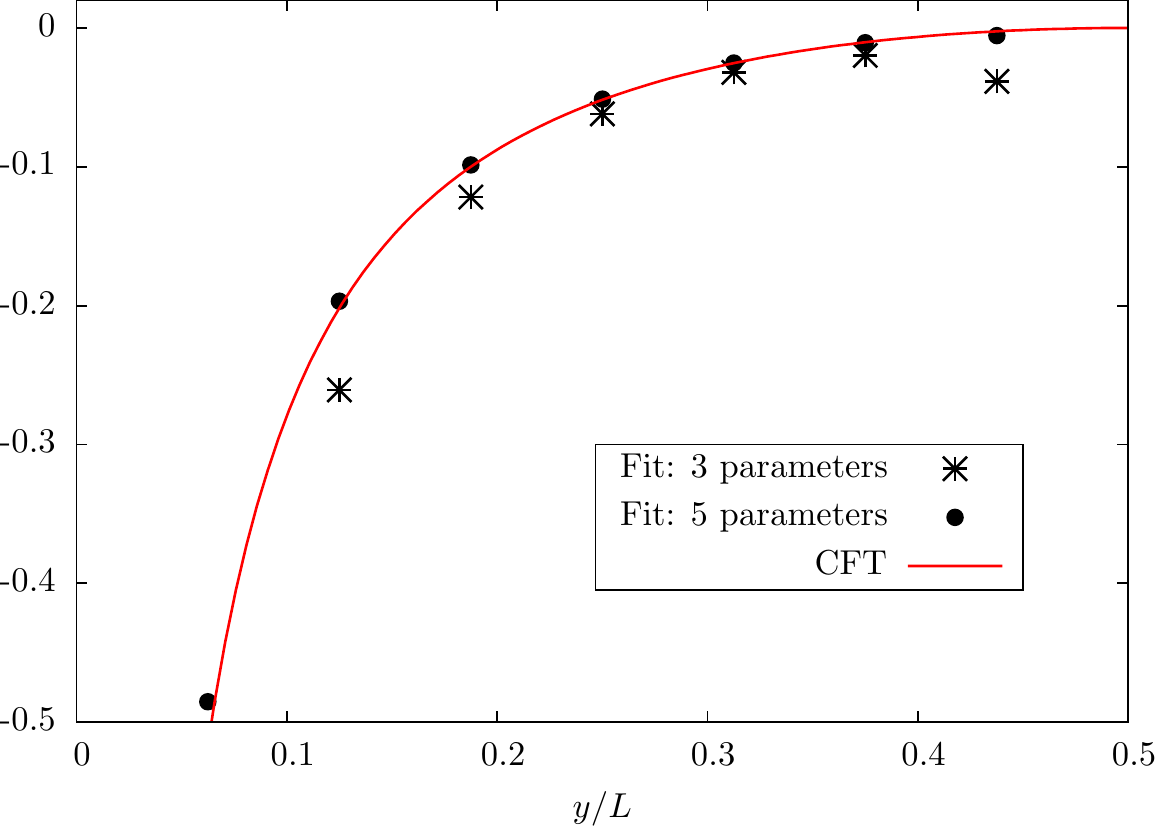}
 \caption{\emph{Left:} Leading term in the one-point function, and comparison with CFT (red curve). \emph{Right:} Subleading $L^{-1}\log L$ term, and comparison with CFT (red curve). The convergence is better if one includes more subleading terms in the fit (black stars). }
 \label{fig:onepoints_pants}
\end{figure}
The results are shown in Fig.~\ref{fig:onepoints_pants}. We use a similar procedure to extract the contributions to the one-point function, but finite-size effects turn out to be much stronger, especially for the $L^{-1}\log L$ term. To remove them, we performed fits with more corrections, of the form $a_1L^{-1}+b_2L^{-2}\log L+a_2L^{-2}+b_3L^{-3}\log L+a_3 L^{-3}$. The agreement with CFT (red) curves becomes then excellent. Notice that this particular form of correction is quite natural in such setups, as it is the same as for the exact calculation presented in Sec.~\ref{sec:lbf_exact}.

\section{Time evolution following a local quench}
\label{sec:time_evolution}
This section is devoted to the study of time-evolution problems, following a local ``cut and glue'' quench. As before, we are looking for global observables that exhibit corner-singularities with angle $2\pi$. We will study here the Loschmidt echo and the entanglement entropy, looking for subleading corrections to the results we obtained in Ref.~\cite{SD_localquench}. It will turn out that the application of our main result is more tricky here, due to interesting subtleties involving the analytic continuation to real time. 

\begin{figure}[htbp]
 \centering
 \includegraphics[width=5.3cm]{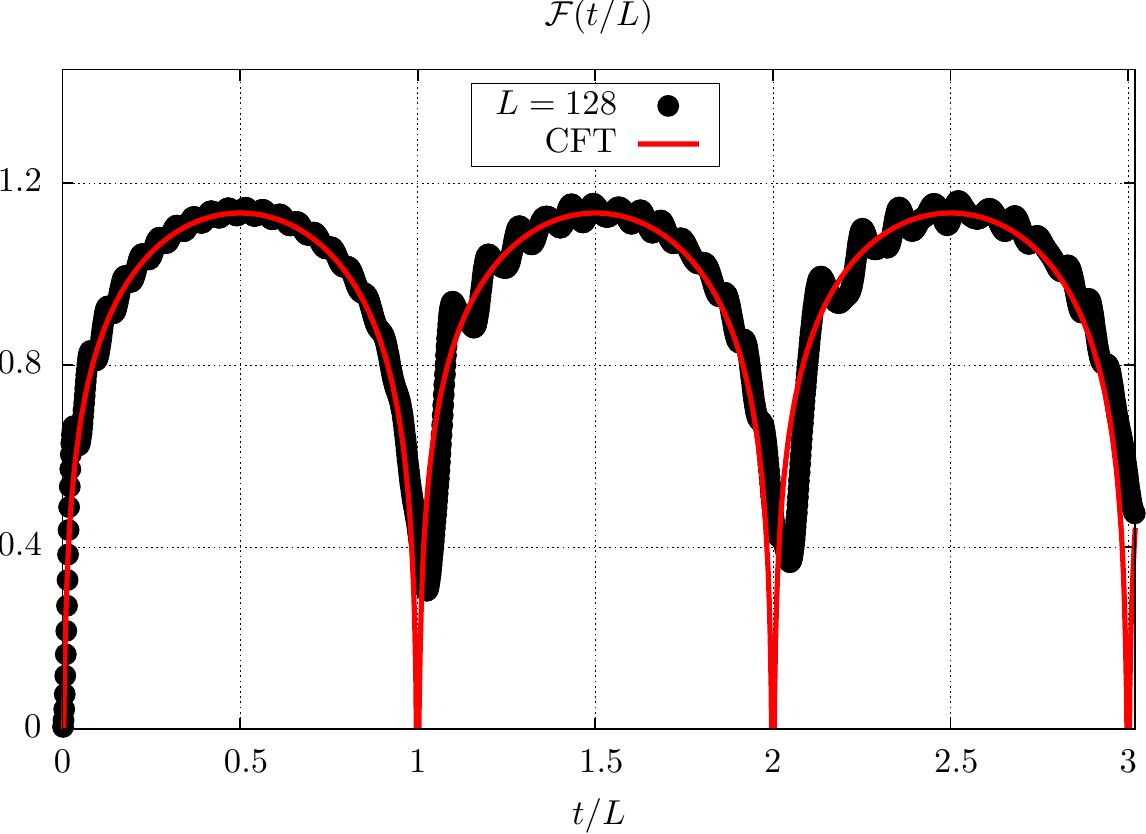}\hfill
 \includegraphics[width=5.3cm]{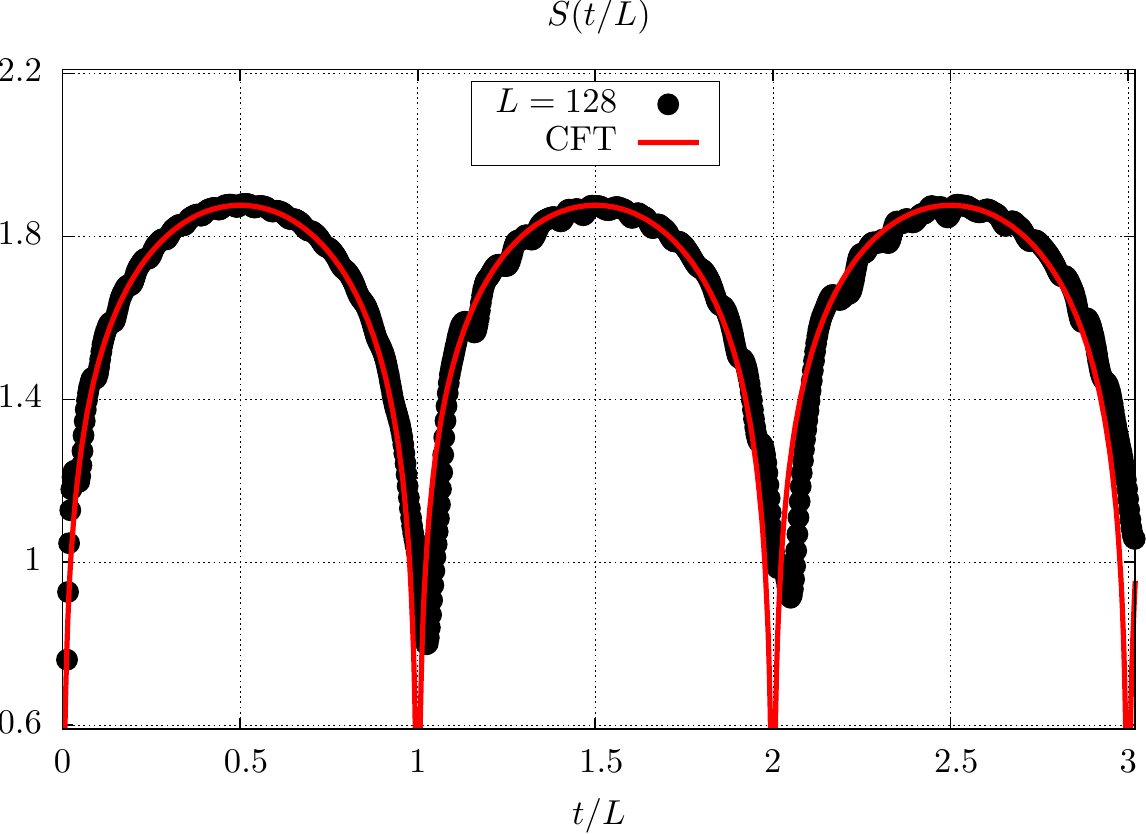}\hfill
  \includegraphics[width=5.3cm]{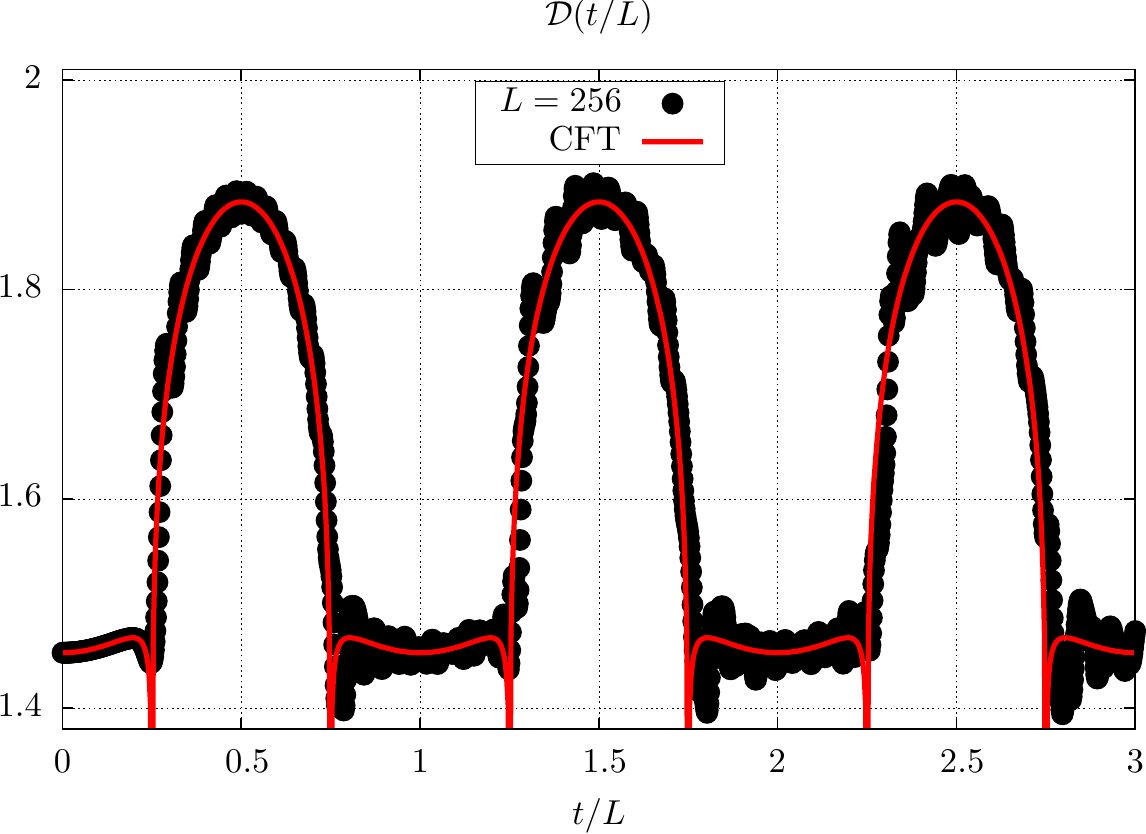}
 \caption{Time evolution of several observables following a local cut and glue quench. \emph{Left}: Logarithmic Loschmidt echo $\mathcal{F}(t)$ for $L=128$.  \emph{Middle}: entanglement entropy $S(t)$ for $L=128$. \emph{Right}: Modified detector Loschmidt echo $\mathcal{D}(t)$, slightly anticipating on Eq.~(\ref{eq:detector_def}), for $L=256$. Each is periodic with period $L/v_F$, as a consequence of the quasiparticle interpretation. We have set $v_F=1$ in the plots.}
 \label{fig:oscillations}
\end{figure}

Let us consider a quantum system prepared in the ground-state $|A\otimes B\rangle$ of the chain cut into two parts. At time $t=0^+$ the two subsystems are joined, so that the initial wave function evolves with the Hamiltonian $H_{A\cup B}$:
\begin{equation}
 |\psi(t)\rangle=e^{itH_{A\cup B}}|A\otimes B\rangle.
\end{equation}
It is now well established that such quenches can be interpreted using a quasiparticle picture \cite{CC_globalquench}. For the local quench these are emitted from the region near the cut between $A$ and $B$, and propagate at the Fermi speed $v_F$ \cite{EislerPeschel,EKPP_localquench,CC_localquench}. They can typically be observed studying correlation functions, but we are interested here in global observables, such as wave function overlaps or the entanglement entropy. In a finite system, the quasiparticles bounce back on the boundaries, and all observables should become periodic in time (see Fig.~\ref{fig:oscillations}). 
\\
Such behavior has been studied in detail in Ref.~\cite{SD_localquench}, for the entanglement entropy (EE) and the logarithmic Loschmidt echo (LLE). Recall that the R\'enyi entanglement entropy (REE) is defined as
\begin{equation}
 S_n=\frac{1}{1-n}\log {\rm tr}_A\, \rho^n\qquad,\qquad \rho={\rm tr}_B \ket{\psi}\bra{\psi}.
\end{equation}
The von Neumann EE is the limit $n\to 1$, $S_1=-{\rm tr}_A \,\rho \log \rho$. Note that the bipartition for the REE coincides with the position of the quench here, this will always be the case in the following. The LLE is defined by 
\begin{equation}\label{eq:lle_def}
 \mathcal{F}(t)=-\log \left|\langle \psi(0)|\psi(t)\rangle\right|^2; 
\end{equation}
it is a natural time-dependent counterpart to the LBF. Interestingly, the leading terms of both quantities can be computed within CFT. We have for two subsystems of the same size
\begin{eqnarray}\label{eq:Soft}
 S(t)&\sim&\frac{c}{3}\log \left|\frac{L}{\pi}\sin \frac{\pi v_F t}{L}\right|,\\\label{eq:foft}
 \mathcal{F}(t)&\sim&\frac{c}{4}\log \left|\frac{L}{\pi}\sin \frac{\pi v_F t}{L}\right|.
\end{eqnarray}
These results have been checked in free fermion systems in Refs.~\cite{Igloi,SD_localquench}, see Fig.~\ref{fig:oscillations} (left and middle) for an illustration. Recently, Eq.~(\ref{eq:Soft}) has also been confirmed numerically in Luttinger liquids away from free fermions \cite{ColluraCalabrese}. 
Note that the similarity between the two formula can be deceiving, as their calculation turns out to be quite different. In particular, the relation $\mathcal{F}(t)=(3/4) S(t)$ does not survive if $L_A\neq L_B$.  
In this section we check for the presence of a $L^{-1}\log L$ correction. We also set, for convenience, $v_F=1$.
\subsection{The Loschmidt echo}
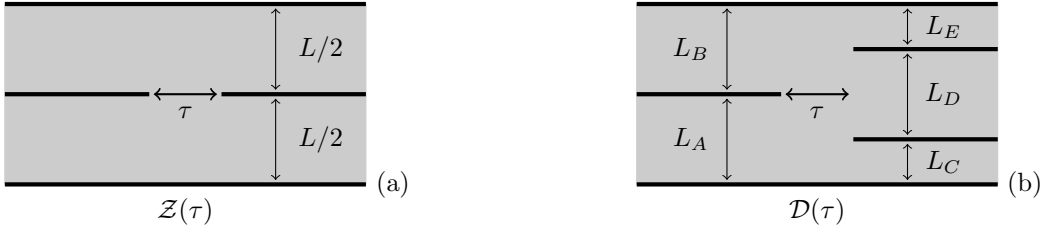
\begin{figure}[ht]
 \centering
\begin{tikzpicture}[scale=1.2]
 \draw [color=black!20,fill=black!20] (0,0) -- (4,0) -- (4,2) -- (0,2) -- (0,0);
  \draw [ultra thick] (0,0) -- (4,0);
  \draw [ultra thick] (0,1) -- (1.6,1);
    \draw [ultra thick] (2.4,1) -- (4,1);
  \draw [ultra thick] (0,2) -- (4,2);
  \draw [<->] (3,0.05) -- (3,0.95);
  \draw (3.5,0.5) node {$L/2$};
  \draw [<->] (3,1.05) -- (3,1.95);
  \draw (3.5,1.5) node {$L/2$};
  \draw [<->,thick] (1.65,1) -- (2.35,1);
  \draw (2,0.8) node {$\tau$};
  \draw (2,-0.3) node {$\mathcal{Z}(\tau)$};
  \draw (4.3,0) node {(a)};
  \begin{scope}[xshift=7cm]
    \draw [color=black!20,fill=black!20] (0,0) -- (4,0) -- (4,2) -- (0,2) -- (0,0);
  \draw [ultra thick] (0,0) -- (4,0);
  \draw [ultra thick] (0,1) -- (1.6,1);
  \draw [ultra thick] (2.4,0.5) -- (4,0.5);
  \draw [ultra thick] (2.4,1.5) -- (4,1.5);
  \draw [ultra thick] (0,2) -- (4,2);
  \draw [<->] (1,0.05) -- (1,0.95);
    \draw (0.6,0.5) node {$L_A$};
  \draw [<->] (1,1.05) -- (1,1.95);
    \draw (0.6,1.5) node {$L_B$};
  \draw [<->] (3,0.55) -- (3,1.45);
  \draw (3.4,0.25) node {$L_C$};
  \draw [<->] (3,0.05) -- (3,0.45);
  \draw (3.4,1) node {$L_D$};
    \draw [<->] (3,1.55) -- (3,1.95);
  \draw (3.4,1.75) node {$L_E$};
  \draw [<->,thick] (1.65,1) -- (2.35,1);
  \draw (2,0.8) node {$\tau$};
  \draw (2,-0.3) node {$\mathcal{D}(\tau)$};
    \draw (4.3,0) node {(b)};
  \end{scope}
\end{tikzpicture}
\caption{Multiple pants geometries representing the (logarithmic) Loschmidt echo (a) and a modified ``detector'' version (b). In each case the horizontal distance between the two-slits is $\tau$. }
\label{fig:double_pants}
\end{figure}

The computation is performed in imaginary time. We need to evaluate 
\begin{eqnarray}
 \mathcal{F}(\tau)&=&-\log \left|\bra{A\otimes B}e^{-\tau H}\ket{A\otimes B}\right|^2\\\label{eq:le_imtau}
 &=&-\log \left|{\rm cst}\times \mathcal{Z}(\tau)\right|^2,
\end{eqnarray}
where $\mathcal{Z}(\tau)$ is the partition function of a ``double-pants'' geometry with spacing $\tau$ between the slits, as shown in Fig.~\ref{fig:double_pants}(a). We assume $L_A=L_B=L/2$ for simplicity here. The constant in (\ref{eq:le_imtau}) acts as a normalization, and is usually fixed when coming back to real time.  
Since the geometry has two $2\pi$ corners, we expect the LLE in imaginary time to scale as
\begin{equation}
 \mathcal{F}(\tau)=\frac{c}{4}\log \left|\frac{L}{\pi}\sinh \frac{\pi \tau}{L}\right|+{\rm cst}+g(\tau)\frac{\log L}{L}+\mathcal{O}(1/L).
\end{equation}
The leading term has been derived in \cite{SD_localquench}, and gives back (\ref{eq:foft}) after analytic continuation. $g(\tau)$ can be obtained from our main result in Sec.~\ref{sec:loglsl_theory}. A possible conformal transformation from the upper-half plane is
\begin{equation}\label{eq:doublepants_mapping}
w(z)=\frac{L}{\pi} {\rm arccoth} \left(\frac{L}{\pi z^2}-{\rm coth} \frac{\tau\pi}{L} \right),
\end{equation}
with the two corners located at $w=0$ and $w=-\tau$. Applying (\ref{eq:loglsl_full}) to the conformal mapping, we get
\begin{equation}
 g(\tau)=\frac{\xi c}{4}\frac{1}{\tanh (\pi \tau/L)}
\end{equation}
for the prefactor of the $L^{-1}\log L$ term. Now we perform the analytic continuation $\tau \to i  t+\epsilon$: 
\begin{equation}
g(it+\epsilon)=\frac{\xi c}{4}\times \frac{\sinh (2\pi \epsilon/L)-i\sin(2\pi t/L)}{\cosh(2\pi \epsilon/L)-\cos(2\pi t/L)}. 
\end{equation}
$\epsilon$ is a UV cutoff of the order of a lattice spacing. It is important to notice that (i) only the real part of this expression contributes to the LLE (see (\ref{eq:lle_def})) and (ii) the result can only be valid for times much larger than the UV cutoff $\epsilon$. So, although $g(\tau)$ was of order one in imaginary time, its contribution becomes of order $L^{-1}$ when coming back to real time. This means that we get, in the end, a correction
\begin{equation}\label{eq:weird}
 \frac{\xi c \pi \epsilon}{2}\times \frac{1}{1-\cos(2\pi t/L)}\times \frac{\log L}{L^2}
\end{equation}
to the LLE. Although the term (\ref{eq:weird}) we just calculated should in principle be there, it is very difficult to identify it, even in a simple free fermion lattice model. This is mainly due to the other fast oscillating subleading corrections (see e.g. Fig.~\ref{fig:oscillations}), which would need to be subtracted\footnote{Note that (\ref{eq:weird}) is not even the leading time-dependent contribution coming from the stress-tensor. A more thorough calculation taking into account all the other non-logarithmic terms yields
$\mathcal{F}(t)=\frac{c}{4}\log \left|\frac{L}{\pi}\sin \frac{\pi t}{L}\right|+cst+\frac{cst'}{L}+\frac{\xi c \pi \cot(\pi t/L)}{8L}+O(L^{-2}\log L)$. Such a contribution is clearly not the only one of order $1/L$, as it has period $L$, and cannot describe the fast $O(1/L)$ oscillations with period $\sim 1$ observed in Fig.~\ref{fig:oscillations} or \cite{SD_localquench}.}.

It is possible to circumvent this difficulty, and show that the logarithmic term we are investigating does appear in real time. To do so, we study a modified version of the LLE, called ``Detector'' in \cite{SD_localquench}. It is defined as 
\begin{equation}\label{eq:detector_def}
 \mathcal{D}(t)=-\log \left|\langle A\otimes B|e^{itH}|C\otimes D\otimes E\rangle\right|^2,
\end{equation}
where $\ket{C\otimes D\otimes E}$ is now the tensor product of three ground-states. 
We take the respective sizes of $C$, $D$ and $E$ to be $L/4$, $L/2$ and $L/4$. See Fig.~\ref{fig:double_pants}(b) for the corresponding space-time geometry. This quantity has $3$ corners with angle $2\pi$, and we expect it to scale as
\begin{equation}
 \mathcal{D}(t)=\frac{3c}{8}\log L+d(t)+cst+e(t)\times \frac{\log L}{L}+O(1/L).
\end{equation}
The constant term is given in the plateau region by \cite{SD_localquench}
\begin{equation}
 d(t)=\frac{c}{16}\left[
 \log \left( 2\cos (2\pi t/L)\right)+2\log \left(1+2\cos(2\pi t/L)-2\sqrt{2}\cos(\pi t/L)\sqrt{\cos(2\pi t/L)}\right)
 \right],
\end{equation}
and agrees with the numerics (see Fig.~\ref{fig:oscillations}, right).
To compute the subleading correction we need the conformal transformation from the upper-half plane to the detector geometry. It is given by
\begin{equation}
 w(z)=\frac{L}{4\pi} \left[\log (1-z^2)-2\log (1-\omega^2 z^2)\right]\qquad,\qquad \tau=\frac{L}{4\pi} \log (4\omega^2(\omega^2-1)).
\end{equation}
Adding the $3$ contributions coming from the three corners in Eq.~(\ref{eq:loglsl_full}), we find
\begin{equation}
 e(\tau)=\frac{\xi c}{8}\left(2+\frac{e^{2\pi \tau/L}+2\sqrt{2}e^{-\pi \tau/L}\sqrt{\cosh \frac{2\pi \tau}{L}}}{\cosh \frac{2\pi \tau}{L}}\right).
\end{equation}
Performing the analytic continuation we get
\begin{equation}\label{eq:loglsl_detector}
 e(t)=\frac{\xi c}{8}\times \left(3+2\sqrt{2}\frac{\cos (\pi t/L)}{\sqrt{\cos(2\pi t/L)}}\right),
\end{equation}
which is of order one, as was hoped. Eq.~(\ref{eq:loglsl_detector}) is much easier to test numerically, because it is less subleading than the analog term for the Loschmidt echo, and the results are not obscured by additional fast oscillation, at least in the first ``plateau'' region $0<t<L/4$ (see Fig.~\ref{fig:oscillations}). In practice, we extract $e(t)$ by studying the linear combination
\begin{equation}\label{eq:detector_linearcomb}
 \left(\mathcal{D}(t)-\mathcal{D}(0)-\left[d(t)-d(0)\right]\right)\times \frac{L}{\log L}.
\end{equation}
\begin{figure}[ht]
 \begin{center}
  \includegraphics[width=10cm]{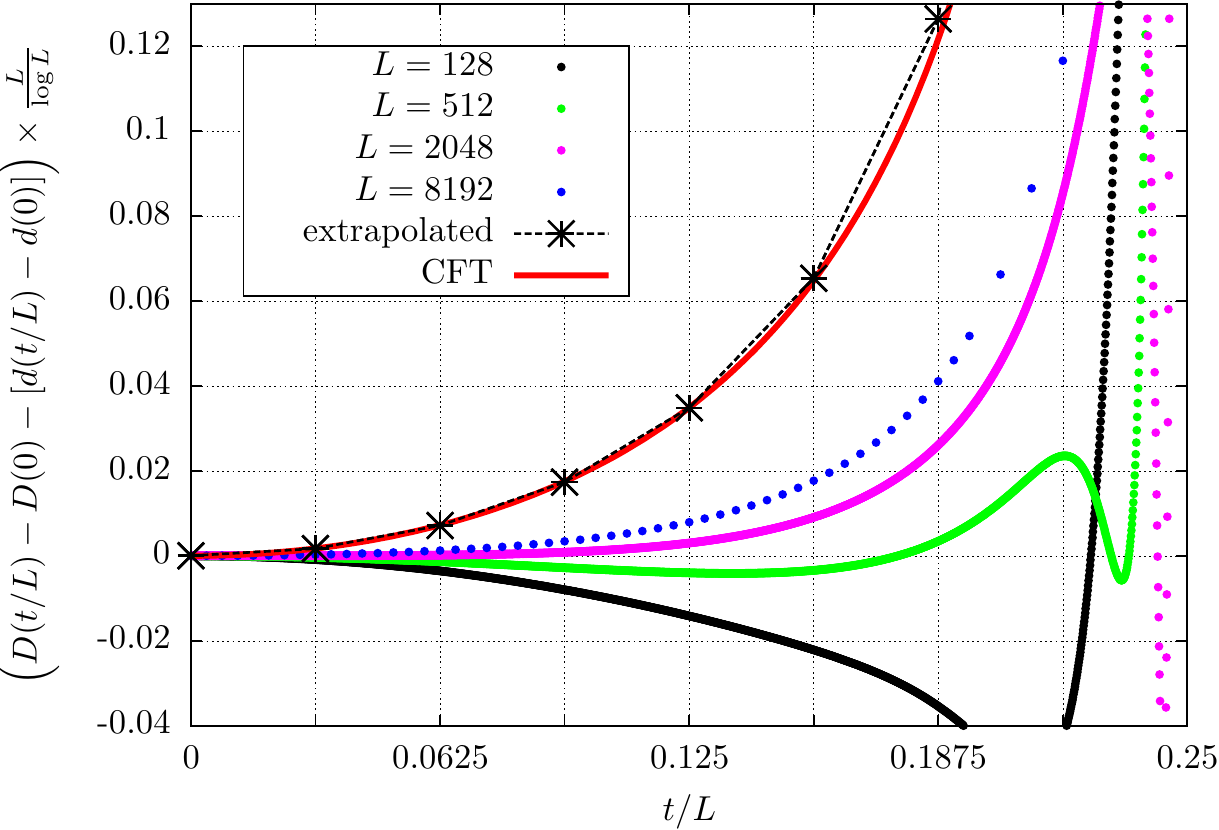}
 \end{center}
\caption{Extraction of the time-dependent $\log L/L$ term of the detector in the plateau region. Big colored dots are the linear combination of Eq.~(\ref{eq:detector_linearcomb}), for system sizes $L=128,512,2048,8192$. This converges very slowly to the CFT prediction (red thick line). However, performing a fit to $b_{-1}+f_{-1}/\log L$ improves the results a lot, and yields a very good agreement with CFT. The results are shown in black stars, for a few values of the reduced time $t/L$.}
\label{fig:detector}
\end{figure}
The results are shown in Fig.~\ref{fig:detector} for an XX chain at half-filling. 
Eq.~(\ref{eq:detector_linearcomb}) should go to (\ref{eq:loglsl_detector}), but the time dependent $L^{-1}$ contribution to the detector translates into a correction of order $1/\log L$ in (\ref{eq:detector_linearcomb}). This makes the convergence very slow, as observed in the figure. However, studying this quantity for various large system sizes at fixed $t/L$ and once again performing fits to $b_{-1} +f_{-1}/\log L$ allows to get much better results. These are the black crosses in the figure, and they agree very well with our CFT prediction.
\subsection{Entanglement entropy}
The subleading corrections to the entanglement entropy in the ground-state are now relatively well understood \cite{Corrections_ee1,Corrections_ee2,Corrections_ee3}, and it is natural to ask what kind of corrections are present after the ``cut and glue'' quench. Here we look for possible corrections of order $O(\log L/L)$ or $O(\log L/L^2)$, coming from the stress-tensor at the boundary. To compute these subleading contributions, we proceed as follows. We use the trick of Calabrese-Cardy \cite{CC_localquench}, namely we observe that the one-point-function of the twist operator on the world-sheet behaves as the one-point function of a usual primary operator in CFT. More precisely, we have
\begin{equation}
{\rm tr}\, \rho^n\propto \braket{\Phi_n(w,\overline{w})},
\end{equation}
where $\Phi_n$ is a primary operator with conformal dimension
\begin{equation}
 h_n=\frac{c}{12}\left(n-\frac{1}{n}\right).
\end{equation}
We evaluate
\begin{equation}\label{eq:ee_toevaluate}
 \braket{\Phi_n(w,\overline{w})}=\frac{\bra{A\otimes B}e^{-H\tau_2}\,\Phi_n(w,\overline{w}) \,e^{-H\tau_1}\ket{A\otimes B}}{\bra{A\otimes B}e^{-H(\tau_1+\tau_2)}\ket{A\otimes B}},
\end{equation}
in imaginary time, and only later will we perform the analytic continuation
\begin{eqnarray}
 \tau_1&\to&\varepsilon+it,\\
 \tau_2&\to&\varepsilon-it.
\end{eqnarray}
Eq.~(\ref{eq:ee_toevaluate}) is nothing but the one point function in the double pants geometry shown in Fig.~\ref{fig:double_pants}(a), with $\tau=\tau_1+\tau_2$. The corresponding mapping $z\mapsto w(z)$ is given by Eq.~(\ref{eq:doublepants_mapping}), with the two corners at $w=0$ and $w=-\tau$. The inverse is
\begin{equation}\label{eq:inverse_conf}
	z(w) \, = \, \sqrt{\frac{L}{\pi}  \sinh \frac{\pi\tau}{L} } \sqrt{\frac{\sinh \frac{\pi w}{L}}{\sinh \frac{\pi(w+\tau)}{L}}} .
\end{equation}
The leading order of the one-point function was computed in \cite{SD_localquench}, and we wish to use our result in Sec.~\ref{sec:corr_onepoint} for the subleading logarithmic correction. There is however an important subtlety that needs to be taken into account: the one-point function of the operator $\Phi_n$ is calculated in a regime where the spacing $\tau=\tau_1+\tau_2$ between the two slits will become very small after the analytic continuation. This is very different from the situation we considered before, and it is not clear whether $\sqrt{L}$ should still be the natural IR cutoff: naively one would now expect $\sqrt{\tau}$ to play this role.

This intuition can be confirmed by an explicit computation of all the contributions from the stress-tensor to the one-point function in this particular geometry. To do so we perturb on the intervals $[\epsilon,\Lambda]$, where $\epsilon$ is a cutoff of the order of the lattice spacing, and $\Lambda\gg L$ (the other perturbation on $[-\Lambda,-\tau-\epsilon]$ can be deduced by symmetry). The corresponding contribution to the one-point function is (see Sec.~\ref{sec:corr_onepoint})
\begin{equation}
 \frac{\delta \braket{\Phi_n(w,\overline{w})}}{\braket{\Phi_n(w,\overline{w})}}=\frac{\xi}{2\pi}\int_{z(\epsilon)}^{z(\Lambda)}dz' \left(\frac{dw'}{dz'}\right)^{-1}
 \frac{\braket{T(z')\Phi_n(z,\overline{z})}}{\braket{\Phi_n(z,\overline{z})}}.
\end{equation}
$z(\epsilon)$ (resp. $z(\Lambda)$) is the image of $\epsilon$ (resp. $\Lambda$) through the inverse conformal mapping (\ref{eq:inverse_conf}). 
Using the Ward identity, we get the integral of a rational fraction, which can be performed. Adding the contribution coming from the other slit we get, after long algebra:
\begin{eqnarray}\fl\nonumber
\frac{\delta \braket{\Phi_n}}{\braket{\Phi_n}}=\frac{\xi\,h_n}{L}&\Bigg\{&
 \frac{-1}{\sinh \frac{\pi \tau}{L}} \left[ \frac{\sinh \frac{\pi \tau_2}{L}}{\sinh \frac{\pi \tau_1}{L}} + \frac{ \sinh \frac{\pi \tau_1}{L} }{\sinh \frac{\pi \tau_2}{L}} \right] \log\left( \frac{1+\tanh \frac{\pi \tau}{L} \coth \frac{\pi \epsilon}{L}}{1+\tanh \frac{\pi \tau}{L}}\right)\;+2-\coth \frac{\pi (\tau_1+\epsilon)}{L}\\
 \fl&-&\coth \frac{\pi (\tau_2+\epsilon)}{L}+\left(\coth \frac{\pi \tau_1}{L}- \coth \frac{\pi \tau_2}{L}\right)\log \left(\frac{e^{-\pi \frac{\tau_2}{L}}\sinh \frac{\pi(\tau_2+\epsilon)}{L}}{e^{-\pi \frac{\tau_1}{L}}\sinh \frac{\pi(\tau_1+\epsilon)}{L}}\right)\Bigg\}+O\left(L^{-2}\right).
\end{eqnarray}
The first term is of order $L^{-1}\log L$ provided $\tau/L$ is of order one. However, after the analytic continuation $\tau_1\to \varepsilon+it$, $\tau_2\to \varepsilon-it$, with $\varepsilon$ another cutoff of the order of a lattice spacing, we observe that it becomes of order $L^{-2}$. Taking into account all contributions, we get
\begin{equation}
\frac{\delta \braket{\Phi_n}}{\braket{\Phi_n}}={\rm cst}+2\pi \xi\,h_n \frac{(2t/L-1)\cot \frac{\pi t}{L}}{L}+O\left(L^{-2}\right).
\end{equation}
In the end, the correction to the entanglement entropy is, using $S=-\lim_{n\to 1}\partial_n {\rm tr} \rho^n$,
\begin{equation}\label{eq:soft_subleading}
 S(t)=\frac{c}{3}\log \left|\frac{L}{\pi}\sin \frac{\pi t}{L}\right|+cst+\frac{cst'}{L}+\frac{\xi \,c\, \pi\, (1-2t/L)\cot \frac{\pi t}{L}}{L}+O(L^{-2}).
\end{equation}
As for the Loschmidt echo, we get that the leading time-dependent correction coming from the stress-tensor is proportional to $L^{-1}$. The only qualitative difference between the two quantities is that the logarithmic term has transformed even more dramatically under the analytic continuation, from $L^{-1}\log L$ to $L^{-2}$. Once again, (\ref{eq:soft_subleading}) is very difficult to test numerically, as it is not the only contribution of order $1/L$. 

Of course, it is a priori not obvious that the analytic continuation method should give correct results, especially when studying subleading contributions. Although it induces several intriguing subtleties in the calculation of the Loschmidt echo as well as the entanglement entropy, we have found no evidence that it could give \emph{incorrect} results in situations were the leading terms are universal. It should however be stressed that this approach, even with more perturbations by local operators, always gives subleading corrections with period $L$; it cannot explain the fast subleading oscillations \cite{EKPP_localquench,SD_localquench,ColluraCalabrese} with period of order one, that can also be observed in Fig.~\ref{fig:oscillations}.  

\section{Conclusion}
\label{sec:conclusion}
In this paper, we have studied possible corrections to the corner free energy of two-dimensional classical systems at criticality. We have focused on the very generic perturbation by the stress tensor at the boundary. In general this perturbation produces corrections of order $L^{-1}$ to the leading universal Cardy-Peschel term. However, when the angle is $\theta=2\pi$, the leading contribution from the stress-tensor becomes of order $L^{-1}\log L$. The prefactor is proportional to the (non  universal) extrapolation length $\xi$, but also the central charge as well as a universal function solely determined by the geometry. We have argued that such a correction is much more interesting than the usual power-laws of the form $L^{-\alpha}$: in a sense the logarithm decouples this term from the other subleading corrections. This $L^{-1}\log L$ term also appears as a correction to the one-point function of an operator near the corner.

We have then applied this result to overlaps of one-dimensional critical wave functions, such as the bipartite fidelity. This was our initial motivation for the study of such corner free energies. Interestingly, it is possible to compare our CFT calculations with numerical evaluations of the fidelity in free fermionic systems. We have found a remarkable agreement between the two. For the special case of a system cut into two subsystems of the same size, it is even possible to obtain the fidelity in closed form for the XX and Ising chains, using a variant of the Cauchy double alternant formula. Performing an asymptotic expansion allowed us to recover the CFT result in this case. We also tested our analytical formula for the one-point function of a primary operator near the corner, and found again a perfect agreement with lattice calculations.

Finally, we also explored the consequences of our result on the time evolution of several global observables, following a local cut and glue quench. It turns out this case requires some care, because of the final analytic continuation to real time. We computed the logarithmic correction to the Loschmidt echo in imaginary time, but found that it becomes of order $L^{-2}\log L$ after analytic continuation. However, certain ``detector'' deformations of the Loschmidt echo preserve the $L^{-1}\log L$ term, and we managed to establish their presence in a lattice model, confirming the validity of our approach. For the entanglement entropy there is a priori an analogous contribution to the one point function, but we showed that it becomes of order $L^{-2}$ in real time. 

There are several interesting directions for future research. First, it would be interesting to study the subleading terms of the form $L^{-k}\log L,k\geq 2$ that we identified in the exact lattice calculations. Such terms appear natural if one thinks of the second order effect of $T(z)$ on the free energy, but there should be some other contributions involved, as they are not semi-universal. A precise determination of the possible structure of logarithmic corrections within CFT is left as an interesting question. It is also possible to study other 1d quantum observables that exhibit $2\pi$ corner singularities in imaginary time. One such example is the emptiness formation probability \cite{Franchini} in critical spin chains described by minimal models, where we expect the presence of a similar logarithmic corrections. A detailed study of this will be presented elsewhere \cite{EFP}. Another interesting problem would be to study the general structure of the subleading corrections following the cut-and-glue quench, to see whether or not the analytic continuation still holds, and determine which operators generate the leading terms. Finally, it would be desirable to numerically check our predictions in interacting systems. The other corrections would presumably take a more complicated form, but we predict that the $L^{-1}\log L$ 
should still be there. For example in an attractive Luttinger liquid, it should still be the leading correction to the free energy, and, as a consequence, to the bipartite fidelity.

\ack
We are grateful to Nick Read for making the observation that the extrapolation length may be interpreted directly as a boundary perturbation
by the stress-tensor, in the context of \cite{DRR}. We also wish to thank Pasquale Calabrese, Viktor Eisler, Paul Fendley, Moshe Goldstein, Gr\'egoire Misguich and Vincent Pasquier for enlightening discussions. This work was  supported by the NSF under the grant DMR/MPS1006549 (JMS), and by a Yale Postdoctoral Prize fellowship (JD).

\pagebreak
\appendix

\section[\;\;\;\;\;\;\;\;\;\;\;\;\;Measures of bipartite entanglement, orthogonality catastrophes, corner free energies, and all that]{Measures of bipartite entanglement, orthogonality catastrophes, corner free energies, and all that}
\label{sec:allthat}
In this appendix, we gather some of our notes on the relations between corner free energies, the bipartite fidelity, X-ray edge singularities and the closely related topic of the measurement of bipartite entanglement; these were our initial motivations for the work presented in this paper. Strictly speaking, this appendix does not contain new results, and most of its content has appeared in papers by other authors, in particular Cardy in \cite{Quench}. It may, however, still be useful to some readers.

\paragraph{}

Recently, there has been some interest in measuring quantities that characterize the bipartite entanglement of many-body systems \cite{Quench,AbaninDemler,Zoller}, such as the R\'enyi entanglement entropies. Recall that these entanglement entropies are defined with respect to a given bipartition of the full Hilbert space $\hilb_A \otimes \hilb_B$ of a quantum system. In this note, the quantum system will always be a $1+1d$ system, such as a spin chain or a tight-binding model of fermions on a line. The subsystem $A$ will be a segment in this line, and $B$ is the complementary subsystem. This spatial partition $A \cup B$ induces a bipartition of the Hilbert space $\hilb_A \otimes \hilb_B$. If the full system is in some given pure state $\left| \psi \right>$, then the subsystem $A$ is described by the reduced density matrix $\rho_A = {\rm tr }_B \left|\psi \right> \left< \psi \right|$, where the partial trace is taken over all the degrees of freedom that lie in the complementary 
 subsystem $B$. The R\'enyi entropies are defined as
\begin{equation}
	S_n \, = \, \frac{1}{1-n} \log \left( {\rm tr}\,  \rho_A^n \right) \, .
\end{equation}
In what follows, $n > 1$ is an integer, although strictly speaking the R\'enyi entropies could also be defined for non-integer $n$; in particular, the limit $n \rightarrow 1$ gives the von Neumann entropy.

The difficulty that arises when one tries to relate the R\'enyi entropies to physical observables comes from the $n^{\rm th}$ power of the reduced density matrix. Physical observables should be linear in $\rho_A$, of the form $\left<\mathcal{O} \right> = {\rm tr \,} \left[ \rho_A \mathcal{O} \right]$, and not of the form ${\rm tr \,} \left[ \rho_A^n \mathcal{O} \right]$.

\subsection[\;\;\;\;\;\;\;\;\;\;\;\;\;Replica trick, and the swap operator]{Replica trick, and the swap operator}
A key observation, which is at the heart of most of the work in this field, is that ${\rm tr \,}\rho_A^n$ can be obtained as a linear expression involving the density matrix of a {\it replicated system}. To see this, imagine that we have $n=2$ replicas of the full system $(\hilb_{A1} \otimes \hilb_{B1})\otimes (\hilb_{A2} \otimes \hilb_{B2})$, in the pure state $\ket{\psi} \otimes \ket{\psi}$. The density matrix that we are going to use is $\rho^{\otimes 2} = \left| \psi \right> \left< \psi \right| \otimes \left| \psi \right> \left< \psi \right|$. One introduces the ``swap'' operator (also called ``switch'' or ``twist'' or ``permutation operator'' in various references), noted $\mathcal{S}_A$, which maps canonically $\hilb_{A1}$ onto $\hilb_{A2}$ and vice-versa, and acts as the identity on $\hilb_{B1} \otimes \hilb_{B2}$. More explicitly, this means that $\mathcal{S}_A$ is linear, and it acts on the basis states as
\begin{equation}
	\mathcal{S}_A \, (\ket{n}_{A1} \otimes \ket{m}_{B1} ) \otimes (\ket{p}_{A2} \otimes \ket{l}_{B2} )  \, = \, (\ket{p}_{A1} \otimes \ket{m}_{B1} ) \otimes (\ket{n}_{A2} \otimes \ket{l}_{B2}) \, .
\end{equation}
The expectation value in the state $\ket{\psi}\otimes \ket{\psi}$ of the swap operator, $ \left< \mathcal{S}_A \right> \, = \, {\rm tr} \,\left[ \mathcal{S}_A \, \cdot\, \rho^{\otimes 2}  \right]$, is equal to
\begin{equation}
\left< \mathcal{S}_A\right> \, = \,{\rm tr }_{\hilb_A} \, \rho_A^2,
\end{equation}
where the right hand side stands for the initial (non-replicated) system.
This observation is easily extended to any integer $n>1$, for a replicated system with $n$ copies. The swap operator, in this case, maps $\hilb_{Ak}$ onto $\hilb_{Ak+1}$, where $k$ labels the $n$ copies (say from $1$ to $n$, modulo $n$). It is thus possible to get all the R\'enyi entropies $S_n$ by computing/measuring the expectation values of these swap operators. This is a trick that is useful analytically (see, e.g., \cite{Holzhey_ee,CalabreseCardy_ee,Doyon}) as well as numerically \cite{swap}. A natural question is whether the swap operator can be implemented physically in a realistic system.

\subsection[\;\;\;\;\;\;\;\;\;\;\;\;\;Action of the swap operator on a local Hamiltonian]{Action of the swap operator on a local Hamiltonian}
\begin{figure}[htbp]
	\begin{tikzpicture}
		\begin{scope}[scale=0.9]
		\filldraw[fill=green!30,draw=white,xshift=0.7cm, yshift=-0.4cm,scale=1.32] (-1,0) arc (180:120:1) -- (-0.5*4,4*0.866) arc (120:180:4);
		\begin{scope}
			\draw[thick] (-4,0) -- (-1,0);
			\foreach \x in {-4,-3.5,...,-1} \filldraw (\x,0) circle (2pt);
			\draw (-4,-0.3) node{$1$};
			\draw (-1,-0.3) node{$L/2$};
			\draw (-3.5,-0.3) node{$2$};
		\end{scope}
		\begin{scope}[rotate = -60]
			\draw[thick] (-4,0) -- (-1,0);
			\foreach \x in {-4,-3.5,...,-1} \filldraw (\x,0) circle (2pt);
			\draw (-4,-0.3) node{$L$};
			\draw (-1.35,-0.7) node{$L/2+1$};
			\draw (-3.5,-0.5) node{$L-1$};
		\end{scope}
			\draw[thick] (-1,0) -- (-0.5,0.866);
		\end{scope}

		\begin{scope}[scale=0.9,rotate=120]
		\filldraw[fill=red!20,draw=white,xshift=0.7cm, yshift=-0.4cm,scale=1.32] (-1,0) arc (180:120:1) -- (-0.5*4,4*0.866) arc (120:180:4);
		\begin{scope}
			\draw[thick] (-4,0) -- (-1,0);
			\foreach \x in {-4,-3.5,...,-1} \filldraw (\x,0) circle (2pt);
		\end{scope}
		\begin{scope}[rotate = -60]
			\draw[thick] (-4,0) -- (-1,0);
			\foreach \x in {-4,-3.5,...,-1} \filldraw (\x,0) circle (2pt);
		\end{scope}
			\draw[thick] (-1,0) -- (-0.5,0.866);
		\end{scope}

		\begin{scope}[scale=0.9,rotate=-120]
		\filldraw[fill=blue!20,draw=white,xshift=0.7cm, yshift=-0.4cm,scale=1.32] (-1,0) arc (180:120:1) -- (-0.5*4,4*0.866) arc (120:180:4);
		\begin{scope}
			\draw[thick] (-4,0) -- (-1,0);
			\foreach \x in {-4,-3.5,...,-1} \filldraw (\x,0) circle (2pt);
		\end{scope}
		\begin{scope}[rotate = -60]
			\draw[thick] (-4,0) -- (-1,0);
			\foreach \x in {-4,-3.5,...,-1} \filldraw (\x,0) circle (2pt);
		\end{scope}
			\draw[thick] (-1,0) -- (-0.5,0.866);
		\end{scope}

	\begin{scope}[xshift=9cm]
		\begin{scope}[scale=0.9]
		\filldraw[fill=green!30,draw=white,xshift=0.7cm, yshift=-0.4cm,scale=1.32] (-1,0) arc (180:120:1) -- (-0.5*4,4*0.866) arc (120:180:4);
		\begin{scope}
			\draw[thick] (-4,0) -- (-1,0);
			\foreach \x in {-4,-3.5,...,-1} \filldraw (\x,0) circle (2pt);
			\draw (-4,-0.3) node{$1$};
			\draw (-1,+0.3) node{$L/2$};
			\draw (-3.5,-0.3) node{$2$};
		\end{scope}
		\begin{scope}[rotate = -60]
			\draw[thick] (-4,0) -- (-1,0);
			\foreach \x in {-4,-3.5,...,-1} \filldraw (\x,0) circle (2pt);
			\draw (-4,-0.3) node{$L$};
			\draw (-1.35,-0.7) node{$L/2+1$};
			\draw (-3.5,-0.5) node{$L-1$};
		\end{scope}
		\end{scope}

		\begin{scope}[scale=0.9,rotate=120]
		\filldraw[fill=red!20,draw=white,xshift=0.7cm, yshift=-0.4cm,scale=1.32] (-1,0) arc (180:120:1) -- (-0.5*4,4*0.866) arc (120:180:4);
		\begin{scope}
			\draw[thick] (-4,0) -- (-1,0);
			\foreach \x in {-4,-3.5,...,-1} \filldraw (\x,0) circle (2pt);
		\end{scope}
		\begin{scope}[rotate = -60]
			\draw[thick] (-4,0) -- (-1,0);
			\foreach \x in {-4,-3.5,...,-1} \filldraw (\x,0) circle (2pt);
		\end{scope}
		\end{scope}

		\begin{scope}[scale=0.9,rotate=-120]
		\filldraw[fill=blue!20,draw=white,xshift=0.7cm, yshift=-0.4cm,scale=1.32] (-1,0) arc (180:120:1) -- (-0.5*4,4*0.866) arc (120:180:4);
		\begin{scope}
			\draw[thick] (-4,0) -- (-1,0);
			\foreach \x in {-4,-3.5,...,-1} \filldraw (\x,0) circle (2pt);
		\end{scope}
		\begin{scope}[rotate = -60]
			\draw[thick] (-4,0) -- (-1,0);
			\foreach \x in {-4,-3.5,...,-1} \filldraw (\x,0) circle (2pt);
		\end{scope}
			\draw[thick] (-0.5,-0.866) -- (-1,0);
			\draw[thick] (0.5,0.866) -- (-0.5,0.866);
			\draw[thick] (1,0) -- (0.5,-0.866);
		\end{scope}
	\end{scope}
	\end{tikzpicture}
	
	\caption{An illustration (for $n=3$ replicas) of the ``swap'' quench discussed by Cardy \cite{Quench}: the Hamiltonian $H_{\rm repl.}$ (left) is instantaneously changed to $H_{\rm repl.}'$ (right). Each shaded area corresponds to one replica of the original spin chain/tight-binding model. Each black line corresponds to an interaction between two sites. The key observation is that the difference $H'_{\rm repl.} - H_{\rm repl.}$ involves only a few local terms in the center of the picture.}
	\label{fig:HHp}
\end{figure}
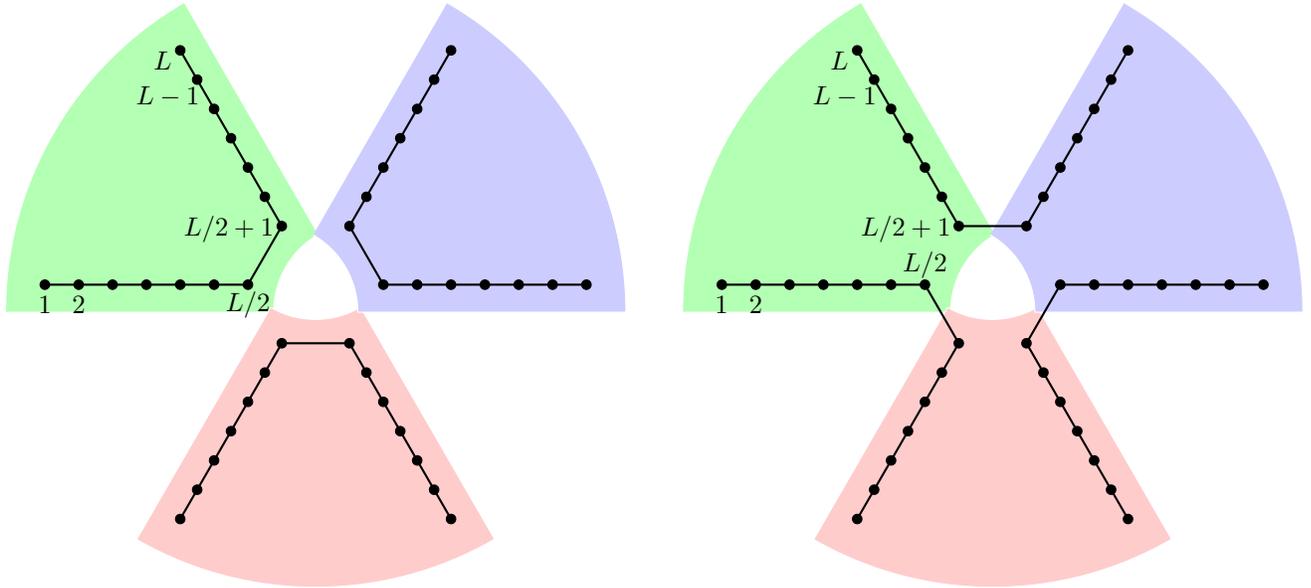

At this point, however, it is not clear that the swap operator may find a reasonable implementation in a concrete physical system, because as it is presented above, it seems to be a highly non-local operator. The next step is to re-express $\left< \mathcal{S}_A \right>$ as an overlap between the ground states of two Hamiltonians that differ only by a few local terms \cite{Quench}. Let us consider a concrete model to illustrate the idea. Take a homogeneous open chain with $L$ sites, populated by free fermions with nearest-neighbor hopping
\begin{equation}
	\label{eq:tightbinding}
	H \, = \, t \sum_{i=1}^{L-1} \left[ c_i^\dagger c_{i+1} + c^\dagger_{i+1} c_{i} \right] \, .
\end{equation}
We chose the subsystem $A$ as the first $L/2$ sites ($i=1,\dots,L/2$, and we assume that $L$ is even). The Hamiltonian $H$ is equal to $H_A + H_B + H^I_{AB}$ where $H_A$ ($H_B$) is the Hamiltonian of the left (right) subsystem, and the interaction term between $A$ and $B$ is $H^I_{AB} = t (c^\dagger_{L/2} c_{L/2+1} + c^\dagger_{L/2+1} c_{L/2})$. The fact that $H_{AB}^I$ is local, namely that it couples degrees of freedom that are located close to the cut between $A$ and $B$ only, is important: as we shall see shortly, it allows to reformulate the expectation value of the swap operator as an overlap between the ground states of two Hamiltonians which differ only by local terms. First, we introduce the $n$ replicas of the system; each replica comes with the Hamiltonian
\begin{eqnarray}
	H_{k} & = & t\, \sum_{i=1}^L \left[ c_{i,k}^\dagger c_{i+1,k} + c_{i+1,k}^\dagger c_{i,k}  \right]  \\
\nonumber	&=& H_{Ak} +  H_{Bk}  +  H^I_{Ak \cup B k} \qquad \qquad k=1,\dots,n , 
\end{eqnarray}
and the Hamiltonian of the full system, acting on the $n$ replicas, is simply
\begin{equation}
	H_{\rm repl.} \, = \, \sum_{k=1}^n H_k \, = \, \sum_{k=1}^n \left[ H_{Ak} +H_{Bk} + H^I_{A\cup B k} \right] \, .
\end{equation}
The ground state of this Hamiltonian is $\ket{\psi_{\rm repl.}} = \ket{\psi} \otimes \dots \otimes \ket{\psi}$. Now let us conjugate $H_{\rm repl.}$ by the swap operator. This defines the ``swapped'' Hamiltonian,
\begin{equation}
	H'_{\rm repl.} \, \equiv \, \mathcal{S}_A \cdot H_{\rm repl.} \cdot \mathcal{S}_A^\dagger ,  
\end{equation}
which has the ground state $\ket{\psi'_{\rm repl.}} = \mathcal{S}_A \ket{\psi_{\rm repl.}}$. The ground state expectation value of the swap operator may thus be viewed as the overlap
\begin{equation}
	\label{eq:swap_expect}
	\left< \mathcal{S}_A \right> \, = \, \left.\left< \psi'_{\rm repl.} \right| \psi_{\rm repl.} \right> .
\end{equation}
Such an overlap is, by definition, the (square root of the) probability of observing the system in the ground state of $H'_{\rm repl.}$ if it is initially prepared in the ground state of $H_{\rm repl.}$. In the recent literature, such out-of-equilibrium problems, where one suddenly changes the Hamiltonian $H \rightarrow H'$, have been dubbed ``quantum quenches'' \cite{CC_globalquench}. In the present case, the two Hamiltonians $H_{\rm repl.}$ and $H'_{\rm repl.}$ differ only by a few local terms,
\begin{equation}
	H'_{\rm repl.} \, = \,\sum_{k=1}^n \left[ H_{Ak} +H_{Bk} \right] \, + \, t  \sum_{k=1}^n \left[ c^\dagger_{L/2,k+1} c_{L/2+1,k} + c^\dagger_{L/2+1,k} c_{L/2,k+1} \right] ,
\end{equation}
so this is one example of a {\it local} ``quantum quench''. It is convenient to picture the two systems described by these Hamiltonians as in Fig. \ref{fig:HHp}. It is clear from the drawing that the difference between $H_{\rm repl.}$ and $H'_{\rm repl.}$ is local in the sense that it involves only terms that are located in the ``center'' of the replicated system (see Fig. \ref{fig:HHp}). Therefore, the problem of measuring the R\'enyi entropy $S_n$ has boiled down to measuring the overlap $\left< \left. \psi_{\rm repl.}' \right| \psi_{\rm repl.} \right>$ in a ``local quench'' $H_{\rm repl.} \rightarrow H_{\rm repl.}'$.
%Although we are mostly interested in $1+1d$ systems here, the generalization of this idea to higher dimensions is straightforward. The swap operator can be defined in the same way, the only difference is that the quench that is obtained in higher dimensions does not involve changing terms located close to a {\it point}, but rather close to a $d-1$-dimensional {\it submanifold}, which is the boundary between $A$ and $B$.

\subsection[\;\;\;\;\;\;\;\;\;\;\;\;\;A simpler version of the ``swap'' quench: the ``cut-and-glue'' quench]{A simpler version of the ``swap'' quench: the ``cut-and-glue'' quench}
Implementing the ``swap'' quench $H_{\rm repl.} \rightarrow H_{\rm repl.}'$ requires a number of skills:
\begin{enumerate}
	\item one needs to create $n$ identical copies of the quantum system,
	\item one also wants to be able to switch on/off {\it one given coupling} at a specific location in the middle of the system, namely to {\it cut} or {\it glue} two given subsystems instantaneously
	\item it is not sufficient to be able to cut or glue {\it two} of the subsystems; one actually needs to cut/glue {\it simultaneously} $n+n$ subsystems, coming from the $n$ replicas.
\end{enumerate}
We note that it has been claimed recently that this type of quantum quench may be realized in cold atoms systems \cite{AbaninDemler, Zoller}. At this point, it is worth emphasizing that, despite the modern vocabulary attached to these problems in the recent literature, local quantum quenches actually have a long history; their study dates back to the 1960's, with such phenomena as the X-ray edge singularities and the Anderson orthogonality catastrophe. This motivates us to look at a simplified version of the above ``swap'' quench, which does not involve replicas, and whose connection with the older class of problems is more transparent.

\paragraph{}
To get a ``baby-version'' of the local quench $H'_{\rm repl.} \rightarrow H_{\rm repl.}$, we simply drop the requirements (i) and (iii). We do not introduce replicas of the original system, but we require the ability (ii) to switch on/off suddenly all the interactions $H^I_{A, B}$ between the two subsystems $A$ and $B$. Then the quench without the replicas is simply
\begin{equation}
	H \, = \, H_A + H_B \; \rightarrow \; H' \, = \,H_A + H_B + H^I_{A , B} \, .
\end{equation}
Again, this is a {\it local} quench, in the sense that $H'- H = H^I_{A,B}$ involves only a few terms that are located close to the cut between $A$ and $B$.
Like in the more complicated case of the ``swap'' quench, one can be interested in measuring the overlap between the two ground states of $H$ and $H'$. Let us refer to this local quench as the ``cut-and-glue'' quench; it is a problem that has been studied in recent papers \cite{CC_localquench, EKPP_localquench, Bipartite_fidelity, SD_localquench}. Let us emphasize that the overlap between the two ground states $H$ and $H'$ is not a R\'enyi entropy like in the case of the ``swap'' quench. It is, by definition, an overlap, dubbed ``bipartite fidelity'' in \cite{Bipartite_fidelity}. Although it is not a R\'enyi entropy, it is a quantity that is interesting in its own right, and it can be used to characterize the entanglement between the two subsystems $A$ and $B$. Measuring this quantity should be simpler than measuring the R\'enyi entropies, since the ``cut-and-glue'' quench is simpler than the ``swap'' quench.
\begin{figure}[htbp]
	\centering
\begin{tikzpicture}[scale=0.9]
	\begin{scope}
		\filldraw[green!30] (-6.3,-0.4) rectangle (5.3,0.4);
		\draw[thick] (-6,0) -- (5,0);
		\foreach \x in {-6,-5,...,5} \filldraw (\x,0) circle (2pt);
		\draw (-6,-0.3) node{$1$};
		\draw (-5,-0.3) node{$2$};
		\draw (5,-0.3) node{$L$};
		\draw (4,-0.3) node{$L-1$};
		\draw (-1.2,-0.3) node{$L/2$};
		\draw (0.2,-0.3) node{$L/2+1$};
	\end{scope}
	\begin{scope}[yshift=-2cm]
		\filldraw[green!30] (-6.3,-0.4) rectangle (5.3,0.4);
		\draw[thick] (-6,0) -- (-1,0);
		\draw[thick] (0,0) -- (5,0);
		\foreach \x in {-6,-5,...,5} \filldraw (\x,0) circle (2pt);
		\draw (-6,-0.3) node{$1$};
		\draw (-5,-0.3) node{$2$};
		\draw (5,-0.3) node{$L$};
		\draw (4,-0.3) node{$L-1$};
		\draw (-1.2,-0.3) node{$L/2$};
		\draw (0.2,-0.3) node{$L/2+1$};
		\draw[<->,thick] (-6,1) -- (-3.5,1) node[below]{$A$} -- (-1,1);
		\draw[<->,thick] (0,1) -- (2.5,1) node[below]{$B$} -- (5,1);
	\end{scope}
\end{tikzpicture}
\caption{The ``cut-and-glue'' quench may be viewed as ``baby-version'' of the quench presented in Fig. \ref{fig:HHp}. It does not involve replicas, but only the ability to switch on/off instantaneously the interactions between $A$ and $B$. The goal is to measure the overlap between the ground states of the Hamiltonians $H'=H_A+H_B +H^I_{A \cup B}$ (top) and $H=H_A+H_B$ (bottom).}
\label{fig:baby-version}
\end{figure}
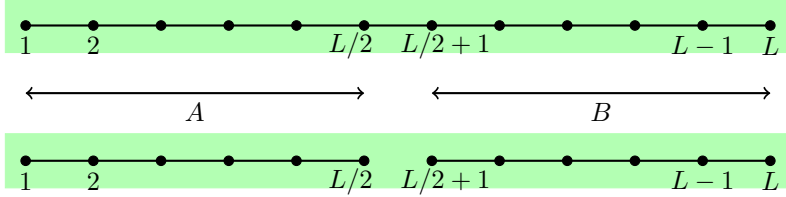

\subsection[\;\;\;\;\;\;\;\;\;\;\;\;\;Orthogonality catastrophes]{Orthogonality catastrophes}
The overlaps between the ground states of two Hamiltonians which differ by a local term is a topic that was pioneered by Anderson in a seminal paper \cite{Anderson}. He showed that the two ground states of a Fermi gas in the presence/absence a localized impurity potential have an overlap which decays algebraically with the system size $L$. This is the ``orthogonality catastrophe'' in its original formulation for a fermionic system. The exponent of the algebraic decay $x$,
\begin{equation}
	\left|  \left< \Psi \left| \Psi' \right>\right. \right| \, \propto \, \left( \frac{a}{L} \right)^{x}, 
\end{equation}
depends only on the low-energy properties of the system. Here $a$ is an UV cutoff (typically, the inverse bandwidth). For free fermion systems, $x$ is given by the phase shift $\delta$ caused by the impurity potential, evaluated at the Fermi momentum: $x \, =\, \frac{1}{2}\left(\delta(k_F)/\pi \right)^2 $. The orthogonality catastrophe in systems with a Fermi surface has been studied extensively since the 1960s. It is not a phenomenon that is restricted to systems with Fermi statistics though; the algebraic decay of the overlap between $\left|\Psi \right>$ and $\left| \Psi' \right>$ may also be interpreted as a signature of the underlying universal critical behavior. After reduction to $s$-wave scattering of the fermions on the impurity potential, and keeping only the radial component of the wave packets, the system can be reformulated as a $1+1d$ semi-infinite gapless system, with a boundary condition related to the impurity potential through the phase shift $\delta (k_F)$.
  The boundary condition is therefore different when the impurity potential is on/off. A precise analysis of these boundary conditions leads to the orthogonality catastrophe and to the exponent $x$. This approach has been developed with great success for a wide variety of quantum impurity problems.

\paragraph{}
The ``cut-and-glue'' quench in a $1+1$d system belongs to this class of impurity problems; it corresponds to the ideal case when the impurity completely decouples the two subsytems. The ``swap'' quench may also be viewed as an impurity problem, but the effect of the impurity is more convoluted, as it switches the couplings between $n$ replicas of the system. The important point here is that both the ``cut-and-glue'' and the ``swap'' quenches in gapless systems are associated with orthogonality catastrophe exponents, $x_{\rm cut/glue}$ and $x_{\rm swap}$.In either case, the overlap between the ground states before and after the quench $H \rightarrow H'$ obey a similar power-law decay with the corresponding exponent $x$. In systems with an energy gap, there is a finite correlation length $\xi$, and the two ground states can differ significantly only in a region of size $\sim \xi$ localized around the the cut between the parts $A$ and $B$. In that case, one expects
$	\left| \left< \Psi \left| \Psi' \right>\right. \right| \, \propto  \, \left( a/\xi \right)^x $
as one approaches the critical point.

\subsection[\;\;\;\;\;\;\;\;\;\;\;\;\;Measurable quantities related to the overlaps]{Measurable quantities related to the overlaps}
Strictly speaking, the (squared) overlap $\left| \left< \Psi \left| \Psi' \right>\right. \right|^2$ is a measurable quantity itself, as it is, by definition, the probability of observing a system---initially prepared in the ground state of $H$---in the ground state of the new Hamiltonian $H'$ immediately after the quench $H \rightarrow H'$. Nevertheless, this probability is usually not directly accessible in realistic many-body systems, especially when they are gapless. In the latter systems, the orthogonality catastrophe exponent is usually more easily accessible than the overlap $\left| \left< \Psi \left| \Psi' \right>\right. \right|$ itself.

%\subsubsection[\;\;\;\;\;\;\;\;\;\;\;\;\;X-ray edge singularity, and absorption spectrum of an auxiliary two-level system:]{X-ray edge singularity, and absorption spectrum of an auxiliary two-level system:}
\subsubsection*{Appendix A.5.1.\, X-ray edge singularity, and absorption spectrum of an auxiliary two-level system:}
it is very well-known that the orthogonality catastrophe exponent appears as a power-law singularity in the X-ray absorption spectra of metals. In this problem, the presence/absence of a localized core electron can be viewed as a two-level system. In the presence of the core electron, the electrons close to the Fermi surface are in the many-body ground state $\left| \psi \right>$. An X-ray photon can remove the core electron and therefore create a localized hole. The Hamiltonian for the electrons close to the Fermi surface is then modified to incorporate the localized potential created by the hole. This is precisely a local quantum quench problem. Since we know that the corresponding orthogonality catastrophe exponent appears in the absorption spectrum of the to-level system, it is a very natural idea to 
generalize this setting to any sort of local quench $H \rightarrow H'$ in gapless systems, including the ``cut-and-glue'' and the ``swap'' quenches. One can imagine that the evolution of the system is governed by a Hamiltonian $H$ or $H'$, depending on the presence/absence of the core electron, which may be represented by a two-level system, with two states $\ket{0}$ and $\ket{1}$, and energies $0$ and $E_0$ respectively. The total Hamiltonian for the full system ([gapless system] $\otimes$ [two-level system]) is
\begin{equation}
	H \otimes \left| 0 \right>\left< 0\right| \, + \, H' \otimes \left|1 \right> \left<1 \right|  \, +\,  E_0 ~ \mathbb{I} \otimes \left|1 \right> \left<1 \right| ,
\end{equation}
such that the gapless system evolves with a Hamiltonian $H$ if the two-level system is in the state $\ket{0}$, and with the Hamiltonian $H'$ if it is in the state $\ket{1}$. If the state of the two-level system changes suddenly, one gets a quantum quench $H \rightarrow H'$. To be more concrete, we consider again the toy-model (\ref{eq:tightbinding}), in the case of the ``cut-and-glue'' quench. We imagine that the hopping term between the sites $i=L/2$ and $i=L/2+1$ is coupled to the two-level system,
\begin{equation}
	\label{eq:Htot_b}
	H_{\rm tot.} \, = \, t \sum_{i \neq L/2} \left[c_i^\dagger c_{i+1} + c^\dagger_{i+1} c_i \right] \otimes \mathbb{I} \; +\; t ~\left[ c_{L/2}^\dagger c_{L/2+1} + c^\dagger_{L/2} c_{L/2+1} \right] \otimes \left| 1\right>\left<1 \right|   \, + E_0 ~ \mathbb{I} \otimes \left|1 \right> \left< 1\right| \, .
\end{equation}
When the two-level system is in the state $\left| 0\right>$, the left and right subsystems are decoupled, while the chain is homogeneous when it is in the state $\left| 1\right>$. 
The correlation function of the two-level system is related to the overlap between the two ground states in the presence/absence of the impurity potential $V$, or, more precisely, to the orthogonality catastrophe exponent. For critical systems with a dynamic exponent $z=1$, on time scales such that $a \ll v t \ll L$ ($v$ is the velocity of the massless excitations, $a$ is an UV cutoff and $L$ is the typical size of the system), one can show that
\begin{eqnarray}
	\label{eq:decay}
	\left< S^-(t) S^+(0) \right> &=& \left< \psi \right| e^{-i H_{\rm tot.} t} \left| \psi \right>  	\; \propto \; e^{-i E_0 t} ~\times  ~  t^{-2x},
\end{eqnarray}
where $S^+ \left| 0 \right> = \left| 1 \right>$, $S^- \left| 1 \right> =\left| 0\right>$ and $S^+ \left|1 \right> = S^-\left|0 \right> =0$. The important point here is that the exponent $x$ in (\ref{eq:decay}) is the {\it same} exponent as in the orthogonality catastrophe above. The absorption spectrum of the two-level system thus has a power-law singularity from which the exponent $x$ can be extracted:
\begin{eqnarray}
	 \int dt ~e^{i E t} ~ \left< S^-(t) S^+(0) \right> \, \propto \, \frac{1}{|E - E_0|^{1-2x}} .
\end{eqnarray}

%\subsubsection[\;\;\;\;\;\;\;\;\;\;\;\;\;Rabi oscillations of the two-level system:]{Rabi oscillations of the two-level system:}
\subsubsection*{Appendix A.5.2.\, Rabi oscillations of the two-level system:} 
The following setting was discussed recently in \cite{AbaninDemler}; it is essentially the same as the one in the previous paragraph, but in a different regime. Consider first the two-level system alone. Imagine that one switches on a small tunneling term $T~ S^+  + T^*~  S^-$ between the two states of the two-level system. If the system is initially ($t=0$) prepared in the state $\left| 0\right>$, then at time $t>0$, the probability of observing it in the state $\left| 1\right>$ is
\begin{equation}
	\label{eq:Rabi}
	P_{\left|0 \right> \rightarrow  \left|1 \right>}(t) \, = \, \frac{|T|^2}{(E_0/2)^2 + |T|^2} \sin^2 \left[t ~\Omega/2 \right] \qquad \qquad \Omega = \sqrt{|T|^2 + (E_0/2)^2}
\end{equation}
Now, consider again the total Hamiltonian (\ref{eq:Htot_b}), and assume that the tunneling amplitude $|T|$ and the shift of the ground state energy $\tilde{E}_0$ are {\it both} much smaller than the energy gaps of $H$ and $H'$. In that limit (and {\it only in that limit}), one can project the system onto the two-dimensional subspace generated by $\left|0 \right> \otimes \left|\psi \right>$ and $ \left|1 \right> \otimes \left|\psi' \right>$. The tunneling term, when it is projected onto this two-dimensional subspace, becomes:
\begin{equation}
	\mathcal{P}_{\rm subsp.} \cdot \left[ T~S^+ + T^* ~S^- \right] \cdot \mathcal{P}_{\rm subsp.} \; = \; T \left< \psi' \left| \psi \right>\right. ~S^+ + (T \left< \psi' \left| \psi \right>  \right.)^* ~S^- \, .
\end{equation}
Thus, in the presence of the chain, the population of the two-level system oscillates as in (\ref{eq:Rabi}), but with the modification $E_0 \rightarrow \tilde{E}_0$ and $T \rightarrow T \left< \psi' \left|\psi \right>\right.$. In principle, this can be used to extract the value of the overlap $\left| \left< \psi '\left| \psi\right> \right.\right|$; if this can be achieved, it is more interesting than being able to measure the exponent $x$ only, since the overlap itself is a more refined quantity than the exponent. However, notice that the restriction of a many-body system to a two-level system is quite a crude approximation, especially in a gapless system, where the energy gap goes as $O(v/L)$. In particular, the shift of the ground state energy between $H$ and $H'$ may be hard to control with such accuracy (for instance, in the case of the ``cut-and-glue'' quench, the main contribution to the shift of the ground state energy comes from the change of boundary conditions at th
 e ends of the subsystems $A$ and $B$, and this is of higher order; there is also a change of the Casimir energy which is itself of order $O(v/L)$; for the ``swap'' quench, although such contributions would ideally cancel, they may remain if the $n$ copies of the system not perfectly identical).

\subsection[\;\;\;\;\;\;\;\;\;\;\;\;\;RG discussion of the perturbation $H \rightarrow H'$]{RG discussion of the perturbation $H \rightarrow H'$}
One-dimensional gapless systems with a dynamic exponent $z=1$ are well described at low energies by field theories that are massless and Lorentz invariant (with the velocity of the low-energy excitations, $v$, which plays the role of the speed of light). It is most convenient to think of the original 1d system as an equal-time slice of a $1+1$-dimensional world-sheet. Both the ``cut-and-glue'' quench and the ``swap'' quench possess a nice interpretation in terms of the geometry of the world-sheet.

\begin{figure}[htbp]
	\centering
	\begin{tikzpicture}[scale=0.8]
		\filldraw[gray!40] (0,-2) -- (3,-2) -- (0,2) -- (-3,2);
		\draw[thick] (0,2) -- (3,-2);
		\draw[thick] (-3,2) -- (0,-2);
		\draw[thick] (0,0) -- (-1.5,2);
		\draw[thin,<-] (-0.1,0.5) -- (1,2) node[above]{defect};
		\draw [thin,->] (-0.5,-2) -- (-3.5,2) node[left]{time};
		\draw [thin,->] (0.5,-2.5) -- (2.5,-2.5) node[below]{space};
	\end{tikzpicture}
	\begin{tikzpicture}[scale=0.8]
		\begin{scope}
		\filldraw[gray!40] (-1.5,0) -- (1.5,0) -- (0,2) -- (-3,2);
		\filldraw[gray!40] (1.5,0) -- (3,-2) -- (0,-2) -- (-0.75,-0.75);
		\draw[thick, dashed] (-1.5,0) -- (1,0);
		\draw[thick] (0,2) -- (3,-2);
		\draw[thick] (-3,2) -- (-1.5,0);
		\draw[thin] (3,-2) -- (0,-2);
		\end{scope}
		\begin{scope}[yshift=1cm,xshift=0.2cm]
		\filldraw[gray!40,thick,smooth] plot coordinates {(-1.5,0) (-1.2,-1.25) (-0.2,-3) (-0.5,-1)};
		\filldraw[gray!40] (0,-2) -- (3,-2) -- (0,2) -- (-3,2);
		\draw[thick] (0,2) -- (3,-2);
		\draw[thick,smooth] plot coordinates {(-3,2) (-1.5,0) (-1.2,-1.25) (-0.2,-3)};
		\draw[thin] (3,-2) -- (0,-2);
		\end{scope}
		\begin{scope}[yshift=2cm,xshift=0.4cm]
		\filldraw[gray!40] (0,-2) -- (3,-2) -- (0,2) -- (-3,2);
		\draw[thick] (0,2) -- (3,-2);
		\draw[thick,smooth] plot coordinates {(-3,2) (-1.5,0) (-1.2,-1.25) (-0.2,-3)};
		\draw[thick,smooth] plot coordinates {(-1.2,-0.25) (0,-2)};
		\draw[thick,dashed] (-1.2,-0.25) -- (0.2,-0.2);
		\draw[thin] (3,-2) -- (0,-2);
		\end{scope}
	\end{tikzpicture}
	\caption{The two world-sheets that correspond to the ``cut-and-glue'' quench and to the ``swap'' quench. For the ``cut-and-glue'' quench, it is a single sheet with a slit. In the case of the ``swap'' quench, there are $n$ sheet corresponding to the $n$ replicas. The top sheet is identified with the bottom one along the dashed line.}
	\label{fig:world-sheets}
\end{figure}
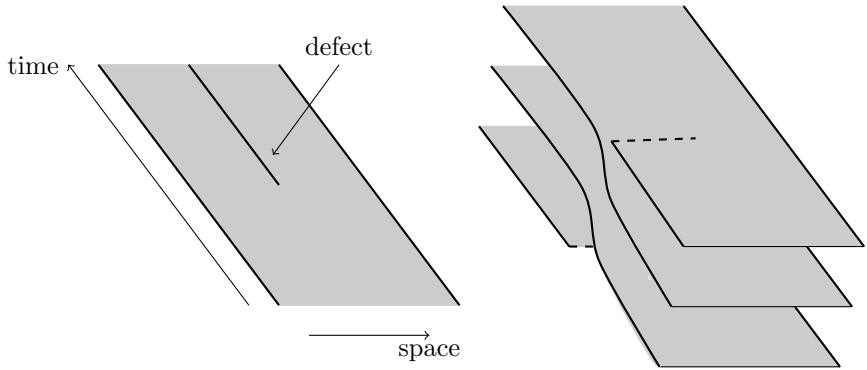

In the field theory language, the difference between the Hamiltonians $H'$ and $H$ corresponds to a local operator, which we note $\lambda~ \phi(x)$ ($\phi(x)$ is some operator in the field theory, and $\lambda$ is a coefficient). This operator perturbs the world-sheet action along a line corresponding to a fixed position $x=x_0$. In euclidean space-time, the action of the theory becomes
\begin{equation}
	S \, \rightarrow \, S \, + \, \lambda \int d\tau \phi(x_0,\tau) .
\end{equation}
This means that, while the original action $S$ describes a homogeneous system on the euclidean world-sheet, there is now a defect along the line $x=x_0$. One needs to discuss whether the operator $\phi(x,\tau)$ is relevant/irrelevant/marginal in the RG sense, when it is integrated along a line. If $\phi(x,\tau)$ is irrelevant along the line, then it cannot induce a significant change of the ground state at large distance, therefore there is no orthogonality catastrophe. On the contrary, when $\phi(x,\tau)$ is relevant, then the system flows towards a new IR fixed point, and the exponent $x$ must be calculated at this IR fixed point. The exponent $x$ is then unique, in the sense that it does not depend on the strength of the perturbation $H'-H$ (the coefficient $\lambda$). When this operator is exactly marginal, however, there is a continuum of possible exponents $x$, which depend on the strength of the perturbation. This is exactly what happens in Anderson's original treatment of the Fermi gas, and this is why there is a continuum of possible values of the exponent $x$ in that case.

Both for the ``cut-and-glue'' and the ``swap'' quenches, one needs to be able to cut some part of the system into two parts. Then the most favorable situation would be a case when the perturbation $H'- H \simeq \lambda \phi(x_0)$ is {\it relevant}. Then, a low energy, the degrees of freedom between the part $A$ and $B$ would be effectively decoupled. A physical situation where this could be realized corresponds to a Luttinger liquid where one switches on a localized potential, which induces some backscattering. The backscattering is known to be always relevant if the interactions between the particles are repulsive. The orthogonality catastrophe exponent $x$ is, however, not completely fixed, even in that case. This is because, although the contribution of the backscattering to the exponent $x$ is $1/16$, as expected in a system with central charge $c=1$ (the general result being $c/16$), it is not the only contribution. There is another contribution due to forward scattering
 , which is always marginal, and therefore depends continuously on a parameter (here, the phase shift at the Fermi point).

On the other hand, one also needs to be able to glue the two parts $A$ and $B$ together, such that the total system $A \cup B$ is homogeneous. In that case, it is the opposite situation which is the most favorable, namely the case when $\phi(x,\tau)$ is {\it irrelevant} along a line. In a Luttinger liquid, this is the case when the interactions are attractive. We thus see that, while the RG flow helps us to realize one of the two operations (either to ``cut'' or to ``glue'' the two subsystems), it makes the other operation difficult. Again, the ``swap'' quench is even more difficult to implement, as it requires not only to be able to perform the ``cut'' and ``glue'' operations, but also to apply them simultaneously on the $n$ copies of the system.

\subsection[\;\;\;\;\;\;\;\;\;\;\;\;\;Orthogonality catastrophe from cone and corner free energies on the world-sheet]{Orthogonality catastrophe from cone and corner free energies on the world-sheet}
The relevance of the world-sheet geometry to the calculation of the entanglement entropy was pointed out already in the early work of Holzhey {\it et al.} \cite{Holzhey_ee}. Essentially, the R\'enyi entropy for integer $n$ is nothing but the ratio of the partition functions of the field theory on two different world-sheets.
One way of seeing that is to write the expectation value of the swap operator in the ground state $\left| \psi \right> \otimes \dots \otimes \left| \psi\right>$ as the overlap $\left.\left.\left< \psi_{\rm repl.}'  \right| \psi_{\rm repl.} \right.\right>$ as in Eq. (\ref{eq:swap_expect}), and then use the fact that $\left| \psi_{\rm repl.} \right>$ (resp. $\left| \psi_{\rm repl.}' \right>$) is the ground state of $H_{\rm repl.}$ (resp. $H_{\rm repl.}' = \mathcal{S}_A \cdot H_{\rm repl.} \cdot \mathcal{S}_A^\dagger$). Since imaginary time evolution with $H_{\rm repl.}$ or $H_{\rm repl.}'$ automatically acts as a projection onto these states after a long time, one can re-write the overlap as
\begin{equation}
	\label{eq:imaginary_time_trick}
	\left| \left.\left. \left< \psi'_{\rm repl.} \right| \psi_{\rm repl.} \right.\right> \right| \, = \, \lim_{\tau \rightarrow +\infty} \left| \frac{\left< out \right| e^{- \tau H'_{\rm repl.}}  e^{- \tau H_{\rm repl.}} \left| in \right>}{\sqrt{\left< out \right| e^{-\tau H_{\rm repl.}'} \left| out \right> \left< in \right| e^{-\tau H_{\rm repl.}} \left| in \right>}} \right|.
\end{equation}
The states $\left| in \right>$ and $\left| out \right>$ can be chosen arbitrarily, as long as they have a non-zero overlap with the ground state of $H_{\rm repl.}$ and $H_{\rm repl.}'$ respectively. In field theory, the numerator in the right-hand side is precisely the partition function on the world-sheet drawn in Fig. \ref{fig:world-sheets} (right), while the denominator is the square root of the partition function on a world-sheet composed of $2\times n$ separated strips (note that the square root compensates the factor 2). This cartoon is very useful to evaluate the exponent of the orthogonality catastrophe associated with the swap,
\begin{equation}
	\label{eq:xswap}
	 \left| \left.\left. \left< \psi'_{\rm repl.} \right| \psi_{\rm repl.} \right.\right> \right| \, \sim \, \left( \frac{a}{L}\right)^{x_{\rm swap}} .
\end{equation}
Indeed, the exponent $x_{\rm swap}$ turns out to be a difference of free energies $f = -\log Z$ on the different world-sheets. There are extensive parts in these free energies, which scale as $L^2$ (bulk free energy) and $L$ (surface free energy) respectively. However, these contributions appear both in the numerator and the denominator of (\ref{eq:imaginary_time_trick}), and they all cancel. The next leading contribution is of order $O(\log L)$ is a conformal field theory; it is the free energy $f_{\rm cone}$ that comes from the conical singularity in the world-sheet. The coefficient of this term is nothing but the orthogonality catastrophe exponent, as can be seen by taking the logarithm of (\ref{eq:xswap}):
\begin{equation}
	x ~ \log (L/a) \, \simeq \, -\log \left| \left.\left. \left< \psi'_{\rm repl.} \right| \psi_{\rm repl.} \right.\right> \right| \, \simeq \, f_{{\rm cone}}.
\end{equation}
It is an interesting exercise in conformal field theory to evaluate $f_{\rm cone}$ for a cone of angle $\theta$; the result was derived in a famous paper of Cardy and Peschel \cite{CardyPeschel}, and it is given by $\frac{c}{12}  \left( \frac{\theta}{2\pi} - \frac{2\pi}{\theta} \right) \log (L/a)$. The swap with $n$ replicas corresponds to a cone of angle $\theta = 2\pi n$. This gives the orthogonality catastrophe exponent \cite{Holzhey_ee,CalabreseCardy_ee}
\begin{equation}
	x_{\rm swap} \, = \, \frac{c}{12} \left(n - \frac{1}{ n} \right) .
\end{equation}
For the world-sheet corresponding to the ``cut-and-glue'' quench, the situation is very similar, except that there is no cone in the bulk, but rather a corner at the tip of the defect line (Fig. \ref{fig:world-sheets}). In that case, the orthogonality catastrophe exponent $x_{\rm cut/glue}$ is identified with the coefficient of the $O(\log L)$ term in the {\it corner free energy}, for a corner of internal angle $\theta = 2\pi$. The evaluation of $f_{\rm corner}$ in a conformal field theory was also done by Cardy and Peschel \cite{CardyPeschel}; we recalled this result in section \ref{sec:cardypeschel}. The orthogonality catastrophe exponent associated to the ``cut-and-glue'' quench is
\begin{equation}
	x_{\rm cut/glue} \, = \, \frac{c}{16}.
\end{equation}

\section[\;\;\;\;\;\;\;\;\;\;\;\;\;The effect of boundary changing operators]{The effect of boundary changing operators}
\label{sec:app_bcc}
In this appendix we consider the effect of boundary changing operators on the universal finite-size function $f(x=\ell/L)$, extending the calculation presented in \cite{Bipartite_fidelity} when there are none. We first derive a formula in the most general case of four different changes in boundary conditions (\ref{sec:gen_bcc}), before specializing to more realistic situations (\ref{sec:abba} and \ref{sec:abab}) relevant to the main text. 
\subsection[\;\;\;\;\;\;\;\;\;\;\;\;\;Leading universal term]{Leading universal term}
\label{sec:gen_bcc}
\begin{figure}[htbp]
 \centering
 \begin{tikzpicture}
 \draw [color=gray!40,fill=gray!40] (0,0) -- (4,0) -- (4,2) -- (0,2) -- (0,0);
  \draw [ultra thick] (0,0) -- (4,0);
  \draw [ultra thick] (0,1.2) -- (2,1.2);
  \draw [ultra thick] (0,2) -- (4,2);
  \draw [<->] (2.6,0.05) -- (2.6,1.15);
  \draw (3.15,0.6) node {$L-\ell$};
  \draw [<->] (2.6,1.25) -- (2.6,1.95);
  \draw (3.15,1.6) node {$\ell$};
  \draw[color=red] (1.7,-0.2) node {$\alpha$};
  \draw[color=red] (1.7,0.95) node {$\beta$};
  \draw[color=red] (1.7,1.4) node {$\gamma$};
  \draw[color=red] (1.7,2.23) node {$\delta$};
  \draw[line width=2pt,color=blue] (0.2,0) -- (3.8,0);
  \draw[line width=2pt,color=blue] (0.2,-0.15) -- (0.2,0.15);
  \draw[line width=2pt,color=blue] (3.8,-0.15) -- (3.8,0.15);
  \draw[line width=2pt,color=blue] (0.2,2) -- (3.8,2);
  \draw[line width=2pt,color=blue] (0.2,1.85) -- (0.2,2.15);
  \draw[line width=2pt,color=blue] (3.8,1.85) -- (3.8,2.15);
  \draw[color=blue] (0.2,-0.4) node {$-\Lambda$};
  \draw[color=blue] (3.8,-0.4) node {$\Lambda$};
  \draw[color=blue] (0.2,2.4) node {$-\Lambda+iL$};
  \draw[color=blue] (3.8,2.4) node {$\Lambda+iL$};
  \begin{scope}[xshift=5.5cm]
  \draw [thick,dotted] (-0.75,-0.5) -- (-0.75,2.5);
  \draw [color=gray!40,fill=gray!40] (0,0) -- (4,0) -- (4,1.15) -- (0,1.15) -- (0,0);
  \draw [color=gray!40,fill=gray!40] (0,1.25) -- (4,1.25) -- (4,2) -- (0,2) -- (0,1.25);
  \draw [ultra thick] (0,0) -- (4,0);
  \draw [ultra thick] (0,1.15) -- (4,1.15);
  \draw [ultra thick] (0,1.25) -- (4,1.25);
  \draw [ultra thick] (0,2) -- (4,2);
  \draw[color=red] (1.7,-0.2) node {$\alpha$};
  \draw[color=red] (1.7,0.9) node {$\beta$};
  \draw[color=red] (1.7,1.45) node {$\gamma$};
  \draw[color=red] (1.7,2.23) node {$\delta$};
  \draw [<->] (2.6,0.05) -- (2.6,1.1);
  \draw (3.15,0.6) node {$L-\ell$};
  \draw [<->] (2.6,1.3) -- (2.6,1.95);
  \draw (3.15,1.6) node {$\ell$};
  \draw[line width=2pt,color=blue] (0.2,0) -- (3.8,0);
  \draw[line width=2pt,color=blue] (0.2,-0.15) -- (0.2,0.15);
  \draw[line width=2pt,color=blue] (3.8,-0.15) -- (3.8,0.15);
  \draw[line width=2pt,color=blue] (0.2,2) -- (3.8,2);
  \draw[line width=2pt,color=blue] (0.2,1.85) -- (0.2,2.15);
  \draw[line width=2pt,color=blue] (3.8,1.85) -- (3.8,2.15);
  \draw[color=blue] (0.2,-0.4) node {$-\Lambda$};
  \draw[color=blue] (3.8,-0.4) node {$\Lambda$};
  \draw[color=blue] (0.2,2.4) node {$-\Lambda+iL$};
  \draw[color=blue] (3.8,2.4) node {$\Lambda+iL$};
  \end{scope}
  \begin{scope}[xshift=11cm]
  \draw [thick,dotted] (-0.75,-0.5) -- (-0.75,2.5);
  \draw [color=gray!40,fill=gray!40] (0,0) -- (4,0) -- (4,2) -- (0,2) -- (0,0);
  \draw [ultra thick] (0,0) -- (4,0);
  \draw [ultra thick] (0,2) -- (4,2);
  \draw [<->] (2.6,0.05) -- (2.6,1.95);
  \draw (3.1,1) node {$L$};
  
  \draw[color=red] (1.7,-0.2) node {$\alpha$};
  \draw[color=red] (1.7,2.23) node {$\delta$};
  \draw[line width=2pt,color=blue] (0.2,0) -- (3.8,0);
  \draw[line width=2pt,color=blue] (0.2,-0.15) -- (0.2,0.15);
  \draw[line width=2pt,color=blue] (3.8,-0.15) -- (3.8,0.15);
  \draw[line width=2pt,color=blue] (0.2,2) -- (3.8,2);
  \draw[line width=2pt,color=blue] (0.2,1.85) -- (0.2,2.15);
  \draw[line width=2pt,color=blue] (3.8,1.85) -- (3.8,2.15);
  \draw[color=blue] (0.2,-0.4) node {$-\Lambda$};
  \draw[color=blue] (3.8,-0.4) node {$\Lambda$};
  \draw[color=blue] (0.2,2.4) node {$-\Lambda+iL$};
  \draw[color=blue] (3.8,2.4) node {$\Lambda+iL$};  
  \end{scope}
 \end{tikzpicture}
\caption{Free energies involved in the calculation of the LBF (Eq.~\ref{eq:freenrj_bcc}), with four different boundary conditions $\alpha,\beta,\gamma,\delta$. \emph{Left:} $f_{A,B}^{\alpha\beta\gamma\delta}$. \emph{Middle:} $f_{A\otimes B}^{\alpha\beta\gamma\delta}=f_A^{\alpha\beta}+f_B^{\gamma\delta}$. \emph{Right:} $f_{A\cup B}^{\alpha\delta}$. The lower boundary is by convention at ${\rm Im}(z)=0.$}
\label{fig:freenrj_bcc}
\end{figure}
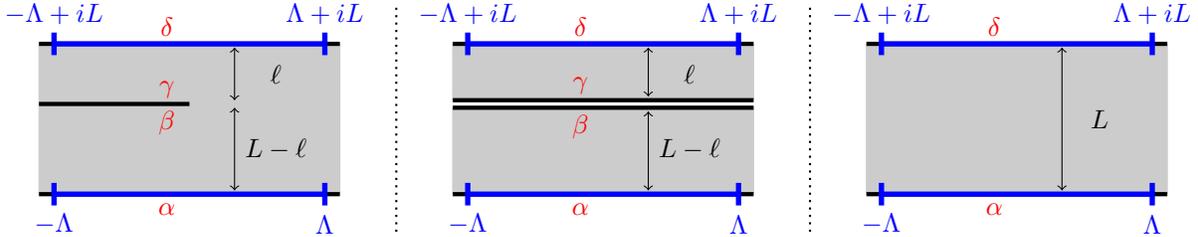

Let us look at the LBF in imaginary time, when the four possible boundary conditions, denoted by $\alpha,\beta,\gamma,\delta$, are different. We have
\begin{equation}\label{eq:freenrj_bcc}
 \mathcal{F}_{\alpha\beta\gamma\delta}=2f_{A,B}^{\alpha\beta\gamma\delta}-f_{A}^{\alpha\beta}-f_B^{\gamma\delta}-f_{A\cup B}^{\alpha\delta}
\end{equation}
which are represented in figure \ref{fig:freenrj_bcc}.

\begin{figure}[htbp]
\centering
 \begin{tikzpicture}
  \draw [color=black!20,fill=black!20] (0,0) -- (5,0) -- (5,2.5) -- (0,2.5) -- (0,0);
  \draw [ultra thick] (0,0) -- (5,0);
  \draw [ultra thick] (0,2.5) -- (5,2.5);
  \draw [ultra thick] (0,1.5) -- (2.5,1.5);
  \draw[color=red] (2.2,-0.2) node {$\alpha$};
  \draw[color=red] (2.2,1.25) node {$\beta$};
  \draw[color=red] (2.2,1.7) node {$\gamma$};
  \draw[color=red] (2.2,2.72) node {$\delta$};
  \draw[line width=2pt,color=blue] (0.3,0) -- (4.7,0);
  \draw[line width=2pt,color=blue] (0.3,-0.15) -- (0.3,0.15);
  \draw[line width=2pt,color=blue] (4.7,-0.15) -- (4.7,0.15);
  \draw[line width=2pt,color=blue] (0.3,2.5) -- (4.7,2.5);
    \draw[line width=2pt,color=blue] (0.3,2.35) -- (0.3,2.65);
  \draw[line width=2pt,color=blue] (4.7,2.35) -- (4.7,2.65);
  \draw[color=blue] (0.3,-0.4) node {$-\Lambda$};
    \draw[color=blue] (4.7,-0.4) node {$\Lambda$};
      \draw[color=blue] (0.3,2.9) node {$-\Lambda+iL$};
    \draw[color=blue] (4.7,2.9) node {$\Lambda+iL$};
 \begin{scope}[xshift=9cm]
  \draw [color=black!20,fill=black!20] (-1,0) -- (7,0) -- (7,3) -- (-1,3) -- (-1,0);
  \draw [ultra thick] (-1,0) -- (7,0);
  \draw [ultra thick] (1.6,-0.15) -- (1.6,0.15);
  \draw [ultra thick] (4.7,-0.15) -- (4.7,0.15);
  \draw [color=red,fill=red] (3,0) circle (2.2pt);
  \draw [color=red,fill=red] (1.6,0) circle (2.2pt);
  \draw [color=red,fill=red] (4.7,0) circle (2.2pt);
  \draw (1.6,0.4) node {$\phi_1(z_1)$};
  \draw (3,0.4) node {$\phi_2(z_2)$};
  \draw (4.7,0.4) node {$\phi_3(z_3)$};
  \draw [color=red,fill=red] (3,3) circle (2.2pt);
  \draw (3,2.6) node {$\phi_4(i\infty)$};
  \draw[<-,very thick,color=green!0!black,xshift=-1.15cm] (-2.5,1.4) .. controls (-2,1.7) and (-1,1.9) .. (-0.2,1.4);
  \draw[color=green!0!black] (-2.5,1.3) node {$w(z)$};
  \draw[line width=2pt,color=blue] (-0.9,0) -- (1.4,0);
  \draw[line width=2pt,color=blue] (-0.9,-0.15) -- (-0.9,0.15);
  \draw[color=blue] (-0.9,-0.4) node {$w^{-1}(\Lambda+iL)$};
  \draw[line width=2pt,color=blue] (1.4,-0.15) -- (1.4,0.15);
  \draw[color=blue] (1.4,-0.4) node {$w^{-1}(-\Lambda+iL)$};
  \draw[line width=2pt,color=blue] (4.9,0) -- (6.9,0);
  \draw[line width=2pt,color=blue] (4.9,-0.15) -- (4.9,0.15);
  \draw[color=blue] (4.9,-0.4) node {$w^{-1}(-\Lambda)$};
  \draw[line width=2pt,color=blue] (6.9,-0.15) -- (6.9,0.15);
  \draw[color=blue] (6.9,-0.4) node {$w^{-1}(\Lambda)$};
 \end{scope}
 \end{tikzpicture}
\caption{\emph{Left}: Asymmetric pants geometry, with four different boundary conditions $\alpha$, $\beta$, $\gamma$, $\delta$. \emph{Right}: Upper Half plane $\mathbb{H}$, and the four boundary changing operators.}
\label{fig:bcc}
\end{figure}
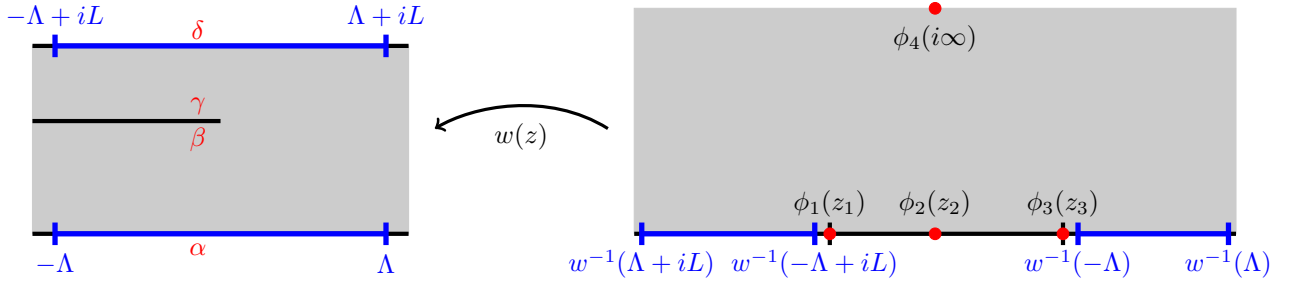

In the following we drop all the greek superscripts, to lighten the notations. Let us now consider the infinitesimal transformation $w\mapsto w+i\delta \ell$ inside the strip, and $w\mapsto w$ otherwise. The corresponding variation in the free energy can be expressed using the $T_{yy}$ component of the stress-energy tensor and the boundary operators. For $f_{A,B}$ we have
\begin{equation}\label{eq:var_freenrj}
 \delta f_{A,B}^{(\Lambda)}=\frac{\delta \ell}{2\pi} \left[\int_{-\Lambda}^{\Lambda} \frac{\langle T_{yy}\phi_{1}\phi_{2}\phi_{3}\phi_{4}\rangle}{\langle \phi_{1}\phi_{2}\phi_{3}\phi_{4}\rangle} dx-\int_{-\Lambda+iL}^{\Lambda+iL} \frac{\langle T_{yy}\phi_{1}\phi_{2}\phi_{3}\phi_{4}\rangle}{\langle \phi_{1}\phi_{2}\phi_{3}\phi_{4}\rangle} dx\right],
\end{equation}
and there are similar expressions for $f_{A\otimes B}^{(\Lambda)}$, $f_{A\cup B}^{(\Lambda)}$. $\phi_1$ (resp. $\phi_2$, $\phi_3$, $\phi_4$) is an operator that changes the boundary condition from $\alpha$ to $\beta$ (resp. $\beta$ to $\gamma$, $\gamma$ to $\delta$, $\delta$ to $\alpha$). It has dimension $h_1$ (resp. $h_2$, $h_3$, $h_4$). Notice that it is necessary to introduce a cutoff $\Lambda$. Indeed all the free energies defining the LBF are infinite in the limit $\Lambda \to \infty$, but the crucial point is that the linear combination
\begin{equation}
 \delta \mathcal{F}=\lim_{\Lambda \to \infty}\left(2\delta f_{A,B}^{(\Lambda)}-\delta f_{A\otimes B}^{(\Lambda)}-\delta f_{A\cup B}^{(\Lambda)}\right)
\end{equation}
remains finite. The variation (\ref{eq:var_freenrj}) can be computed using $T_{yy}=T(w)+\bar{T}(\bar{w})$ and the inverse conformal transformation to the upper-half plane, see Fig.~\ref{fig:bcc}. 
The conformal transformation that maps the upper half-plane to the asymmetric pants geometry is given by
\begin{equation}\label{eq:pants_transformation}
 w(z)=\frac{L}{\pi} \left(x\log [z+1]+(1-x)\log \left[z\frac{x}{1-x}-1\right]\right),
\end{equation}
and the transformation law of the stress tensor reads
\begin{equation}
 T(w)\left(\frac{dw}{dz}\right)^2=T(z)-\frac{c}{12}\{w(z),z\},
\end{equation}
where $\{w(z),z\}$ is the schwarzian derivative. Using this, Eq.~(\ref{eq:var_freenrj}) becomes 
\begin{equation}\label{eq:ugly_integral}
 \delta f_{A,B}^{(\Lambda)}=\frac{\delta \ell}{\pi}\left[ \int_{w^{-1}(-\Lambda)}^{w^{-1}(\Lambda)}\left(v_c(z)+v_h(z)\right)dz - \int_{w^{-1}(-\Lambda+iL)}^{w^{-1}(\Lambda+iL)}\left(v_c(z)+v_h(z)\right)dz\right],
\end{equation}
with 
\begin{eqnarray}
 v_c(z)&=&-\frac{c}{12}\left(\frac{dw}{dz}\right)^{-1}\{w(z),z\},\\
 v_h(z)&=&\left(\frac{dw}{dz}\right)^{-1} \frac{\left\langle T (z)\phi_{1}(z_1)\phi_{2}(z_2)\phi_{3}(z_3)\phi_{4}(z_4)\right\rangle}{\langle \phi_{1}(z_1)\phi_{2}(z_2)\phi_{3}(z_3)\phi_{4}(z_4)\rangle}.
\end{eqnarray}
The bcc operators are inserted at the points $z_1=-1$, $z_2=0$, $z_3=1/x-1$ and at $z_4\equiv i \infty$ in the $z$ plane. Without them $v_h(z)=0$ and the calculation boils down to that in Ref.~\cite{Bipartite_fidelity}. $v_h(z)$ can be found using the conformal Ward identity for chiral operators
\begin{equation}\label{eq:ward}
 \left\langle T(z)\prod_{\alpha}\phi_\alpha(z_\alpha)\right\rangle=
 \left(
 \sum_\alpha \frac{h_\alpha}{(z-z_\alpha)^2}+\frac{1}{z-z_\alpha}\frac{\partial}{\partial z_\alpha}
 \right)\left\langle \prod_{\alpha}\phi_\alpha(z_\alpha)\right\rangle,
\end{equation}
 and the general expression for the chiral four-point function
\begin{equation}\label{eq:fourpoint_gen}
 \left\langle \phi_1(z_1)\phi_2(z_2)\phi_3(z_3)\phi_4(z_4) \right\rangle=\left(\prod_{1\leq i <j\leq 4} z_{ij}^{H/3-h_i-h_j}\right)\times \Upsilon\left(\zeta\right)
 \quad,\quad \zeta=\frac{z_{12}z_{34}}{z_{13}z_{24}}.
\end{equation}
Here $z_{ij}=|z_i-z_j|$, $H=\sum_{i=1}^{4} h_i$ and $\Upsilon$ is a universal function of the anharmonic ratio $\zeta$. The second integral in (\ref{eq:ugly_integral}) can be deduced from the first by symmetry, replacing $x$ by $1-x$ and exchanging  the roles of $h_1$ and $h_3$. Computing the first integral and repeating the same procedure for $f_{A\cup B}^{(\Lambda)}$ and $f_{A\otimes B}^{(\Lambda)}$ cancels out all the diverging cutoff-dependent terms, as should be. We get 
\begin{eqnarray}\fl\nonumber
\frac{\delta \mathcal{F}}{\delta x}&=&
\frac{1}{1-x}\left[\left(\frac{c}{8}-2h_1+h_2\right)\frac{(x-2)}{3}-\frac{2h_3}{3}(1+x)+\frac{2h_4}{3}(2x-1)\right]+\left[
\frac{2h_1}{x}-2h_4 x+\frac{c(x^2-1)}{12x}\right]\log (1-x)
\\\fl
&+&\left[2h_4 x-\frac{2h_3 x}{(1-x)^2}+\frac{c(2-x)x^2}{12(1-x)^2}\right]\log x-2x\frac{\Upsilon^\prime(x)}{\Upsilon(x)}-\bigg\{ x\to 1-x
\;;\;h_1 \leftrightarrow h_3\bigg\},
\end{eqnarray}
after long algebra. We have introduced the notation
\begin{equation}
 \bigg\{ x\to 1-x
\;;\;h_1 \leftrightarrow h_3\bigg\}
\end{equation}
for the action of replacing $x$ by $1-x$ and swapping $h_1$ and $h_3$ in all the preceding terms. After integration this gives us the universtal scaling function $f(x)$ for the LBF. The final result reads
\vspace{0.2cm}
\begin{equation}\fl
\boxed{
 \begin{array}{lll}
  f(x)&=&\displaystyle{\left[\frac{c}{24}\left(2x-1+\frac{2}{x}\right)+\left(\frac{4}{3}-\frac{2}{x}\right)h_1+\frac{h_2}{3}-\frac{2h_3}{3}+\left(\frac{4}{3}-2x\right)h_4\right]\log (1-x)}-\log \Upsilon(x)\\&&\\
 &+&\displaystyle{\bigg\{x\to 1-x\;;\;h_1 \leftrightarrow h_3\bigg\}}
 \end{array}
 }\label{eq:f_general}
\end{equation}
\vspace{0.2cm}
Of course this function depends on the precise form of $\Upsilon$, and $f(x)$ is sensible to the refined details of the CFT. A simple case of interest is that of vertex operators
\begin{equation}
 \left\langle V_{\alpha_1}(z_1)V_{\alpha_2}(z_2)V_{\alpha_3}(z_3)V_{\alpha_4}(z_4) \right\rangle=\prod_{1\leq i <j\leq 4} z_{ij}^{\alpha_i\alpha_j},
\end{equation}
where the $\alpha_i$ are the charges, submitted to the neutrality condition $\sum_{i=1}^{4} \alpha_i=0$. This form can be identified with (\ref{eq:fourpoint_gen}), setting $h_i=\alpha_i^2/2$ and
\begin{equation}
 \Upsilon(\zeta)=\zeta^{-H/3+1/2(\alpha_1+\alpha_2)^2}\left(1-\zeta\right)^{-H/3+1/2(\alpha_2+\alpha_3)^2}.
\end{equation}
We get
\begin{eqnarray}\nonumber
 f(x)&=&\left[\frac{c}{24}\left(2x-1+\frac{2}{x}\right)+\left(1-\frac{1}{x}\right)\alpha_1^2-\frac{\alpha_2^2}{2}-2\alpha_2\alpha_3-\alpha_3^2+(1-x)\alpha_4^2\right]
 \log (1-x)\\\label{eq:f_vertex}
 &+&\bigg\{x\to 1-x \;;\;\alpha_1 \leftrightarrow \alpha_3\bigg\}
\end{eqnarray}
The case without any boundary condition changes \cite{Bipartite_fidelity} can be recovered by setting all the charges $\alpha_i=0$ in (\ref{eq:f_vertex}), or, equivalently, setting all the dimensions $h_i=0$ and $\Upsilon(\zeta)=1$ in (\ref{eq:f_general}).
\subsection[\;\;\;\;\;\;\;\;\;\;\;\;\;Subleading semi-universal term]{Subleading semi-universal term}
Using the results presented in \ref{sec:gen_bcc}, the calculation of the $L^{-1}\log L$ becomes straightforward. It suffice to notice that $\braket{T(w)}$ is modified, in the presence of boundary condition changing operators, to $\braket{T(w)\phi_1\phi_2\phi_3\phi_4}/\braket{\phi_1\phi_2\phi_3\phi_4}$. Then all the arguments given in Sec.~\ref{sec:loglsl_theory} apply, and we find
 \begin{equation}\label{eq:loglsl_full_bcc}\fl
\Delta F=\left(\sum_c e^{i\alpha_c}\,  {\rm Res}\, \left[\left(\frac{dw}{dz}\right)^{-1}\left\{ \frac{c}{12}\{w(z),z\}-\frac{\braket{T(z)\phi_1(z_1)\phi_2(z_2)\phi_3(z_3)\phi_4(z_4)}}{\braket{\phi_1(z_1)\phi_2(z_2)\phi_3(z_3)\phi_4(z_4)}}\right\},z=z_c\right]\right)\times \frac{\xi}{2 \pi} \log L,
\end{equation}
which may be evaluated using the Ward identity (\ref{eq:ward}). For the pants geometry we find
\begin{equation}
 \Delta F=\left[h_{4}-\frac{c}{24}+\left(\frac{c}{24}-h_{1}\right)\frac{1}{x}+\left(\frac{c}{24}-h_{3}\right)\frac{1}{1-x}
 \right]\times \xi\,\frac{\log L}{L}.
\end{equation}
Note that this expression does not depend on the function of the anharmonic ratio $\Upsilon(\zeta)$.
\subsection[\;\;\;\;\;\;\;\;\;\;\;\;\;Special cases]{Special cases}
\paragraph{Special case abba}
\label{sec:abba}
We now specialize to the case abba, with only two boundary changing operators inserted. This is relevant to Sec.~\ref{sec:lbf_isingbcc}. We find, setting $\alpha_2=\alpha_4=0$ and $\alpha_1^2=\alpha_3^2=2h$: 
\begin{equation}
 f(x)=\left[\frac{c}{24}\left(2x-1+\frac{2}{x}\right)-\frac{2h}{x}\right]\log (1-x)\;+\;\bigg\{x \to 1-x\bigg\}
\end{equation}
\paragraph{Special case abab}
\label{sec:abab}
Here we specialize to the case of $4$ bcc operators of the vertex type, relevant to Sec.~\ref{sec:lbf_filling}. Setting $\alpha_1=\alpha_3=-\alpha_2=-\alpha_4=\sqrt{2h}$, we get 
\begin{equation}
 f(x)=\left[\left(\frac{c}{24}-h\right)\left(2x-1+\frac{2}{x}\right)+4h\right]\log (1-x)+\;\bigg\{x \to 1-x\bigg\}
\end{equation}
In the XXZ chain $h=h_\rho=\frac{R^2}{2} \left(\frac{\delta}{\pi}\right)^2$, where $R$ is the compactification radius and $\delta$ the phase shift at the Fermi surface\cite{XXZ_bethenum2}. In the XX limit $R=1$ and $h_\rho=1/2(\rho-1/2)^2$.

\pagebreak
\section[\;\;\;\;\;\;\;\;\;\;\;\;\;LBF for an XX chain cut into two equal parts: an exact result]{LBF for an XX chain cut into two equal parts: an exact result}
\label{sec:exact}

In this appendix we recall and expand the exact result of \cite{PhDStephan}, for the LBF of an open XX chain cut in two subsystems of the same size. A summary is presented in \ref{sec:summary}. After diagonalization (\ref{sec:diag}), we show how the fidelity can be computed as a Cauchy-type determinant, and expressed in closed form (\ref{sec:cauchy}). The asymptotic expansion is explained in \ref{sec:asymptotic}. Finally, we give in \ref{sec:other_examples} two other examples to which the method presented here can be applied. 
\subsection[\;\;\;\;\;\;\;\;\;\;\;\;\;Summary of the asymptotic expansion]{Summary of the asymptotic expansion}
\label{sec:summary}
Before proceeding with the derivation, let us first present the exact result for the asymptotic expansion of the LBF. For any integer $p\geq 1$, we have
\begin{equation}
 \mathcal{F}(L)=\sum_{k=0}^{p-1}\left(\gamma_k\log L+\mu_k\right)L^{-k}+\mathcal{O}\left(\frac{1}{L^p}\right).
\end{equation}
The leading terms are given by
\begin{eqnarray}\label{eq:gamma0}
 \gamma_0&=&\frac{1}{8}+h_\rho\\
 \gamma_1&=&\frac{1}{8}-3 h_\rho
\end{eqnarray}
$h_{\rho}=\frac{1}{2}\left(\rho-1/2\right)^2$ is the dimension of the (phase shift) boundary changing operator discussed in the text. In CFT, $\gamma_0$ is determined from the Cardy-Peschel formula, while $\gamma_1$ is a consequence of our main result. In this particular geometry $\gamma_k$ is given by
\begin{equation}\label{eq:gammak}
 \gamma_k=(-1)^{k+1}\left(2^{k+1}-1\right)\frac{\gamma_1}{3}=\left(\frac{1}{24}-h_\rho\right)(-1)^{k+1}\left(2^{k+1}-1\right),
\end{equation}
for $k\geq 1$. We were also able to compute the (non-universal) constant term, which is given by 
\begin{eqnarray}\fl\nonumber
 \mu_0&=&\frac{\zeta(3)}{2\pi^2}+\frac{1}{4}\left(1+2\rho-2\rho^2\right)\log \sin (\pi \rho)-\rho \log \sin \left(\frac{\pi \rho}{2}\right)
 -\frac{1}{2}\left(2\rho^2-2\rho+1\right)\log 2-\frac{1}{4}\left(2\rho^2-6\rho+3\right)\log \pi\\
 \fl &-&\frac{1}{2\pi}\left[2\,{\rm Cl}_2(\pi \rho)+\left(1-2\rho\right){\rm Cl}_2(2\pi \rho)+\frac{1}{\pi}\,{\rm Cl}_3(2\pi\rho)\right]-2\log \left[G\left(1+\frac{\rho}{2}\right)G\left(\frac{1+\rho}{2}\right)\right]\label{eq:mu0}
\end{eqnarray}
where $\zeta$ is the Riemann zeta function, $G$ the Barnes G function, and the ${\rm Cl}_n$ are the $n-$th Clausen functions defined by
\begin{equation}
 {\rm Cl}_n(x)=\left\{\begin{array}{ccc}
                       \displaystyle{\sum_{k=1}^{\infty}\frac{\sin (kx)}{k^n}}&,& n \;{\rm even}\\\\
                       \displaystyle{\sum_{k=1}^{\infty}\frac{\cos (kx)}{k^n}}&,& n \;{\rm odd}
                      \end{array}
 \right.
\end{equation}
At half filling $\rho=1/2$, the constant can be shown to take the simpler form
\begin{equation}
 \mu_0=\frac{9}{2}\log A+\frac{1}{2}\log \left(\frac{\Gamma(3/4)}{\Gamma(1/4)}\right)-\frac{1}{8}\log \pi -\frac{3}{8}-\frac{C}{\pi}+\frac{7}{8}\frac{\zeta(3)}{\pi^2}\simeq -0.126058180038976
\end{equation}
$C$ is Catalan's constant and $A$ is the Glaisher-Kinkelin constant. 
\subsection[\;\;\;\;\;\;\;\;\;\;\;\;\;Diagonalization]{Diagonalization}\label{sec:diag}
We consider an XX chain of length $L$, in a transverse magnetic field. The Hamiltonian reads
\begin{equation}
 H_{A\cup B}=-\sum_{j=1}^{L-1}\left(\sigma_j^x\sigma_{j+1}^x+\sigma_j^y\sigma_{j+1}^y\right)-h\sum_{j=1}^{L}\sigma_j^z.
\end{equation}
Upon performing a Jordan-Wigner transformation
\begin{eqnarray}
 \sigma_j^x+i\sigma_j^y&=&c_j^\dag \exp\left(i\pi \sum_{l=1}^{j-1}c_j^\dag c_j\right)\\
 \sigma_j^z&=&2c_j^\dag c_j-1,
\end{eqnarray}
it may be rewritten as
\begin{equation}
 H_{A\cup B}=-\sum_{j=1}^{L} \left(c_{j+1}^\dag c_j +c_{j}^\dag c_{j+1}\right)-h\sum_{j=1}^{L}\left(2c_j^\dag c_j-1\right),
\end{equation}
and then diagonalized by a standard techniques\cite{lieb1961two}. Introducing the new set of fermions
\begin{equation}
 d_m^\dag=\left(\frac{2}{L+1}\right)^{1/2}\sum_{j=1}^L \sin (j\phi_m) c_j^\dag,
\end{equation}
with the quasimomenta given by
\begin{equation}
 \phi_m=\frac{m\pi }{L+1}\quad,\quad m=1,2,\ldots,L,
\end{equation}
allows to rewrite $H_{A\cup B}$ as
\begin{equation}
 H_{A \cup B}={\rm cst}+\sum_{m=1}^{L} \epsilon_m d_m^\dag d_m\qquad,\qquad \epsilon_m=-\cos(\phi_m)-h.
\end{equation}
The ground-state may be obtained by filling the momenta with negative energy:
\begin{equation}
 |A\cup B\rangle =d_1^\dag d_2^\dag \ldots d_N^\dag |0\rangle
\end{equation}
$N$ is the number of fermions and depends on $h$. At zero magnetic field and we have $N=L/2$ (half-filling). In general we assume a constant filling fraction $\rho=N/L$. Let us repeat the same procedure for the cut Hamiltonian
\begin{equation}\label{eq:XX_cut}
 H_{A\otimes B}=H_A+H_B,
\end{equation}
with
\begin{eqnarray}
H_A&=&-\sum_{j=1}^{L_A-1} \left(c_{j+1}^\dag c_j+c_j^\dag c_{j+1}\right)-h\sum_{j=1}^{L_A}(2c_j^\dag c_j-1)
\\
H_B&=&-\sum_{j=L_A+1}^{L} \left(c_{j+1}^\dag c_j+c_j^\dag c_{j+1}\right)-h\sum_{j=L_A+1}^{L}(2c_j^\dag c_j-1).
\end{eqnarray}
$H_{A\otimes B}$ is identical to $H_{A\cup B}$ except for the local hoping term $-\big(c_{L_A+1}^\dag c_{L_A}+h.c\big)$ that connects subsystem $A$ and $B$ in $H_{A\cup B}$. The two terms in Eq.~(\ref{eq:XX_cut}) can be diagonalized separately, introducing the momenta
\begin{equation}
 \theta_k^{\Omega}=\frac{k\pi}{L_\Omega+1} \quad ,\quad k=1,2,\ldots,L_\Omega
\end{equation}
for either $\Omega=A$ or $\Omega=B$. The two new set of fermions are
\begin{eqnarray}
a_k^\dag&=&\left(\frac{2}{L_A+1}\right)^{1/2}\sum_{j=1}^{L_A} \sin \left(j\theta_k^A \right) c_j^\dag\\
b_k^\dag&=&\left(\frac{2}{L_B+1}\right)^{1/2}\sum_{j=L_A+1}^{L} \sin \left(j\theta_k^B \right) c_{j}^\dag
\end{eqnarray}
The ground state is then given by
\begin{equation}
 |A\otimes B\rangle=a_1^\dag a_2^\dag\ldots a_{N_A}^\dag b_1^\dag b_2^\dag \ldots b_{N_B}^\dag |0\rangle
\end{equation}
where $N_\Omega=\rho L_\Omega$ is the number of fermions in subsystem $\Omega$. To avoid any unnecessary complications, we will always assume $\rho L_\Omega$ to be an integer. For example at half filling ($\rho=1/2$) this means both $L_A$ and $L_B$ have to be even.  
\subsection[\;\;\;\;\;\;\;\;\;\;\;\;\;A simple variation on the Cauchy determinant identity]{A simple variation on the Cauchy determinant identity}
\label{sec:newcauchy}
We take a small detour to derive a Cauchy-type determinant identity that will be useful to us later. The celebrated Cauchy determinant formula is given by
\begin{equation}
\det_{1\leq i,j \leq n} \left(\frac{1}{x_i+y_j}\right)=
\frac{\prod_{1\leq i<j\leq n} (x_i-x_j) \prod_{1\leq i<j\leq n} (y_i-y_j)}{\prod_{1\leq i,j\leq n} (x_i+y_j)}.
\end{equation}
It can be understood by noticing that both sides of the equality are rational fractions that have the same poles and zeroes. In the following we will need to evaluate a determinant with a similar structure. Consider a $n\times n$ matrix with $n$ even and matrix elements
\begin{equation}\label{eq:newcauchyform}
 m_{ij}=\left\{
  \begin{array}{ccc}
   \displaystyle{\frac{(-1)^i}{x_i+y_j}}&,&1\leq i \leq n/2\\&&\\
   \displaystyle{\frac{(-1)^j}{x_{i-n/2}+y_j}}&,&n/2+1\leq i\leq n
  \end{array}
 \right.
\end{equation}
The determinant can be expressed as a product of two Cauchy determinants, using elementary row-column manipulations. However, it is much more elegant to evaluate it directly, using a similar reasoning:
\begin{equation}\label{eq:newcauchy}
 \det_{1\leq i,j\leq n} \left(m_{ij}\right)=2^{n/2}\times\frac{\prod_{1\leq i<j\leq n/2} \left(x_i-x_j\right)^2\prod_{1\leq i<j\leq n}^{(i-j)\,{\rm even}}\left(y_i-y_j\right)}
 {\prod_{i=1}^{n/2}\prod_{j=1}^{n}\left(x_i+y_j\right)}.
\end{equation}
The poles are obvious. Then, the determinant has double zeroes for $x_i=x_j$, $i\neq j$ and simple zeroes for $y_i=y_j$, $i\neq j$ \emph{provided} $(-1)^i=(-1)^j$. The remaining $2^{n/2}$ factor is explained by the possible choice of two $x_i$ for a given $i=1,\ldots,n/2$. Despite of its simplicity, we have found no mention of the identity (\ref{eq:newcauchy}) in the literature. 
\subsection[\;\;\;\;\;\;\;\;\;\;\;\;\;A closed-form formula for the fidelity]{A closed-form formula for the fidelity}
\label{sec:cauchy}
Using the results of the previous section, the overlap we are looking for can be expressed as a determinant:
\begin{eqnarray}\label{eq:wick1}
 \langle A\otimes B|A\cup B\rangle&=&\langle 0|a_1 a_2\ldots a_{N_A}b_1 b_2 \ldots b_{N_B}d_1^\dag d_2^\dag\ldots d_N^\dag|0\rangle\\\label{eq:wick2}
 &=&\det_{1\leq k,l\leq N} \left(M_{kl}\right).
\end{eqnarray}
Eq.~(\ref{eq:wick2}) follows from Eq.~(\ref{eq:wick1}) by applying the Wick theorem. The matrix elements are, after some algebra, given by
\begin{eqnarray}\label{eq:matelements}
 M_{kl}=\langle 0|a_k d_l^\dag|0\rangle=(-1)^k z^A \frac{\alpha_k^A \beta_l^A}{u_k^A-v_l}\qquad\qquad\quad\;\,,\qquad 1\leq k\leq N_A\\
 M_{kl}=\langle 0|b_{k-N_A} d_l^\dag|0\rangle=(-1)^l z^B \frac{\alpha_{k-N_A}^B \beta_l^B}{u_{k-N_A}^B-v_l}\qquad,\qquad N_A< k\leq N,
\end{eqnarray}
with
\begin{eqnarray}
 z^\Omega&=&\left[(L+1)(L_\Omega+1)\right]^{-1/2}\\
 \alpha_k^\Omega&=&\sin \theta_k^\Omega\\
 \beta_l^\Omega&=&\sin(\phi_l[L_\Omega+1])\\
 u_k^\Omega&=&\cos \theta_k^\Omega\\
 v_l&=&\cos \phi_l.
\end{eqnarray}
This form allows for fast numerical evaluation using standard linear algebra routines. Even more interesting, a remarkable simplification occurs when the two subsystems have the same size $L_A=L_B=L/2$: we have $z^A=z^B=z$, and the same goes for the $\alpha_k^\Omega$, $\beta_m^\Omega$ and $u_k^\Omega$, so that we can remove all $\Omega=A,B$ superscripts. In this case we can use the determinant identity (\ref{eq:newcauchy}) of \ref{sec:newcauchy}, and get
\begin{equation}\label{eq:exact_bf}
\boxed{
D_N=2^{N/2} z^N \frac{\prod_{k=1}^{N/2}(\alpha_k)^2 \prod_{l=1}^{N}\beta_l\prod_{1\leq k<l\leq N/2}(u_k-u_l)^2 \prod_{1\leq k<l\leq N}^{(k-l)\,{\rm even}}(v_k-v_l)}{\prod_{k=1}^{N/2}\prod_{l=1}^{N}(u_k-v_l)}}
\end{equation}
Our final result for the LBF therefore reads
\begin{equation}\label{eq:exact_lbf}
 \mathcal{F}=\lbf=-\log \left(D_N^2\right).
\end{equation}
\subsection[\;\;\;\;\;\;\;\;\;\;\;\;\;Extracting the asymptotic expansion]{Extracting the asymptotic expansion}\label{sec:asymptotic}
It is straightforward, if somewhat cumbersome, to extract the asymptotic expansion of Eqs.~(\ref{eq:exact_lbf}) and (\ref{eq:exact_bf}). Let us explain a possible method on the most complicated term, at general filling fraction $\rho=N/L$.
\begin{eqnarray}\label{eq:most_complicated}
-\log B&=&-\log \left(\prod_{k=1}^{N/2}\prod_{l=1}^{N}(u_k-v_l)\right)\\\label{eq:most_complicated2}
&=&-\sum_{k=1}^{N/2}\sum_{l=1}^{N} \log \left|2\sin \left(\frac{\theta_k+\phi_l}{2}\right)\sin \left(\frac{\theta_k-\phi_l}{2}\right)\right|
\end{eqnarray}
Let us now cut (\ref{eq:most_complicated2}) in two 
\begin{equation}
-\log B=-\log B_{\rm reg}-\log B_{\rm lin},
\end{equation}
using $\log (\sin x)=\log \left({\rm sinc}\, x\right)+\log x$. Since $\log \left({\rm sinc} \, x\right)$ is analytic, one can apply twice the Euler-Maclaurin formula
\footnote{$\sum_{i=a}^{b} f(i)\sim \int_a^b f(x)\,dx+\frac{f(a)+f(b)}{2}+\sum_{k=1}^\infty \frac{B_{2k}}{2k!}\left[f^{(2k-1)}(b)-f^{(2k-1)}(a)\right]$ where $f^{(n)}$ is the $n-$th derivative of $f$ and the $B_{2k}$ are the Bernoulli numbers, defined through the exponential generating function $x (e^x-1)^{-1}=\sum_{k=1}^{\infty}\frac{B_k}{k!}x^k$. The first few are given by $B_2=1/6$, $B_4=-1/30$, $B_6=1/42$.} 
at any order, and check that the expansion takes the regular form
\begin{equation}\label{eq:regular}
 -\log B_{\rm reg}=\sum_{k=0}^\infty b_{2-k}^\prime L^{2-k}.
\end{equation}
The equality is meant in the sense of asymptotic series. 
 Having performed this expansion we are left with the ``linearized'' term
\begin{eqnarray}
 B_{{\rm lin}}&=&\prod_{k=1}^{N/2}\prod_{l=1}^{N}(\theta_k+\phi_l)(\theta_k-\phi_l)/4\\
 &=&\left[\frac{\pi }{2(L+1)}\right]^{N(N+1/2)}\;\prod_{i=1}^{N/2} \;\frac{\Gamma\left(N+1+2i-\frac{2i}{L+2}\right)\Gamma\left(N+1-2i+\frac{2i}{L+2}\right)}{(\pi/2)\Gamma\left(1-\frac{2i}{L+2}\right)\Gamma\left(1+\frac{2i}{L+2}\right)}
\end{eqnarray}
$\Gamma$ is Euler's Gamma function. The denominator can be handled at arbitrary order using the Euler-Maclaurin formula, and generates only terms of the form (\ref{eq:regular}). The prefactor is easy. We are then left with
\begin{eqnarray}\label{eq:as_example}
 B_{\rm lin}^\prime&=&\prod_{i=1}^{N/2}\Gamma\left(N+1-2i+\textstyle{\frac{2i}{L+2}}\right)\Gamma\left(N+1+2i-\textstyle{\frac{2i}{L+2}}\right)\\
 &=&\prod_{i=1}^{N/2}\Gamma\left(2i-1+\rho+2\textstyle{\frac{1-\rho-i}{L+2}}\right)\Gamma\left(N+1+2i-\textstyle{\frac{2i}{L+2}}\right)\\\fl
 &=&C\times D,
\end{eqnarray}
with
\begin{eqnarray}
 C&=&\frac{\pi^{-\frac{N}{2}}G(N+\frac{3}{2})G(N+2)G(\frac{N+\rho+1}{2})G(\frac{N+\rho+2}{2})}{2^{-N(N+\rho/2)}G(\frac{N}{2}+2)G(\frac{N+3}{2})G(\frac{\rho+1}{2})G(\frac{\rho+2}{2})}\\
 D&=&\prod_{i=1}^{N/2}\frac{\Gamma\left(2i-1+\rho+2\frac{1-\rho-i}{L+2}\right)\Gamma\left(N+1+2i-\frac{2i}{L+2}\right)}{\Gamma\left(2i-1+\rho\right)\Gamma\left(N+1+2i\right)},
\end{eqnarray}
where we have introduced the Barnes $G$-function. $C$ can be obtained by induction, using the functional relation $G(z+1)=\Gamma(z)G(z)$ and the duplication formula $\Gamma(z)\Gamma(z+1/2)=2^{1-2z}\sqrt{\pi}\,\Gamma(2z)$. This is useful because the asymptotic expansion of $G(z)$ for large $z$ is known:
\begin{equation}\label{eq:barnes_as}
 \log G(z+1)=\frac{1}{2}z^2\log z -\frac{3}{4}z^2+\frac{\log (2\pi)}{2}z-\frac{1}{12}\log z+\frac{1}{12}-\log A+\sum_{k=1}^{\infty}\frac{\nu_k}{z^{2k}},
\end{equation}
where $A$ is the Glaisher-Kinkelin constant. The $\nu_k$ are known but unimportant to us. Hence
\begin{equation}
 -\log C=\left(c_2 L^2+c_1L+c_0\right)\log L +\sum_{k=0}^{\infty} c_{2-k}^\prime L^{2-k}
\end{equation}
The last remaining term is
\begin{equation}
 -\log D=-\sum_{i=1}^{N/2}\log \left(\frac{\Gamma\left(2i-1+\rho+2\frac{1-\rho-i}{L+2}\right)\Gamma\left(N+1+2i-\frac{2i}{L+2}\right)}{\Gamma\left(2i-1+\rho\right)\Gamma\left(N+1+2i\right)}\right)
\end{equation}
We wish to use once again the Euler-Maclaurin formula. 
Because of the $L+2$ denominators, it is much more convenient to work as a function of $M=L+2$. It is also easy to see that the integration will generate a denominator proportional to $M-1$. The integral of $\log \Gamma(u)$ is the generalized polygamma function $\psi^{(-2)}(u)$. Its asymptotic expansion is simply obtained by integrating the Stirling formula
\begin{equation}
 \log \Gamma(z+1)=z\log z-z+\frac{1}{2}\log (2\pi z)+\sum_{k=0}^\infty \frac{a_k}{z^{2k+1}}
\end{equation}
We get 
\begin{equation}
 -(M-1)\log D=\left(d_2M^{2}+d_1 M+d_0+d_{-1}M^{-1}\right)\log M+\sum_{k=0}^{\infty} d_{2-k}^\prime M^{2-k}.
\end{equation}
Notice that the power series prefactor of $\log M$ terminates at $M^{-1}$. However, it does \emph{not} in terms of $L$. Indeed expanding $M^{-1}=(L+2)^{-1}$ and $(M-1)^{-1}=(L+1)^{-1}$ in power series, we get
\begin{equation}\fl \label{eq:someexpansion}
 -\log D=\left(d_2 L+3d_2+d_1+\sum_{k=1}^{\infty} \frac{(-1)^k}{L^k}\left[\left(2^{k-1}-1\right)d_{-1}-d_0-d_1-d_2\right]\right)\log L+\sum_{k=0}^{\infty} d^{\prime\prime}_{2-k}L^{2-k},
\end{equation}
with, in particular, $d_2=\rho^2/2$, $d_1=-3\rho^2/2$, $d_0=(1-6\rho+14\rho^2)/8$ and $d_{-1}=(-1+6\rho-6\rho^2)/6$. Eq.~(\ref{eq:someexpansion}) already gives all the $L^{-k}\log L$ ($k\geq 1$) terms advertized in \ref{sec:summary}.
In principle it is also possible to keep track of the $L^{2-k}$ prefactors, although these become rapidly gruesome. We were able to do so up to order $k=2$. In the end $-\ln B$ admits the asymptotic series
\begin{equation}
 -\ln B= \left(\sum_{k=0}^{\infty}b_{1-k} L^{1-k}\right)\log L +\sum_{k=0}^{\infty} b^\prime_{2-k} L^{2-k} 
\end{equation}
with
\begin{eqnarray}
 b_1&=&0\\
 b_0&=&\left(1+6\rho-6\rho^2\right)/24\\
 b_{-p}&=&(-1)^{p+1}(2^{p+1}-1) \left(1-6\rho+6\rho^2\right)/24\qquad,\qquad p\geq 1\\
 b^\prime_2&=&\frac{1}{2}\rho^2 \log 2+\frac{\zeta(3)-{\rm Cl}_3(2\pi \rho)}{2\pi^2}\\
 b^\prime_1&=&\frac{3(\zeta(3)-{\rm Cl}_3(2\pi\rho))}{2\pi^2}+\frac{(2-3\rho){\rm Cl}_2(2\pi \rho)-3{\rm Cl}_2(\pi \rho)}{2\pi}\\\nonumber
 b^\prime_0&=&\frac{(3-4\rho){\rm Cl}_2(2\pi \rho)-4{\rm Cl}_2(\pi \rho)}{2\pi}+\frac{24\zeta(3)-{\rm Cl}_3(2\pi \rho)}{\pi^2}-\frac{23-78\rho+54\rho^2}{24}\log \sin \pi \rho\\\fl\;
(1-2\rho)\log \sin \frac{\pi\rho}{2}&-&\frac{1}{6}+\frac{3+2\rho-8\rho^2}{4}\log 2+\frac{5-18\rho+6\rho^2}{24}\log \pi+\log\left[A^2 G\left(1+\textstyle{\frac{\rho}{2}}\right)G\left(\textstyle{\frac{1+\rho}{2}}\right)\right].
\end{eqnarray}
Performing the same expansion on all the other terms in (\ref{eq:exact_bf}), we observe that all the terms proportional to $L^2,L\log L,L$ cancel each other, as should be. We finally obtain
\begin{equation}
\mathcal{F}(L)=\sum_{k=0}^{\infty}\frac{\gamma_k\log L+\mu_k}{L^k},
\end{equation}
with the $\gamma_k$ given by
\begin{eqnarray}
 \gamma_0&=&\frac{1}{8}+h_\rho,\\
 \gamma_k&=&\frac{(-1)^{k+1}\left(2^{k+1}-1\right)}{3}\left(\frac{1}{8}-3h_\rho\right)\quad,\quad k\geq 1
\end{eqnarray}
and $\mu_0$ by Eq.~(\ref{eq:mu0}). Notice that it is possible to put the $L^{-k}\log L$ corrections in a more compact form. Introducing the effective length
\begin{equation}
 L_e=\frac{(L+1)(L+2)}{L+2/3},
\end{equation}
the series expansion reads
\begin{equation}
 \mathcal{F}(L_e)=\gamma_0 \log L_e+\mu_0+\gamma_1 \frac{\log L_e}{L_e}+\sum_{k=1}^{\infty} \frac{\mu_k^\prime}{L_e^{k}}.
\end{equation}

%\subsection[\;\;\;\;\;\;\;\;\;\;\;\;\;Asymptotic expansion for general $x$]{Asymptotic expansion for general $x$}
%Combining the CFT prediction with the exact result at $x=1/2$ allows to get the full asymptotic expansion of the LBF at general $x$ and $\rho$. We have
%\begin{eqnarray}\fl\nonumber
% \mathcal{F}(x,L)&=&\left(\frac{1}{8}+h\right)\log L+\left[\left(\frac{1}{24}-h\right)\left(2x+\frac{2}{x}-1\right)+4h\right]\log (1-x)
% +\left[\left(\frac{1}{24}-h\right)\left(1-2x+\frac{2}{1-x}\right)+4h\right]\log x\\
% \fl&+&\mu_0-\frac{1}{3}\log 2+\left(\frac{1}{24}-h\right)\left(\frac{1}{x(1-x)}-1\right)\times \frac{\log L}{L}\;+\;\mathcal{O}\left(\frac{1}{L}\right).
%\end{eqnarray}

\subsection[\;\;\;\;\;\;\;\;\;\;\;\;\;Related examples]{Related examples}
\label{sec:other_examples}
It is possible to derive similar exact formulae in two other geometries, that we present here.
\paragraph{Open-periodic overlap:}
The first possibility is to consider the overlap between two full chains of length $L$, one with open boundary conditions, and the other with periodic boundary conditions. We have
\begin{equation}
 \langle {\rm open}|{\rm periodic}\rangle=\det_{1\leq k,l\leq N} \left(\sqrt{\frac{2}{L(L+1)}}\sin \left[\frac{l\pi}{L+1}\right]
 \frac{\cos\left[\frac{(2k-1-N)\pi}{2L}-\frac{m\pi}{2}\right]}{\cos\left[\frac{(2k-1-N)\pi}{L}\right]-\cos\left[\frac{l\pi}{L+1}\right]}\right)
\end{equation}
This determinant can also be put in the form (\ref{eq:newcauchyform}), using the symmetry $k\to N+1-k$. Evaluating this in explicit form, and performing an asymptotic expansion similar to that described in \ref{sec:asymptotic}, we get
\begin{eqnarray}\nonumber
 -\log \left|\langle {\rm open}|{\rm periodic}\rangle\right|^2&=&\left(\frac{1}{8}+h_\rho\right)\log L+\mu_0-\left(\frac{1}{8}+h_\rho\right)\frac{\log L}{L}+\sum_{k=1}^{\infty}\frac{\mu_k^{\prime\prime}}{L^{k}}\\
 &-&\left(\frac{1}{24}-h_\rho\right)\sum_{k=2}^{\infty} (-1)^{k} \frac{\log L}{L^k}
\end{eqnarray}
\paragraph{Fidelity in a periodic system:}
This was the other situation considered in \cite{Bipartite_fidelity}. When the periodic system of length $L$ is cut into two open chains of size $L/2$, it is also possible to put the fidelity under the form (\ref{eq:newcauchyform}). After asymptotic expansion we find
\begin{equation}
 \tilde{\mathcal{F}}(L)=\left(\frac{1}{4}+2h_\rho\right)\log L-4h_\rho \frac{\log L}{L} +\frac{1-24h_\rho}{6}\sum_{k=2}^{\infty}(-2)^{k-1} \frac{\log L}{L^k}
 +\sum_{k=0}^{\infty}\frac{\mu_k^{\prime\prime\prime}}{L^{k}}
\end{equation}
Notice that the $L^{-1}\log L$ correction vanishes at half-filling, contrary to the other two examples.
\pagebreak
\section*{References}
\bibliography{biblio.bib}{}

\providecommand{\href}[2]{#2}\begingroup\raggedright\begin{thebibliography}{10}

\bibitem{Baxter}
R.~Baxter, ``{Exactly solved models in statistical mechanics, Academic Press,
  London},''  (1982) .

\bibitem{zinnjustin}
J.~Zinn-Justin, {\em Quantum field theory and critical phenomena}, vol.~142.
\newblock Clarendon Press Oxford, 2002.

\bibitem{Affleck_c}
I.~Affleck, ``{Universal term in the free energy at a critical point and the
  conformal anomaly},''
  \href{http://dx.doi.org/10.1103/PhysRevLett.56.746}{{\em Phys. Rev. Lett}
  {\bfseries 56} 746(1986) }.

\bibitem{Cardy_c}
H.~W.~J. Blote, C.~J. L., and M.~P. Nightingale, ``{Conformal invariance, the
  central charge, and universal finite-size amplitudes at criticality},''
  \href{http://dx.doi.org/10.1103/PhysRevLett.56.742}{{\em Phys. Rev. Lett}
  {\bfseries 56} 742(1986) }.

\bibitem{Zamo_c}
A.~Zamolodchikov, ``Irreversibility of the flux of the renormalization group in
  a 2d field theory,'' {\em JETP lett} {\bfseries 43} no.~12, (1986) .

\bibitem{CardyA}
J.~Cardy, ``{Is there a $c$-theorem in four dimensions?},'' {\em Phys. Lett.
  B.} {\bfseries 215} 749(1988) .

\bibitem{KomargodskiSchwimmer}
Z.~Komargodski and A.~Schwimmer, ``On renormalization group flows in four
  dimensions,'' \href{http://dx.doi.org/10.1007/JHEP12(2011)099}{{\em Journal
  of High Energy Physics} {\bfseries 2011} 1(2011) }.

\bibitem{Klebanov}
D.~Jafferis, I.~Klebanov, S.~Pufu, and B.~Safdi, ``Towards the f-theorem: N = 2
  field theories on the three-sphere,''
  \href{http://dx.doi.org/10.1007/JHEP06(2011)102}{{\em Journal of High Energy
  Physics} {\bfseries 2011} 1(2011) },
  \href{http://arxiv.org/abs/1103.1181}{{\ttfamily arXiv:1103.1181}}.

\bibitem{BPZ}
A.~Belavin, A.~Polyakov, and A.~Zamolodchikov, ``Infinite conformal symmetry in
  two-dimensional quantum field theory,''
  \href{http://dx.doi.org/10.1016/0550-3213(84)90052-X}{{\em Nuclear Physics B}
  {\bfseries 241} 333(1984) }.

\bibitem{Holzhey_ee}
C.~Holzhey, F.~Larsen, and F.~Wilczek, ``{Geometric and renormalized entropy in
  conformal field theory},''
  \href{http://dx.doi.org/10.1016/0550-3213(94)90402-2}{{\em Nucl. Phys. B}
  {\bfseries 424} 443(1994) },
  \href{http://arxiv.org/abs/hep-th/9403108}{{\ttfamily arXiv:hep-th/9403108}}.

\bibitem{Vidal_ee}
G.~Vidal, J.~I. Latorre, E.~Rico, and A.~Kitaev, ``{Entanglement in Quantum
  Critical Phenomena},''
  \href{http://dx.doi.org/10.1103/PhysRevLett.90.227902}{{\em Phys. Rev. Lett.}
  {\bfseries 90} 227902(2003) },
  \href{http://arxiv.org/abs/quant-ph/0211074}{{\ttfamily
  arXiv:quant-ph/0211074}}.

\bibitem{CalabreseCardy_ee}
P.~Calabrese and J.~Cardy, ``{Entanglement entropy and quantum field theory},''
  \href{http://dx.doi.org/10.1088/1742-5468}{{\em J. Stat. Mech.} {\bfseries
  P06002} (2004) }, \href{http://arxiv.org/abs/hep-th/0405152}{{\ttfamily
  arXiv:hep-th/0405152}}.

\bibitem{ZanardiPaunkovic}
P.~Zanardi and N.~Paunkovi{\'c}, ``Ground state overlap and quantum phase
  transitions,'' \href{http://dx.doi.org/10.1103/PhysRevE.74.031123}{{\em
  Physical Review E} {\bfseries 74} no.~3, 031123(2006) },
  \href{http://arxiv.org/abs/quant-ph/0512249}{{\ttfamily
  arXiv:quant-ph/0512249}}.

\bibitem{venuti2009universal}
L.~C. Venuti, H.~Saleur, and P.~Zanardi, ``Universal subleading terms in
  ground-state fidelity from boundary conformal field theory,''
  \href{http://dx.doi.org/10.1103/PhysRevB.79.092405}{{\em Physical Review B}
  {\bfseries 79} 092405(2009) },
  \href{http://arxiv.org/abs/0807.0104}{{\ttfamily arXiv:0807.0104}}.

\bibitem{Bipartite_fidelity}
J.~Dubail and J.-M. St{\'e}phan, ``{Universal behavior of a bipartite fidelity
  at quantum criticality},'' \href{http://dx.doi.org/10.1088/1742-5468}{{\em J.
  Stat. Mech.} L03002(2011) }, \href{http://arxiv.org/abs/1010.3716}{{\ttfamily
  arXiv:1010.3716}}.

\bibitem{zanardi2012entanglement}
P.~Zanardi and L.~C. Venuti, ``Entanglement susceptibility: Area laws and
  beyond,'' \href{http://arxiv.org/abs/1205.2507}{{\ttfamily arXiv:1205.2507}}.

\bibitem{CardyPeschel}
J.~Cardy and I.~Peschel, ``{Finite-size dependence of the free energy in
  two-dimensional critical systems},''
  \href{http://dx.doi.org/10.1016/0550-3213(88)90604-9}{{\em Nucl. Phys. B.}
  {\bfseries 300} 377(1988) }.

\bibitem{Rectangle}
R.~Bondesan, J.~Dubail, J.~L. Jacobsen, and H.~Saleur, ``{Conformal boundary
  state for the rectangular geometry},''
  \href{http://dx.doi.org/10.1016/j.nuclphysb.2012.04.021}{{\em Nucl. Phys. B.}
  {\bfseries 862} 553(2012) }, \href{http://arxiv.org/abs/1110.6861}{{\ttfamily
  arXiv:1110.6861}}.

\bibitem{Rectangle2}
R.~Bondesan, J.~L. Jacobsen, and H.~Saleur, ``{Rectangular amplitudes,
  conformal blocks, and applications to loop models},''
  \href{http://dx.doi.org/10.1016/j.nuclphysb.2012.10.018}{{\em Nucl. Phys. B.}
  {\bfseries 867} 913(2013) }, \href{http://arxiv.org/abs/1207.7005}{{\ttfamily
  arXiv:1207.7005}}.

\bibitem{Jesper}
J.~L. Jacobsen, ``{Bulk, surface and corner free-energy series for the
  chromatic polynomial on the square and triangular lattices},''
  \href{http://dx.doi.org/10.1088/1751-8113/43/31/315002}{{\em J. Phys. A:
  Math. Theor.} {\bfseries 43} 315002(2010) },
  \href{http://arxiv.org/abs/1005.3609}{{\ttfamily arXiv:1005.3609}}.

\bibitem{Jesper2}
E.~Vernier and J.~L. Jacobsen, ``{Corner free energies and boundary effects for
  Ising, Potts and fully packed loop models on the square and triangular
  lattices},'' \href{http://dx.doi.org/10.1088/1751-8113/45/4/045003}{{\em J.
  Phys. A: Math. Theor.} {\bfseries 45} 045003(2012) },
  \href{http://arxiv.org/abs/1110.2158}{{\ttfamily arXiv:1110.2158}}.

\bibitem{WuIzmailianGuo}
X.~Wu, N.~Izmailian, and W.~Guo, ``Finite-size behavior of the critical ising
  model on a rectangle with free boundaries,''
  \href{http://dx.doi.org/10.1103/PhysRevE.86.041149}{{\em Phys. Rev. E}
  {\bfseries 86} 041149(2012) },
  \href{http://arxiv.org/abs/1207.4540}{{\ttfamily arXiv:1207.4540}}.

\bibitem{KlebanVassileva}
P.~Kleban and I.~Vassileva, ``Free energy of rectangular domains at
  criticality,'' \href{http://dx.doi.org/10.1088/0305-4470/24/14/027}{{\em J.
  Phys. A: Math. Gen.} {\bfseries 24} 3407(1999) }.

\bibitem{CardyIgloi}
I.~A. Kov\'acs, F.~Igl\'oi, and J.~Cardy, ``Corner contribution to percolation
  cluster numbers,'' \href{http://dx.doi.org/10.1103/PhysRevB.86.214203}{{\em
  Phys. Rev. B} {\bfseries 86} 214203(2012) },
  \href{http://arxiv.org/abs/1210.4671}{{\ttfamily arXiv:1210.4671}}.

\bibitem{Diehl}
H.~Diehl, ``{Field-theoretic approach to critical behaviour at surfaces},''
  {\em Phase transitions and critical phenomena, Ed. Domb and Lebowitz,
  Academic Press, New York} {\bfseries 10} (1986) .

\bibitem{CC_globalquench}
P.~Calabrese and J.~Cardy, ``{Time Dependence of Correlation Functions
  Following a Quantum Quench},''
  \href{http://dx.doi.org/10.1103/PhysRevLett.96.136801}{{\em Phys. Rev. Lett.}
  {\bfseries 96} 136801(2006) },
  \href{http://arxiv.org/abs/cond-mat/0601225}{{\ttfamily
  arXiv:cond-mat/0601225}}.

\bibitem{DRR}
J.~Dubail, N.~Read, and E.~H. Rezayi, ``Edge-state inner products and
  real-space entanglement spectrum of trial quantum hall states,''
  \href{http://dx.doi.org/10.1103/PhysRevB.86.245310}{{\em Phys. Rev. B}
  {\bfseries 86} 245310(2012) },
  \href{http://arxiv.org/abs/1207.7119}{{\ttfamily arXiv:1207.7119}}.

\bibitem{Kondoscreening}
E.~S. Sorensen, M.-S. Chang, N.~Laflorencie, and I.~Affleck, ``{Impurity
  Entanglement Entropy and the Kondo Screening Cloud},''
  \href{http://dx.doi.org/10.1088/1742-5468/2007/01/L01001}{{\em J. Stat.
  Mech.} L01001(2006) },
  \href{http://arxiv.org/abs/cond-mat/0606705}{{\ttfamily
  arXiv:cond-mat/0606705}}.

\bibitem{Eriksson_Johannesson_Kondo}
E.~Eriksson and H.~Johannesson, ``Impurity entanglement entropy in kondo
  systems from conformal field theory,''
  \href{http://dx.doi.org/10.1103/PhysRevB.84.041107}{{\em Phys. Rev. B}
  {\bfseries 84} 041107(2011) }.
  \url{http://link.aps.org/doi/10.1103/PhysRevB.84.041107}.

\bibitem{Corrections_ee2}
E.~Eriksson and H.~Johannesson, ``{Corrections to scaling in entanglement
  entropy from boundary perturbations},''
  \href{http://dx.doi.org/10.1088/1742-5468}{{\em J. Stat. Mech.} P02008(2011)
  }, \href{http://arxiv.org/abs/1011.0448}{{\ttfamily arXiv:1011.0448}}.

\bibitem{EFP}
J.-M. St{\'e}phan, ``{Emptiness formation probability, toeplitz determinants,
  and conformal field theory},'' {\em in preparation} (2013) .

\bibitem{SD_localquench}
J.-M. St{\'e}phan and J.~Dubail, ``{Local quantum quenches in critical
  one-dimensional systems: entanglement, the Loschmidt echo, and light-cone
  effects},'' \href{http://dx.doi.org/10.1088/1742-5468}{{\em J. Stat. Mech.}
  P08019(2011) }, \href{http://arxiv.org/abs/1105.4846}{{\ttfamily
  arXiv:1105.4846}}.

\bibitem{BigYellowBook}
P.~Di~Francesco, P.~Mathieu, and D.~S{\'e}n{\'e}chal, {\em Conformal Field
  Theory}.
\newblock Springer Verlag, 1997.

\bibitem{Cardy1984}
J.~Cardy, ``{Conformal invariance and suface critical behaviour},''
  \href{http://dx.doi.org/10.1016/0550-3213(84)90241-4}{{\em Nucl. Phys. B.}
  {\bfseries 240} 514--532(1984) }.

\bibitem{Cardy1989}
J.~Cardy, ``{Boundary conditions, fusion rules and the Verlinde formula},''
  \href{http://dx.doi.org/10.1016/0550-3213(89)90521-X}{{\em Nucl. Phys. B.}
  {\bfseries 324} 581--596(1989) }.

\bibitem{ahlfors}
L.~V. Ahlfors, {\em Complex analysis: an introduction to the theory of analytic
  functions of one complex variable}.
\newblock 1979.

\bibitem{Weston1}
R.~Weston, ``Correlation functions and the boundary qkz equation in a fractured
  xxz chain,'' \href{http://dx.doi.org/10.1088/1742-5468/2011/12/P12002}{{\em
  J. Stat. Mech.} P12002(2011) },
  \href{http://arxiv.org/abs/1110.2032}{{\ttfamily arXiv:1110.2032}}.

\bibitem{Weston2}
R.~Weston, ``Exact and scaling form of the bipartite fidelity of the infinite
  xxz chain,'' \href{http://dx.doi.org/10.1088/1742-5468/2012/04/L04001}{{\em
  J. Stat. Mech.} L04001(2012) },
  \href{http://arxiv.org/abs/1203.2326}{{\ttfamily arXiv:1203.2326}}.

\bibitem{EggertAffleck}
S.~Eggert and I.~Affleck, ``{Magnetic impurities in half-integer-spin
  Heisenberg antiferromagnetic chains},''
  \href{http://dx.doi.org/10.1103/PhysRevB.46.10866}{{\em Phys. Rev. B}
  {\bfseries 46} 10866(1992) }.

\bibitem{lieb1961two}
E.~Lieb, T.~Schultz, and D.~Mattis, ``{Two soluble models of an
  antiferromagnetic chain},''
  \href{http://dx.doi.org/10.1016/0003-4916(61)90115-4}{{\em Annals of Physics}
  {\bfseries 16} no.~3, 407(1961) }.

\bibitem{PeschelSchotte}
I.~Peschel and K.~D. Schotte, ``{Time correlations in quantum spin chains and
  the X-ray absorption problem},''
  \href{http://dx.doi.org/10.1007/BF01485827}{{\em Zeitschrift f{\"u}r Physik
  B} {\bfseries 54} 305(1984) }.

\bibitem{Turban}
L.~Turban, ``{Exactly solvable spin-1/2 quantum chains with multispin
  interactions},'' \href{http://dx.doi.org/10.1016/0375-9601(84)90751-5}{{\em
  Phys. Lett. A} {\bfseries 104} 435(1984) }.

\bibitem{Cardy_bccoperatorcontent}
J.~Cardy, ``{Effect of boundary conditions on the operator content of
  two-dimensional conformally invariant theories},''
  \href{http://dx.doi.org/10.1016/0550-3213(86)90596-1}{{\em Nucl. Phys. B}
  {\bfseries 275} 200(1986) }.

\bibitem{Fermiedge2}
I.~Affleck, ``{Boundary condition changing operations in conformal field theory
  and condensed matter physics},''
  \href{http://dx.doi.org/10.1016/S0920-5632(97)00411-8}{{\em Nucl. Phys. B:
  Proc. Suppl} {\bfseries 58} 35(1997) },
  \href{http://arxiv.org/abs/hep-th/9611064}{{\ttfamily arXiv:hep-th/9611064}}.

\bibitem{Fermiedge3}
A.~M. Zagoskin and I.~Affleck, ``{Fermi edge singularities: Bound states and
  finite-size effects},'' \href{http://dx.doi.org/10.1088/0305-4470}{{\em J.
  Phys. A: Math. Gen.} {\bfseries 30} 5743(1997) },
  \href{http://arxiv.org/abs/cond-mat/9704248}{{\ttfamily
  arXiv:cond-mat/9704248}}.

\bibitem{XXZ_bethenum2}
S.~Qin, M.~Fabrizio, M.~Oshikawa, and I.~Affleck, ``{Impurity in a Luttinger
  liquid away from half-filling: a numerical study},''
  \href{http://dx.doi.org/10.1103/PhysRevB.56.9766}{{\em Phys. Rev. B.}
  {\bfseries 56} 9766(1997) },
  \href{http://arxiv.org/abs/cond-mat/9705269}{{\ttfamily
  arXiv:cond-mat/9705269}}.

\bibitem{XXZ_bethe1}
N.~M. Bogoliubov, A.~G. Izergin, and V.~E. Korepin, ``{Critical exponents for
  integrable models},''
  \href{http://dx.doi.org/10.1016/0550-3213(86)90579-1}{{\em Nucl. Phys. B.}
  {\bfseries 275} 687(1986) }.

\bibitem{XXZ_bethenum1}
F.~Woynarovich, H.~P. Eckle, and T.~T. Truong, ``{Non-analytic finite-size
  corrections in the one-dimensional Bose gas and Heisenberg chain },''
  \href{http://dx.doi.org/10.1088/0305-4470/22/18/035}{{\em J. Phys. A: Math.
  Gen.} {\bfseries 22} 4027(1989) }.

\bibitem{XXZ_bethenum3}
D.~C. Cabra, A.~Honecker, and P.~Pujol, ``{Magnetization Plateaux in N-Leg Spin
  Ladders},'' \href{http://dx.doi.org/10.1103/PhysRevB.58.6241}{{\em Phys. Rev.
  B.} {\bfseries 58} 6241(1998) },
  \href{http://arxiv.org/abs/cond-mat/9802035}{{\ttfamily
  arXiv:cond-mat/9802035}}.

\bibitem{OshikawaAffleck1}
M.~Oshikawa and I.~Affleck, ``{Defect lines in the Ising model and boundary
  states on orbifolds},''
  \href{http://dx.doi.org/10.1103/PhysRevLett.77.2604}{{\em Phys. Rev. Lett.}
  {\bfseries 77} 2604(1996) },
  \href{http://arxiv.org/abs/hep-th/9606177}{{\ttfamily arXiv:hep-th/9606177}}.

\bibitem{OshikawaAffleck2}
M.~Oshikawa and I.~Affleck, ``{Boundary conformal field theory approach to the
  critical two-dimensional Ising model with a defect line},''
  \href{http://dx.doi.org/10.1016/S0550-3213(97)00219-8}{{\em Nucl. Phys. B}
  {\bfseries 495} 533(1997) },
  \href{http://arxiv.org/abs/cond-mat/9612187}{{\ttfamily
  arXiv:cond-mat/9612187}}.

\bibitem{PhDStephan}
J.-M. St{\'e}phan, {\em {Intrication dans des syst{\`e}mes quantiques {\`a}
  basse dimension}}.
\newblock PhD thesis, 2011.
\newblock
  \url{http://ipht.cea.fr/Docspht//articles/t11/231/public/these-stephan.pdf}.

\bibitem{Franchini}
F.~Franchini and A.~G. Abanov, ``{Asymptotics of Toeplitz determinants and the
  emptiness formation probability for the XY spin chain},''
  \href{http://dx.doi.org/10.1088/0305-4470/38/23/002}{{\em J. Phys. A:
  Math.Gen.} {\bfseries 38} 5069(2005) },
  \href{http://arxiv.org/abs/cond-mat/0502015}{{\ttfamily
  arXiv:cond-mat/0502015}}.

\bibitem{EislerPeschel}
V.~Eisler and I.~Peschel, ``Evolution of entanglement after a local quench,''
  \href{http://dx.doi.org/10.1088/1742-5468/2007/06/P06005}{{\em J. Stat.
  Mech.} P06005(2007) },
  \href{http://arxiv.org/abs/cond-mat/0703379}{{\ttfamily
  arXiv:cond-mat/0703379}}.

\bibitem{EKPP_localquench}
V.~Eisler, D.~Karevski, T.~Platini, and I.~Peschel, ``{Entanglement evolution
  after connecting finite to infinite quantum chains},''
  \href{http://dx.doi.org/10.1088/1742-5468}{{\em J. Stat. Mech.} P01023(2008)
  }, \href{http://arxiv.org/abs/0711.0289}{{\ttfamily arXiv:0711.0289}}.

\bibitem{CC_localquench}
P.~Calabrese and J.~Cardy, ``{Entanglement and correlation functions following
  a local quench: a conformal field theory approach},''
  \href{http://dx.doi.org/10.1088/1742-5468}{{\em J. Stat. Mech.} P10004(2007)
  }, \href{http://arxiv.org/abs/0708.3750}{{\ttfamily arXiv:0708.3750}}.

\bibitem{Igloi}
F.~Igl\'oi, Z.~Szatm\'ari, and Y.-C. Lin, ``Entanglement entropy with localized
  and extended interface defects,''
  \href{http://dx.doi.org/10.1103/PhysRevB.80.024405}{{\em Phys. Rev. B}
  {\bfseries 80} 024405(2009) },
  \href{http://arxiv.org/abs/0903.3740}{{\ttfamily arXiv:0903.3740}}.

\bibitem{ColluraCalabrese}
M.~Collura and P.~Calabrese, ``{Entanglement evolution across defects in
  critical anisotropic Heisenberg chains},''
  \href{http://arxiv.org/abs/1302.4274}{{\ttfamily arXiv:1302.4274}}.

\bibitem{Corrections_ee1}
J.~Cardy and P.~Calabrese, ``{Unusual corrections to scaling in entanglement
  entropy},'' \href{http://dx.doi.org/10.1088/1742-5468}{{\em J. Stat. Mech.}
  P04023(2010) }, \href{http://arxiv.org/abs/1002.4353}{{\ttfamily
  arXiv:1002.4353}}.

\bibitem{Corrections_ee3}
J.~C. Xavier and F.~C. Alcaraz, ``{Finite-size corrections of the entanglement
  entropy of critical quantum chains},''
  \href{http://dx.doi.org/10.1103/PhysRevB.85.024418}{{\em Phys. Rev. B}
  {\bfseries 85} 024418(2012) },
  \href{http://arxiv.org/abs/1111.6577}{{\ttfamily arXiv:1111.6577}}.

\bibitem{Quench}
J.~Cardy, ``{Measuring Entanglement Using Quantum Quenches},''
  \href{http://dx.doi.org/10.1103/PhysRevLett.106.150404}{{\em Phys. Rev.
  Lett.} {\bfseries 106} 150404(2011) },
  \href{http://arxiv.org/abs/1012.5116}{{\ttfamily arXiv:1012.5116}}.

\bibitem{AbaninDemler}
D.~A. Abanin and E.~Demler, ``{Measuring entanglement entropy of a generic
  many-body system with a quantum switch},''
  \href{http://dx.doi.org/10.1103/PhysRevLett.109.020504}{{\em Phys. Rev.
  Lett.} {\bfseries 109} 020504(2012) },
  \href{http://arxiv.org/abs/1204.2819}{{\ttfamily arXiv:1204.2819}}.

\bibitem{Zoller}
A.~J. Daley, H.~Pichler, J.~Schachenmayer, and P.~Zoller, ``Measuring
  entanglement growth in quench dynamics of bosons in an optical lattice,''
  \href{http://dx.doi.org/10.1103/PhysRevLett.109.020505}{{\em Phys. Rev.
  Lett.} {\bfseries 109} 020505(2012) },
  \href{http://arxiv.org/abs/1205.1521}{{\ttfamily arXiv:1205.1521}}.

\bibitem{Doyon}
O.~A. Catro-Alvaredo and B.~Doyon, ``{Permutation operators, entanglement
  entropy, and the XXZ chain in the limit $\Delta \rightarrow -1$},''
  \href{http://dx.doi.org/10.1088/1742-5468/2011/P02001}{{\em J. Stat. Mech.}
  {\bfseries 1102} P02001(2011) },
  \href{http://arxiv.org/abs/1011.4706}{{\ttfamily arXiv:1011.4706}}.

\bibitem{swap}
M.~B. Hastings, I.~Gonz{\'a}lez, A.~B. Kallin, and R.~G. Melko, ``{Measuring
  Renyi Entanglement Entropy in Quantum Monte Carlo Simulations},''
  \href{http://dx.doi.org/10.1103/PhysRevLett.104.157201}{{\em Phys. Rev.
  Lett.} {\bfseries 104} no.~15, 157201(2010) },
  \href{http://arxiv.org/abs/1001.2335}{{\ttfamily arXiv:1001.2335}}.

\bibitem{Anderson}
P.~W. Anderson, ``{Infrared catastrophe in Fermi gases with local scattering
  potentials},'' \href{http://dx.doi.org/10.1103/PhysRevLett.18.1049}{{\em
  Phys. Rev. Lett.} {\bfseries 18} 1049--1051(1967) }.

\end{thebibliography}\endgroup
\bibliographystyle{utphys}

\end{document}